\pdfoutput=1

\documentclass[11pt,twoside,a4paper,cmspaper,final,collab]{cms-tdr}
\def\svnVersion{300876}\def\cmsCernNo{2013-037}\def\cmsCernDate{\today}\def\cmsMessage{Published in the Journal of Instrumentation as \href{http://dx.doi.org/10.1088/1748-0221/10/08/P08010}{\doi{10.1088/1748-0221/10/08/P08010.}}}

\begin{document}\cmsNoteHeader{EGM-14-001}

\hyphenation{had-ron-i-za-tion}
\hyphenation{cal-or-i-me-ter}
\hyphenation{de-vices}
\RCS$Revision: 300630 $
\RCS$HeadURL: svn+ssh://svn.cern.ch/reps/tdr2/papers/EGM-14-001/trunk/EGM-14-001.tex $
\RCS$Id: EGM-14-001.tex 300630 2015-08-19 07:09:34Z seez $
\newlength\cmsFigWidth
\ifthenelse{\boolean{cms@external}}{\setlength\cmsFigWidth{0.85\columnwidth}}{\setlength\cmsFigWidth{0.4\textwidth}}
\ifthenelse{\boolean{cms@external}}{\providecommand{\cmsLeft}{top}}{\providecommand{\cmsLeft}{left}}
\ifthenelse{\boolean{cms@external}}{\providecommand{\cmsRight}{bottom}}{\providecommand{\cmsRight}{right}}

\newcommand{\mgg}{\ensuremath{m_{\gamma\gamma}}\xspace}
\newcommand{\mH}{\ensuremath{m_{\PH}}\xspace}
\newcommand{\ptgg}{\ensuremath{p_{\mathrm{T}}^{\gamma\gamma}}\xspace}
\newcommand{\Hgg}{\ensuremath{\PH\to\Pgg\Pgg}\xspace}
\newcommand{\Zee}{\ensuremath{\cPZ\to\Pep\Pem}\xspace}
\newcommand{\Zmm}{\ensuremath{\cPZ\to\Pgmp\Pgmm}\xspace}
\newcommand{\Zmmg}{\ensuremath{\cPZ\to\Pgmp\Pgmm\Pgg}\xspace}
\newcommand{\HZg}{\ensuremath{\PH\to\cPZ\Pgg}\xspace}
\newcommand{\Wenu}{\ensuremath{\PW\to\Pe\Pgn}\xspace}
\newcommand{\ttH}{\ensuremath{\cPqt\cPaqt\PH}\xspace}
\newcommand{\Xnot}{\,\ensuremath{X_0}\xspace}
\newcommand{\sigee}{\ensuremath{\sigma_{\eta\eta}}\xspace}
\newcommand{\covep}{\ensuremath{q_{\eta\phi}}\xspace}
\newcommand{\Etwo}{\ensuremath{E_{2\times2}/E_{5\times5}}\xspace}
\newcommand{\sige}{\ensuremath{\sigma_{\eta}}\xspace}
\newcommand{\sigp}{\ensuremath{\sigma_{\phi}}\xspace}
\newcommand{\ptg}{\ensuremath{p^{\gamma}_\mathrm{T}}\xspace}
\renewcommand{\PYTHIA}{\textsc{pythia}\xspace}

\cmsNoteHeader{EGM-14-001}

\title{Performance of photon reconstruction and identification with the CMS detector in proton-proton collisions at $\sqrt{s}=8\TeV$}

\date{\today}

\abstract{
A description is provided of the performance of the CMS detector for photon reconstruction and identification
in proton-proton collisions at a centre-of-mass energy of 8\TeV at the CERN LHC.
Details are given on the reconstruction of photons from energy deposits in the electromagnetic calorimeter (ECAL)
and the extraction of photon energy estimates.
The reconstruction of electron tracks from photons that convert to electrons in the CMS tracker is also described,
as is the optimization of the photon energy reconstruction and its accurate modelling in simulation,
in the analysis of the Higgs boson decay into two photons.
In the barrel section of the ECAL, an energy resolution of about 1\% is achieved for unconverted or
late-converting photons from $\PH\to\gamma\gamma$ decays.
Different photon identification methods are discussed and their corresponding
selection efficiencies in data are compared with those found in simulated events.
}

\hypersetup{%
pdfauthor={CMS Collaboration},%
pdftitle={Performance of photon reconstruction and identification with the CMS detector in proton-proton collisions at sqrt(s) = 8 TeV},%
pdfsubject={CMS},%
pdfkeywords={CMS, physics, photon, egamma}}

\maketitle
\newpage
\section{Introduction}
\label{sec:intro}

This paper describes the reconstruction and identification of photons with
the CMS detector~\cite{Chatrchyan:2008aa} in data taken in
proton-proton collisions at $\sqrt{s}=8\TeV$ during the 2012 CERN LHC running period.
Particular emphasis is put on the use of photons in the observation and measurement of the diphoton decay of the Higgs boson~\cite{legacy-paper}.
For this decay mode, the energy resolution has significant impact on the sensitivity of the search and
on the precision of measurements made in the analysis.
The uncertainties related to the photon energy scale are the dominant contributions to the systematic uncertainty in the Higgs boson mass,
$\mH~=~124.70\pm0.31\stat\pm0.15\syst\GeV$, measured in Ref.~\cite{legacy-paper}.
The procedure employed to optimize the photon energy estimation and its accurate
modelling in the simulation is described.
This procedure relies on the large sample of recorded Z boson decays to dielectrons, whose showers are reconstructed as photons, and on simulation to model differences in detector response to electrons and photons.

The reconstruction of photons from the measured energy deposits in the electromagnetic calorimeter
(ECAL)~\cite{ECAL-TDR} and the extraction of a photon energy estimate is described, as well as the association
of the electron tracks to clusters in the ECAL for photons that convert in the tracker.
A large fraction of the energy deposited in the detector by all proton-proton interactions arises from photons
originating in the decay of neutral mesons, and these electromagnetic showers provide a substantial background to signal photons.
The use and interest of photons as signals or signatures in measurements and searches
is therefore mainly focussed on those with high transverse momentum where this background is less severe.
Photon selection methods used for the $\Hgg$ channel and other analyses are described, together with
measurements of the selection efficiency.
The efficiency measured in data is compared with that found in simulated events.

The paper starts with brief descriptions of the CMS detector (Section~\ref{sec:detector}), paying particular attention to geometrical details of the electromagnetic calorimeter that are important for shower reconstruction, and of the data and simulated event samples used (Section~\ref{sec:dat-sim}).
Section~\ref{sec:photon-reco} describes photon reconstruction in CMS: clustering of the shower energy deposited in the ECAL crystals, correction of the cluster energy and fine tuning of the calibration, photon energy resolution, and uncertainties in the photon energy scale.
Section~\ref{sec:photon-conversion} describes the reconstruction of the electron tracks
resulting from photons that undergo conversion before reaching the ECAL.
Section~\ref{sec:photon-id} discusses the separation of prompt photons from energy deposits originating
from the decay of neutral mesons, describing two identification algorithms, and giving results on their performance.
The main results are summarized in Section~\ref{sec:summary}.

\section{CMS detector}
\label{sec:detector}

The central feature of the CMS apparatus is a superconducting solenoid of 6\unit{m} internal diameter, providing a magnetic field of 3.8\unit{T}.
Within the superconducting solenoid volume are a silicon pixel and strip tracker, a lead tungstate crystal electromagnetic calorimeter,
and a brass/scintillator hadron calorimeter (HCAL), each one composed of a barrel and two endcap sections.
Muons are measured in gas-ionization detectors embedded in the steel flux-return yoke outside the solenoid.
Extensive forward calorimetry complements the coverage provided by the barrel and endcap detectors.
A more detailed description of the CMS detector can be found in Ref.~\cite{Chatrchyan:2008aa}.

The pseudorapidity coordinates, $\eta$, of detector elements are measured with respect to
the coordinate system origin at the centre of the detector,
whereas the pseudorapidity of reconstructed particles and jets is measured with respect to the interaction vertex from which they
originate.
The transverse energy, denoted by $\ET$, is defined as the product of energy and $\sin\theta$, with $\theta$ being measured
with respect to the origin of the coordinate system.

Charged-particle trajectories are measured by the silicon pixel and strip tracker, with full azimuthal coverage within $\abs{\eta}<2.5$.
Consisting of 1\,440 silicon pixel detector modules and 15\,148 silicon strip detector modules,
totalling about 10 million silicon strips and 60 million pixels,
the silicon tracker provides an impact parameter resolution of ${\approx}15\mum$ and a transverse momentum, $\pt$, resolution
of about 1.5\% for charged particles with $\pt=100\GeV$~\cite{TrackerPaper}.

The total amount of material between the interaction point and the ECAL,
in terms of radiation lengths ($\Xnot$),
raises from $0.4\Xnot$ close to $\eta=0$ to almost $2\Xnot$ near $\abs{\eta}=1.4$, before falling to about $1.3\Xnot$ around $\abs{\eta}=2.5$.
The probability of photon conversion before reaching the ECAL is thus large and, since the resulting electrons ($\Pep\Pem$ pairs) emit bremsstrahlung
in the material, the electromagnetic shower of some photons starts to develop in the tracker.
The electrons are deflected by the 3.8\unit{T} magnetic field, resulting in multiple electromagnetic showers in the ECAL.

The ECAL is a homogeneous and hermetic calorimeter made of lead tungstate, $\mathrm{PbWO}_4$, scintillating crystals.
The high density ($8.28\unit{g}\unit{cm}^{-3}$), short radiation length (8.9\mm), and small
Moli\`{e}re radius (23\mm) of the $\mathrm{PbWO}_4$ crystals enabled the construction of a compact calorimeter with fine lateral granularity.
The central barrel covers $\abs{\eta}<1.48$ with the inner surface located at a radius of 1290\mm.
The endcaps cover $1.48<\abs{\eta}<3.00$ and are located at $\abs{z}>3154\mm$.
A preshower detector consisting of two planes of silicon sensors interleaved with a total of $3\Xnot$ of lead is located
in front of the endcaps and covers $1.65<\abs{\eta}<2.60$.

The ECAL barrel is made of 61\,200 trapezoidal crystals with front face transverse sections of about $22\times22\mm^2$,
giving a granularity of 0.0174 in $\eta$ and $\phi$.
The crystals have a length of 230\mm  ($25.8\Xnot$).
Each half-barrel is formed by 18 barrel supermodules each covering 20\de\ in $\phi$ and containing $85\times20=1700$ crystals.
The crystals of a half-barrel may be viewed as positioned in a regular rectangular grid in $(\eta,\phi)$ space (which wraps round on itself in $\phi$),
and indexed by $85\times360$ integer pairs.
The supermodules are composed of four modules.
Within the modules there are submodules each containing two rows of five crystals.
The void between adjacent crystals within the same submodule is 350\mum wide.
The void between adjacent crystals in adjacent submodules is 550\mum wide.
The voids between adjacent crystals separated by module and supermodule boundaries are about 6\mm wide.
The module boundaries occur at $\abs{\eta}= 0$, 0.435, 0.783, and 1.131, and the supermodules boundaries
occur every $20\de$ in $\phi$.
The geometry is quasi-projective, with almost all the crystal axes tilted by an angle of $3\de$
with respect to the line from the coordinate origin in both the $\eta$ and $\phi$ directions, and only the void at $\eta=0$ points to the
origin---the $3\de$ tilt relative to the $\eta$ direction is introduced progressively for the first five rings of crystals away from this boundary.

The ECAL endcaps are made of 14\,648 trapezoidal crystals (7324 each) with a front face transverse section
of $28.6\times28.6\mm^2$, and a length of 220\mm ($24.7\Xnot$).
The crystals are grouped in $5\times5$ crystal structural units, with the crystals in adjacent units being separated by a void of 2\mm.
The voids between adjacent crystals within the $5\times5$ units are 350\mum wide.
Each endcap is constructed as two half-disks.
The crystals are installed within a quasi-projective geometry pointing 1300\mm beyond the
centre of the detector, giving tilts of $2\de$ to $8\de$ relative to the direction of the coordinate origin.

\section{Data and simulated event samples}
\label{sec:dat-sim}

The results presented here use data corresponding to an integrated luminosity of 19.7\fbinv taken at a centre-of-mass
energy of 8\TeV.

The Monte Carlo (MC) simulation of the response of the CMS detector employs a detailed description of it,
and uses \GEANTfour version~9.4 (patch 03)~\cite{Agostinelli:2002hh}.
The simulated events include the presence of multiple $\Pp\Pp$ interactions taking place in each bunch crossing
weighted to reproduce the distribution of the number of such interactions in data.
The presence of signals from multiple $\Pp\Pp$ interactions in each recorded event is known as pileup.
Interactions taking place in a preceding or a following bunch crossing, \ie within a window of ${\pm}50\unit{ns}$ around
the triggering bunch crossing, are included.
The interactions used to simulate pileup are generated with
\PYTHIA 6.426~\cite{Sjostrand:2006za}, the same version that is used for other purposes as described below.

Samples of simulated Higgs boson events produced in gluon-gluon and vector-boson fusion processes are
obtained using the next-to-leading-order
matrix-element generator \POWHEG~(version 1.0)~\cite{powheg1,powheg2,powheg3,powheg-ggH,powheg-VBF}
interfaced with $\PYTHIA$.
For the associated Higgs boson production with W and Z bosons, and with $\ttbar$ pairs, \PYTHIA is used alone.

Direct-photon production in $\GAMJET$ processes is simulated using \PYTHIA alone.
Nonresonant diphoton processes involving two prompt photons are simulated
using $\SHERPA\,1.4.2$~\cite{Gleisberg:2008ta}.
The $\SHERPA$ samples are found to give a good description of diphoton continuum events accompanied by one or two jets.
To complete the description of the diphoton background in the $\Hgg$ channel, the remaining processes where one of the photon candidates arises from misidentified jet fragments are simulated with \PYTHIA.
The cross sections for these processes are scaled to match their values measured in
data, using the $K$-factors at 8\TeV that were obtained at 7\TeV~\cite{Chatrchyan:2011qt,CMS-jj}.

Simulated samples of $\Zee$ and $\Zmmg$ events, generated with
\MADGRAPH 5.1~\cite{Alwall:2014hca}, \SHERPA, and \POWHEG~\cite{powheg-Zjj}, are used for some tests,
for comparison with data, and for the derivation of energy scale corrections in data and resolution corrections in the simulations.

The simulation of the ECAL response has been tuned to match
test beam results, and uses a detailed simulation of the 40\unit{MHz} digitization based on an accurate model of the
signal pulse as a function of time.
The effects of electronics noise, fluctuations due to the number of photoelectrons, and the amplification process of the photodetectors
are included.
The simulation also includes a spread of the single-channel response
corresponding to the estimated intercalibration precision, an additional 0.3\%
constant term to account for longitudinal nonuniformity
of light collection, and the few nonresponding channels identified in data.
The measured intercalibration uncertainties range from 0.35\% in most of the barrel, to 0.9\% at the end of the fourth barrel module,
and 1.6\% in most of the region covered by the endcaps with a steep rise for $\abs{\eta}>2.3$.

As a general rule, for the simulation of data taken at 7 and 8\TeV, the response variation with time is not simulated.
However, for the simulation of photon signals and Z-boson background samples used for data-MC
comparisons of the photon energy scale, energy resolution, and photon selection, two refinements are implemented:
the changes in the energy-equivalent noise in the electromagnetic calorimeter during the data-taking period are simulated, and
a significantly increased time window (starting 300\unit{ns} before the triggering bunch crossing) is used to simulate
out-of-time pileup.
These refinements improve the agreement between data and simulated events, seen when
comparing distributions of shower shape variables,
and they provide improved corrections to the energy measurement.

\section{Photon reconstruction}
\label{sec:photon-reco}

Photons for use as signals or signatures in measurements and searches,
rather than for use in the construction of jets or missing transverse energy,
are reconstructed from energy deposits in the ECAL using algorithms that constrain the clusters to
the size and shape expected for electrons and photons with $\pt\gtrsim15\GeV$.
The algorithms do not use any hypothesis as to whether the particle originating from the interaction
point is a photon or an electron, consequently electrons from $\Zee$ events,
for which pure samples with a well defined invariant mass can be selected,
can provide excellent measurements of the photon trigger, reconstruction, and identification efficiencies, and of
the photon energy scale and resolution.
The reconstructed showers are generally limited to a fiducial region excluding the last two crystals at each end of the
barrel ($\abs{\eta}<1.4442$).
The outer circumferences of the endcaps are obscured by services passing between the barrel and the endcaps, and
this area is removed from the fiducial region by excluding the first ring of trigger towers of the endcaps ($\abs{\eta}>1.566$).
The fiducial region terminates at $\abs{\eta}=2.5$ where the tracker coverage ends.

The photon reconstruction proceeds through several steps. Sections~\ref{sec:intercalib}, \ref{sec:photon-clustering}, and~\ref{sec:regression} cover the
intercalibration of the individual channels, the clustering of recorded energy signals resulting
from showers in the calorimeter, and the energy assignment to a cluster.
Section~\ref{sec:fine-tune} discusses the procedure used in the $\Hgg$ analysis to (i) obtain corrections for fine-tuning the photon energy
assignment in data, and (ii) tune the resolution of simulated photons reconstructed in MC
samples.
Section~\ref{sec:resolution} examines the resulting photon resolution in data and in simulation.
Section~\ref{sec:E-uncertainty} discusses the estimation of the uncertainty in the energy scale after implementing the corrections obtained in Section~\ref{sec:fine-tune}.

\subsection{Calibration of individual ECAL channels}\label{sec:intercalib}
The calorimeter signals in data must be
calibrated and corrected for several detector effects~\cite{Chatrchyan:2013dga}.
The crystal transparency is continuously monitored during data taking by measuring the response
to light from a laser system, and the observed changes are corrected for when the events are reconstructed.
The relative calibration of the individual channels is achieved using the $\phi$-symmetry
of the energy deposited by pileup and the underlying event,
the invariant mass measured in two photon decays of $\Pgpz$ and $\Pgh$ mesons,
and the momentum measured by the tracker for isolated electrons from $\PW$
and $\cPZ$ boson decays.

\subsection{Clustering}
\label{sec:photon-clustering}

Clustering of ECAL shower energy is performed on intercalibrated, reconstructed signal amplitudes.
The clustering algorithms collect the energy from radiating electrons and converted photons that gets spread in the $\phi$ direction
by the magnetic field.
These algorithms are described in detail in Ref.~\cite{electron-paper},
and evolved from fixed matrices of $5\times5$ crystals, which provide the best reconstruction of unconverted photons,
by allowing extension of the energy collection in the $\phi$ direction, to form ``superclusters''.
Clusters are built starting from a ``seed crystal'': one containing a signal corresponding to a transverse energy greater than those of all its
immediate neighbours and above a predefined threshold.
In the barrel, where the crystals are arranged in an $(\eta, \phi)$ grid, the clusters have a fixed width of five crystals
centred on the seed crystal, in the $\eta$ direction.
In the $\phi$ direction, adjacent strips of five crystals are added if their summed energy is above another predefined threshold.
Further clusters, aligned in $\eta$, may be seeded and added to the original, ``seed'', cluster if they lie within
an extended $\phi$ window (seed crystal $\pm17$ crystals)---under the control of a further predefined threshold.
Clustering in the endcaps uses fixed matrices of $5\times5$ crystals.
After a seed cluster has been defined, further $5\times5$ matrices are added if
their centroid lies within a small $\eta$ window and within a $\phi$ distance roughly equivalent to the 17 crystals span used in the barrel.
The $5\times5$ matrices are allowed to partially overlap one another.
For unconverted photons, the superclusters resulting from both the barrel and endcap algorithms are usually simply $5\times5$ matrices.

\begin{figure}[hbtp]
  \begin{center}
    \includegraphics[width=0.55\linewidth]{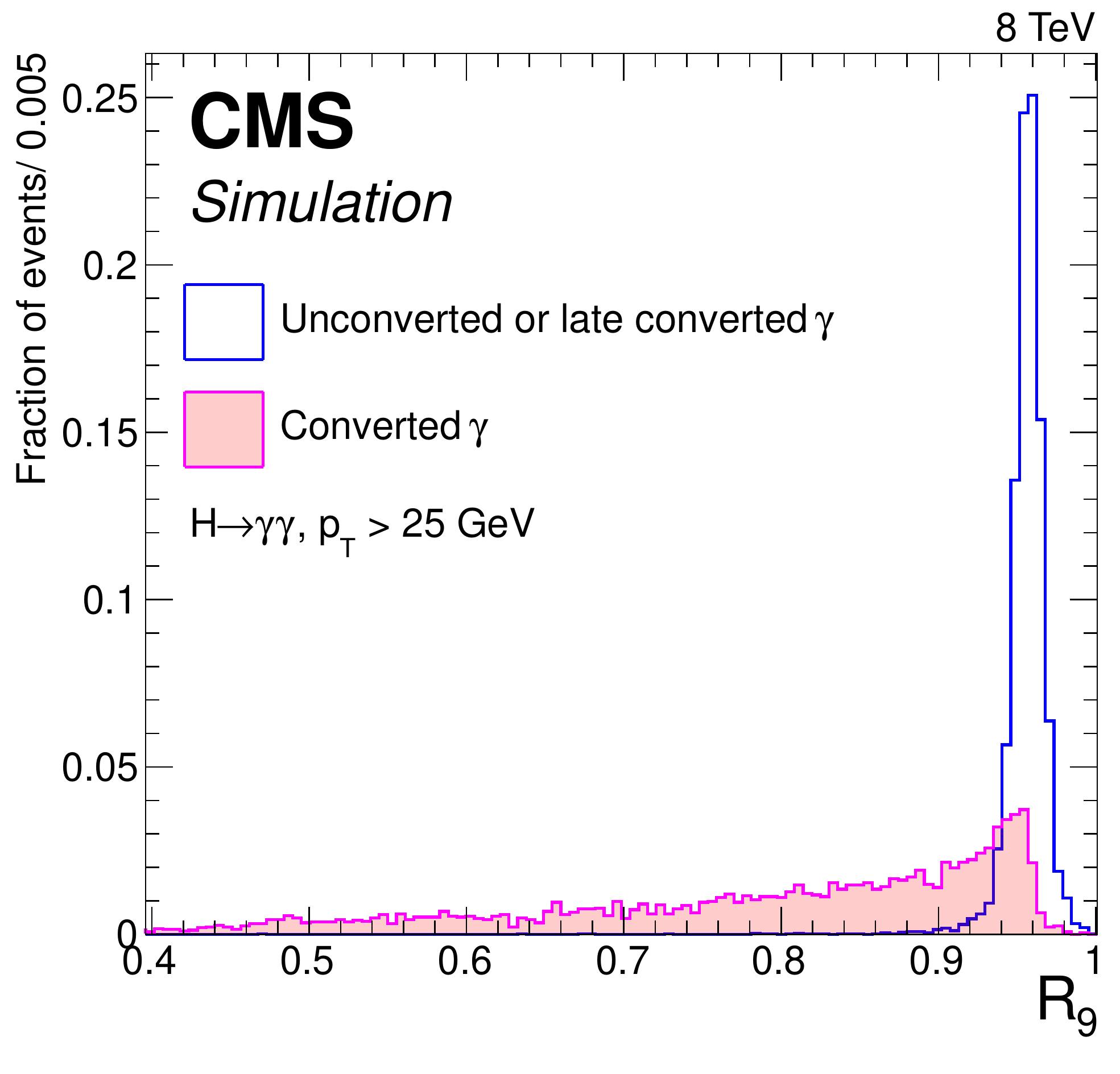}
    \caption{Distributions of the $\RNINE$ variable for photons in the ECAL barrel that convert
in the material of the tracker before a radius of 85\cm (solid filled histogram), and those
that convert later, or do not convert at all before reaching the ECAL (outlined histogram).
  }
    \label{fig:R9}
  \end{center}
\end{figure}

The $\RNINE$~variable is defined as the energy sum of the $3\times3$ crystals centred on the most energetic crystal in the
supercluster divided by the energy of the supercluster.
The showers of photons that convert before reaching the calorimeter have wider transverse profiles and lower values of $\RNINE$
than those of unconverted photons.
Figure~\ref{fig:R9} shows the $\RNINE$ distribution for photons in the ECAL barrel that convert in the material of the tracker before a radius of 85\cm, and those
that convert later, or do not convert at all before reaching the ECAL.
The events are simulated Higgs boson diphoton decays, $\Hgg$, and the photons are required to satisfy $\pt>25\GeV$.
Both histograms are normalized to unity.
Despite being an imperfect indicator of whether a photon converts before reaching the ECAL,
$\RNINE$ is strongly correlated with the photon energy resolution degradation
due to the spreading of showers initiated in the tracker, induced by the magnetic field.
Based on such information, the simplest energy estimation for photons is made by summing the energy in the supercluster for barrel (endcap) photons with
$\RNINE<0.94$ ($\RNINE<0.95$), and summing the energy in a $5\times5$ crystal matrix for the remaining ``unconverted'' photons.
Signals recorded in the preshower detector are included in the region $\abs{\eta} > 1.65$.

\subsection{Correction of cluster energy}
\label{sec:regression}
Significant improvements in energy resolution are obtained by correcting the initial sum of energy deposits forming
the supercluster for the variation of shower containment
in the clustered crystals and for the shower losses of photons that convert before reaching the calorimeter.
The main mechanisms resulting in systematic variation of the fraction of the initial energy contained
in the clustered crystals, ranked in approximate order of increasing severity, are

\renewcommand{\labelenumi}{(\roman{enumi})}
\begin{enumerate}
\item variation of longitudinal depth at which the shower passes through the off-pointing intercrystal voids (causing variation of longitudinal containment),
\item variation of shower location with respect to the lateral granularity (causing variation of lateral containment),
\item variation in the amount of energy absorbed before reaching the ECAL for showers starting before the ECAL,
\item variation in the extent to which the energy of showers starting before the ECAL is clustered, and,
\item if the shower passes through an intermodule void, the variation of longitudinal depth at which the shower passes through it.
\end{enumerate}

The direction of a shower crossing any of the voids between adjacent crystals (detailed in Section~\ref{sec:detector})
makes an angle of about 3\de\ relative to the crystal sides.
The result is a loss of crystal depth seen by the shower.
For a 350\mum void the loss of depth is small: $0.35\mm/\sin{3\de}\approx6.7\mm$ (about $0.75\Xnot$).
For the 6\mm intermodule voids the loss of depth is equal to about half a crystal length.
The effect of such a reduction of calorimeter thickness depends on the shower development at the depth at which the void is crossed.

Corrections as a function of $\eta$, $\ET$, $\RNINE$, and the lateral extension of the cluster in $\phi$, have been obtained from the observed losses in simulated events, and used in many data analyses~\cite{EWK-11-009, EXO-11-038, SMP-13-001, SUS-12-001, SUS-12-014, SUS-13-014, SMP-13-001}.
Corrections have also been extracted from data, using photons from final state radiation in dimuon decays of Z bosons~\cite{EWK-11-009},
although limits on precision start to be severe for $\pt>30\GeV$ since the steeply falling \pt spectrum of these photons limits the number available.

To obtain the best possible energy resolution for the $\Hgg$ analysis~\cite{legacy-paper} the energy measurement is obtained using a multivariate regression technique.
The $\Hgg$ analysis uses events containing pairs of photons with an invariant mass in the range $100<\mgg<180\GeV$,
with the threshold on the lowest $\pt$ photon set at $\mgg/4$.
This corresponds to $\pt>25\GeV$ for all photons used in the analysis, and $\pt\gtrsim30\GeV$ for photons used in the estimation of the mass of the Higgs boson at 125\GeV.
The photon energy response is parameterized by a function with a Gaussian core and two power law tails,
an extended form of the Crystal Ball function~\cite{CrystalBall}.
The regression provides an estimate of the parameters of the function for a single photon, and consequently a prediction of the probability distribution of the ratio of true energy to uncorrected energy.
The corrected photon energy is taken from the most probable value of this distribution.
The input variables are the $\eta$ coordinate of the supercluster, the $\phi$ coordinate of barrel superclusters,
and a collection of shower shape variables: \RNINE~of the supercluster,
the energy weighted $\eta$-width and $\phi$-width of the supercluster,
and the ratio of the energy in the HCAL behind the supercluster and the energy of the supercluster.
In the endcap, the ratio of preshower energy to raw supercluster energy is also included.

Additional information is included for the seed cluster of the supercluster: the relative energy and position of the seed cluster,
the local covariance matrix of the magnitude of the crystal energy signals,
and a number of energy ratios of crystal matrices of different sizes defined with respect to the position of the seed crystal.
These variables provide information on the likelihood and location of a photon
conversion and the degree of showering in the material between the interaction vertex and the calorimeter,
and together with their correlation with the $\eta$ and $\phi$ position of the supercluster,
drive the magnitude of containment correction predicted by the regression.
In the barrel, the $\eta$ and $\phi$ indices of the seed crystal, as well as the position of the seed cluster with respect to the seed crystal are also included.
These variables, together with the seed cluster energy ratios, provide information on the amount of energy that is likely to be contained in the cluster,
or lost in the intermodule voids, and drive the corrections for local containment predicted by the regression.
Although the variations of local containment and the losses due to showering that starts in the tracker material are different effects,
the corrections are allowed to be correlated in the regression to account for the fact that a showering photon is not incident on the ECAL at a single point,
and is consequently less affected by variations of local containment.

Finally, the number of primary vertices and the median transverse energy density $\rho$~\cite{Cacciari:2007fd} in the event are included
in order to allow for the correction of residual systematic effects due to the average amount of pileup in the event.

\begin{figure}[hbtp]
\begin{center}
 \includegraphics[width=0.49\textwidth]{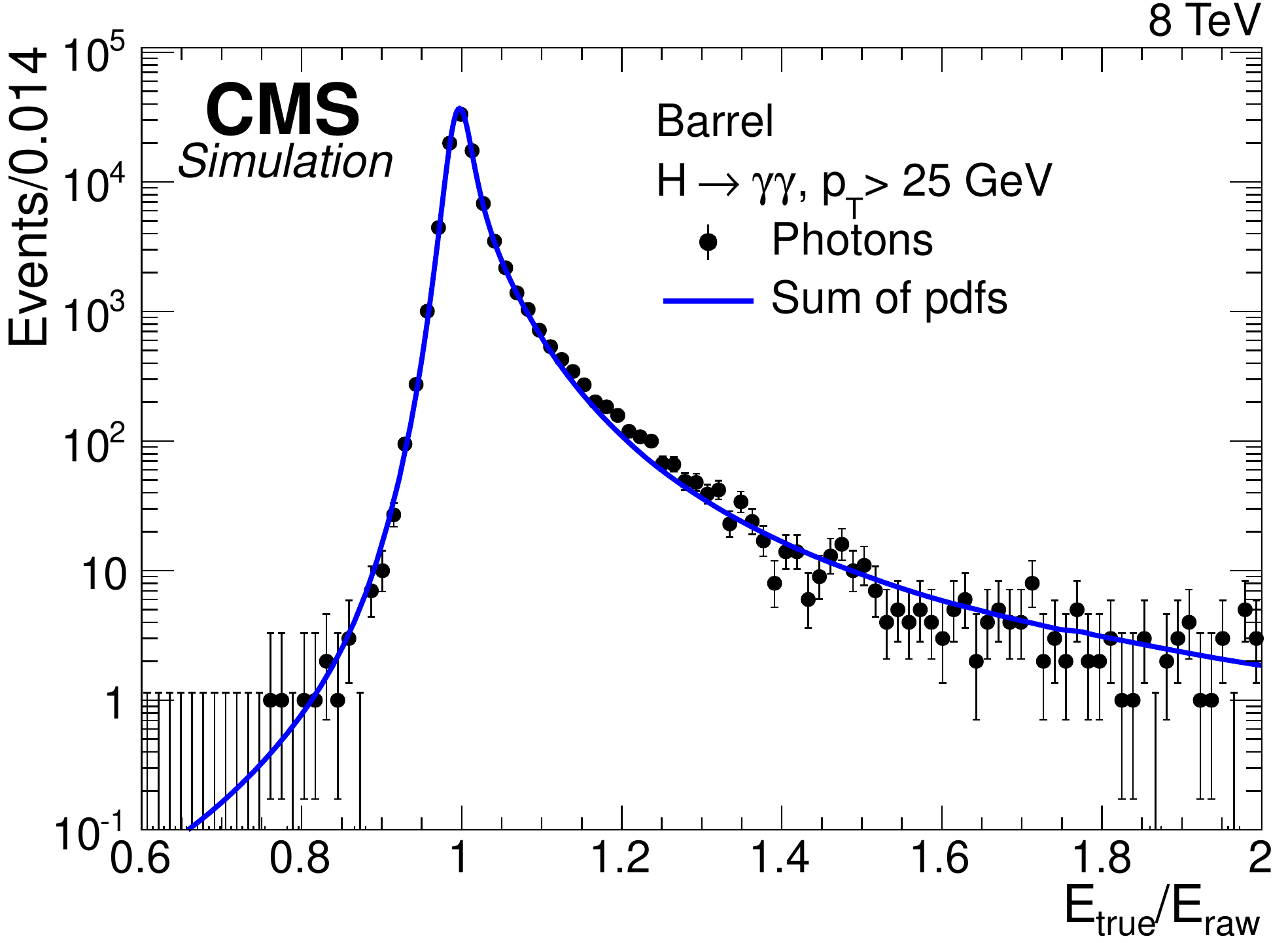}
 \includegraphics[width=0.49\textwidth]{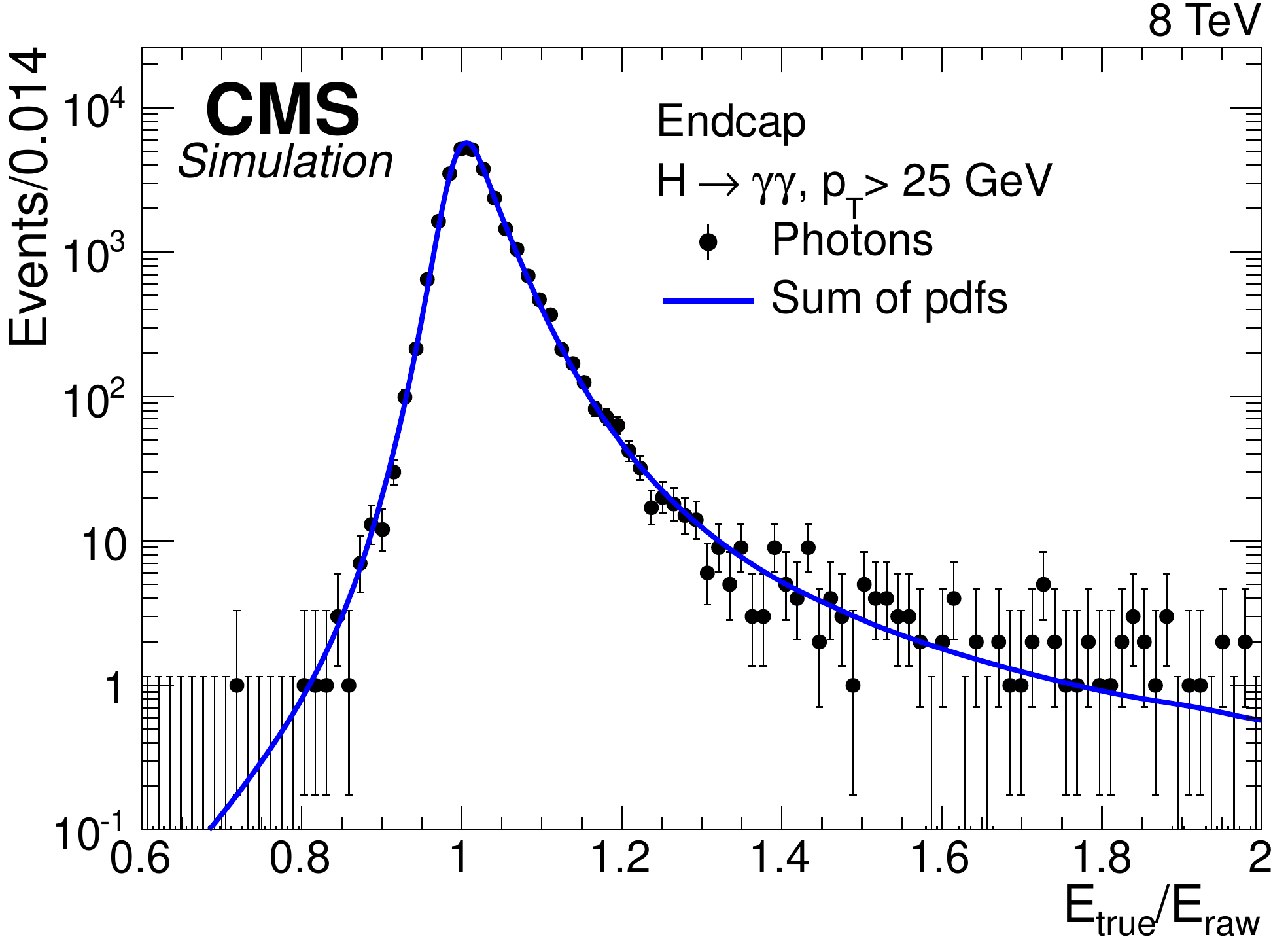}
 \caption{Comparison of the distribution of the inverse response, $E_\text{true}/E_\text{raw}$, in simulated events (points with error bars) with the sum of the
pdfs predicted by the regression (curve).
The comparison is made using a set of simulated photons independent of the training sample, in the (left) ECAL barrel and (right) endcap.}
 \label{fig:edist}
\end{center}
\end{figure}

The semiparametric regression is trained to predict the true energy of the photon, $E_\text{true}$,
given the uncorrected supercluster energy.
The uncorrected energy, $E_\text{raw}$, is taken as the sum of individual crystal energies in a supercluster.
After training, the regression predicts the full probability density function (pdf) for the inverse response, $E_\text{true}/E_\text{raw}$, for each individual photon.
In Fig.~\ref{fig:edist} the sum of predicted distributions for photons with $\pt>25\GeV$ in simulated $\Hgg$ events
is compared to the observed distribution of $E_\text{true}/E_\text{raw}$.
The agreement is excellent, although there are deviations, \eg in the barrel at $E_\text{true}/E_\text{raw}\approx1.2$,
that are larger than can be explained by the statistical uncertainties, and although at $E_\text{true}/E_\text{raw}\approx1.2$ the probability is down by more than two orders of magnitude from the peak
the deviation points to the existence of systematic effects in the event-by-event estimate of the tails of the energy response.
The prediction of the pdf for the inverse response is used in the $\Hgg$ analysis to estimate the mass resolution of individual diphoton systems, which assists in the classification of diphoton events, and is shown here for information.
The energy of photon superclusters is taken to be the most probable value of the pdf, and the performance of this specific assignment, which is probed by the assessment of the resolution in Section 4.5, is therefore independent of the details of the pdf.

\subsection{Fine tuning of calibration and simulated resolution}
\label{sec:fine-tune}
In the $\Hgg$ analysis the final calibration of the energy measurement in data
and the modelling of the energy resolution in simulation were fine-tuned.
Electron showers from rather pure samples (the background contribution is $<$0.1\%) of $\cPZ$
bosons decaying to electrons were reconstructed as photons,
using only the information in the ECAL and without using any information from the tracker.
The dielectron invariant mass was then calculated using the vertex position obtained from the electron tracks,
and its distribution compared to that obtained in simulated events.

The corrections required are small.
They comprise a correction to the energy scale for the data, and a correction
to the energy resolution of the MC simulation (achieved by
adding a Gaussian distributed random contribution to the energy reconstructed in simulated events).
Before the fine-tuning the data have already been corrected for variations of crystal transparency, and the individual crystals have been intercalibrated.
The simulation of the showers in the ECAL includes these uncertainties.
The increase of the energy-equivalent noise during the data-taking period is also simulated.
The noise variation is due to a gradual increase of the leakage current in the silicon avalanche
photodiodes used in the ECAL barrel region, and due to response loss in
the endcap, with the amount of variation depending on $\eta$.

Three explanations have been suggested for the need of an additional smearing of the energy estimate
in simulated events to achieve complete agreement with the data.
The slightly worse energy resolution may be explained by

\renewcommand{\labelenumi}{(\roman{enumi})}
\begin{enumerate}
\item the presence of more tracker material in the detector, between the interaction point and the ECAL, than in the simulation,
\item underestimation of the uncertainty in the individual crystal calibration---although it would be difficult to reconcile a significant underestimation with the fact that the individual crystal calibration uncertainties have been obtained by detailed comparisons among different methods of intercalibration,
\item residual differences between the actual ECAL geometry and the one implemented in the simulation
so that the energy correction estimates, obtained by multivariate regression from simulated events, are suboptimal for data.
\end{enumerate}

Measurements (discussed in Section~\ref{sec:E-uncertainty}) show that there is, indeed, more tracker material present in the detector
than is simulated, and this results in worse energy resolution for photons that convert in the tracker, and an increase in their number.
This fact, however, does not account for all the observed resolution discrepancies, which include the need to worsen the simulated
resolution of showers for which the $\RNINE$ variable has a high value (corresponding to photons that convert late or not at all).
The other two factors listed above represent further contributions in addition to that from mismodelling of tracker material, although their
relative magnitude is not known~\cite{Chatrchyan:2013dga}.
While additional intercalibration errors would increase the constant term in the fractional energy resolution,
the contributions of the other effects have an energy dependence.
As described below, the applied smearing is allowed to have an energy-dependent component.

The supercluster energy scale is tuned and corrected by varying the scale
in the data to match that observed in simulated events.
Two procedures have been used to obtain these corrections: the
``fit method'' and the ``smearing method''.
The fit method uses an analytic fit to the Z boson invariant mass peak, with a convolution of
a Breit--Wigner distribution with a Crystal Ball function.
Distributions obtained from data and from simulated events are fitted separately and the results are compared to extract a scale offset.
The Breit--Wigner width is fixed to that of the Z boson: $\Gamma_\cPZ=2.495\GeV$~\cite{Agashe:2014kda}.
The parameters of the Crystal Ball function, which gives a reasonable description of the calorimeter
resolution effects and of bremsstrahlung losses in front of the calorimeter,
are left free in the fit.
The smearing method uses the simulated Z-boson invariant mass shape as a probability density function in a maximum likelihood fit.
All the known detector effects, reconstruction inefficiencies, and the Z-boson kinematics are taken into account in the simulation.
The residual discrepancy between data and simulation is described by an energy smearing function.
A Gaussian smearing applied to the simulated response has been found to be adequate to describe the data in all the categories of events examined.
A larger number of electron shower categories can be handled by the smearing method as compared to the fit method.

The procedure implemented to fine-tune the energy scale has three steps for the barrel, and two steps for the endcap calorimeters.
In each step, the parameters defining the scale and the width are both allowed to float in the fit, and corrections to the scale are extracted. Only in the final step, the third step for the barrel and the second step for the endcaps, are energy smearing corrections extracted for application to simulated events.

\begin{figure}[hbtp]
  \begin{center}
    \includegraphics[width=0.8\linewidth]{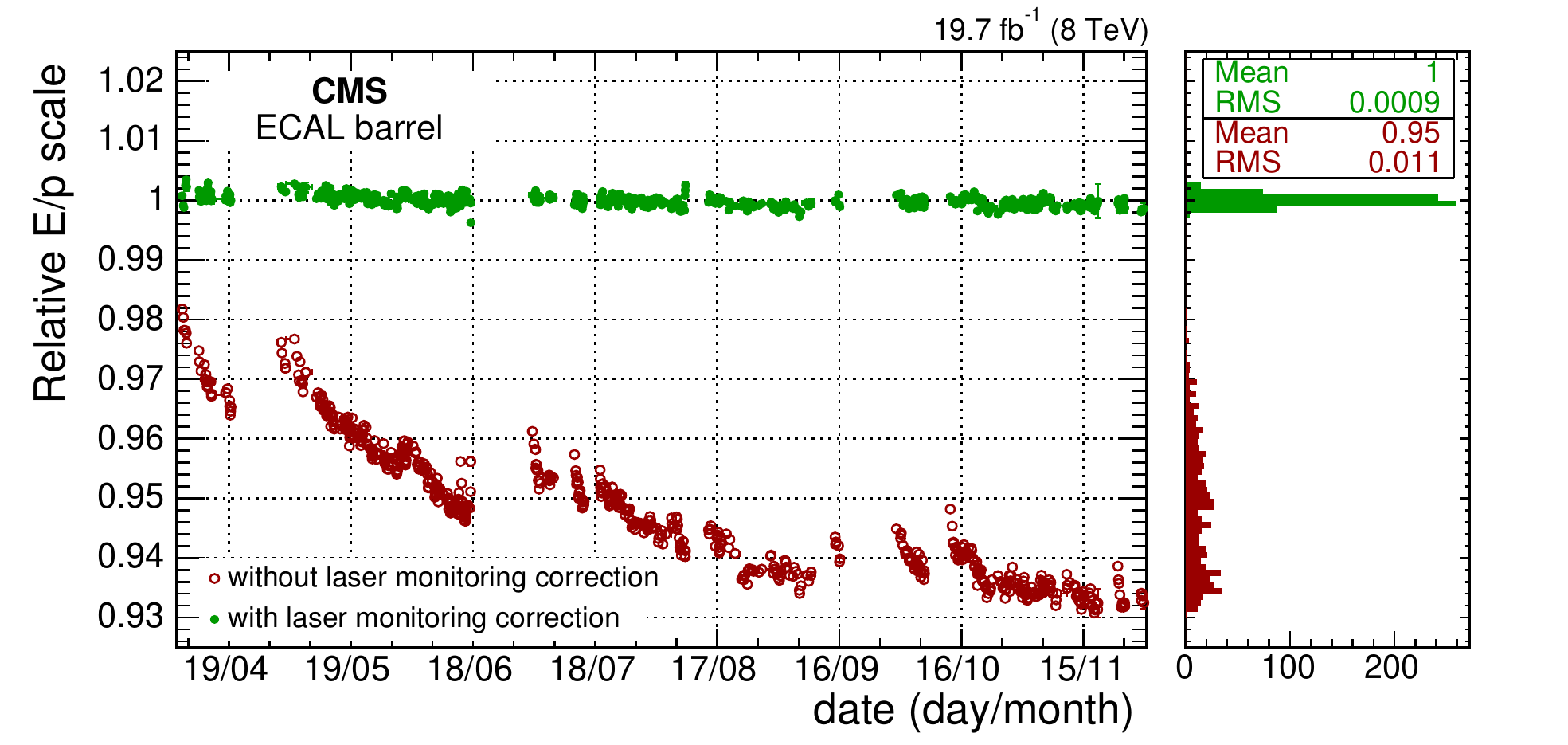}
    \includegraphics[width=0.8\linewidth]{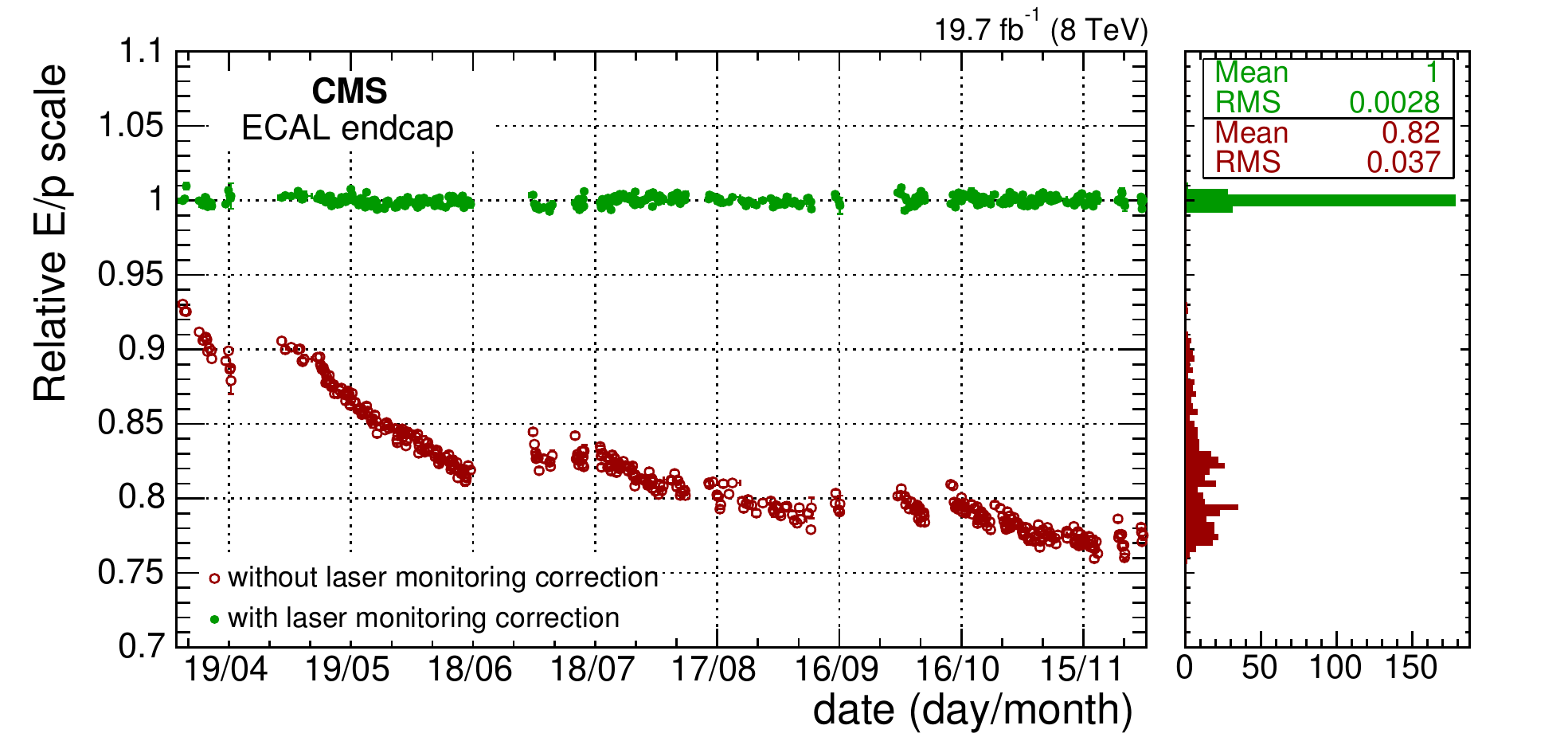}
    \caption{Ratio of the energy measured by the ECAL over the momentum measured by the tracker, $E/p$, for electrons selected from $\Wenu$ decays, as a function of the date at which they were recorded. The ratio is shown both before (red points), and after (green points), the application of transparency corrections obtained from the laser monitoring system, and for both the barrel (upper plot) and the endcaps (lower plot). Histograms of the values of the measured points, together with their mean and RMS values are shown beside the main plots.
  }
    \label{fig:Eop}
  \end{center}
\end{figure}

The first step corrects for possible time dependencies during data taking by extracting, with the fit method, the scale correction to be applied to the data
for each data-taking epoch (51 epochs defined by ranges of run numbers), and for each region in absolute pseudorapidity (4 bins, two in the barrel and two in the endcaps).
This step was originally introduced to account for possible imperfections in the transparency corrections.
However the transparency corrections obtained from the laser monitoring system during 8\TeV data taking are of quality such that there is very little variation to correct.
This can be seen from Fig.~\ref{fig:Eop}, which shows the ratio of the energy measured by the ECAL over the momentum measured by the tracker, $E/p$, for electrons selected from $\Wenu$ decays, as a function of the date at which they were recorded.
The magnitudes of the energy scale corrections extracted in the first step of the fine-tuning procedure are thus small, generally $<0.1\%$ in the barrel and $<0.2\%$ in the endcaps.

The second step derives corrections for effects mainly related to the material in front of the calorimeter, and uses the smearing method. Showers are classified in two $\RNINE$ bins in each of two barrel and two endcap pseudorapidity regions, yielding eight shower categories.
Combining different pairs of shower categories, 36 $\Zee$ invariant mass distributions are constructed for both data and simulated events.
The shower energies in simulated events are modified by applying a Gaussian multiplicative random factor with a mean value $1+\Delta P$ and a standard deviation $\Delta\sigma$.
The method maximizes the likelihood of the fit between the invariant mass distributions as a function of the 16 parameters ($\Delta P$ and $\Delta \sigma$ for each shower category), for the full $\Zee$ data sample, including events where the two showers are in different categories. The energy scale discrepancies found in this step are shown in Table~\ref{tab:scalecor} together with their uncertainties.
The corrections that must be applied to the data are the reciprocals of these values.

\begin{table}[hbtp]
    \caption{\label{tab:scalecor}Energy scale discrepancies, and associated statistical uncertainties, found in the second step of
the fine-tuning procedure. The corrections that must be applied to the data are the reciprocals of these values.
      }
  \begin{center}
    \begin{tabular}{l l|c c}
\multicolumn{2}{l|}{Category} & Scale deviation & Uncertainty \\
\hline
$|\eta| < 1$, & $\RNINE\geq0.94$ & 1.0021 & 0.42\ten{-4} \\
& $\RNINE<0.94$ & 0.9993 & 0.33\ten{-4} \\
\hline
$1< |\eta| < 1.44$, & $\RNINE\geq0.94$ & 1.0097 & 2.06\ten{-4} \\
& $\RNINE<0.94$ & 0.9987 & 0.63\ten{-4} \\
\hline
$1.57< |\eta| < 2$, & $\RNINE\geq0.94$ & 1.0058 & 2.27\ten{-4} \\
& $\RNINE<0.94$ & 0.9989 & 1.05\ten{-4} \\
\hline
$2< |\eta| < 2.5$, & $\RNINE\geq0.94$ & 1.0023 & 1.26\ten{-4} \\
& $\RNINE<0.94$ & 0.9973 & 1.52\ten{-4} \\
    \end{tabular}
    \end{center}
\end{table}

\begin{figure}[hbtp]
  \begin{center}
    \includegraphics[width=0.65\linewidth]{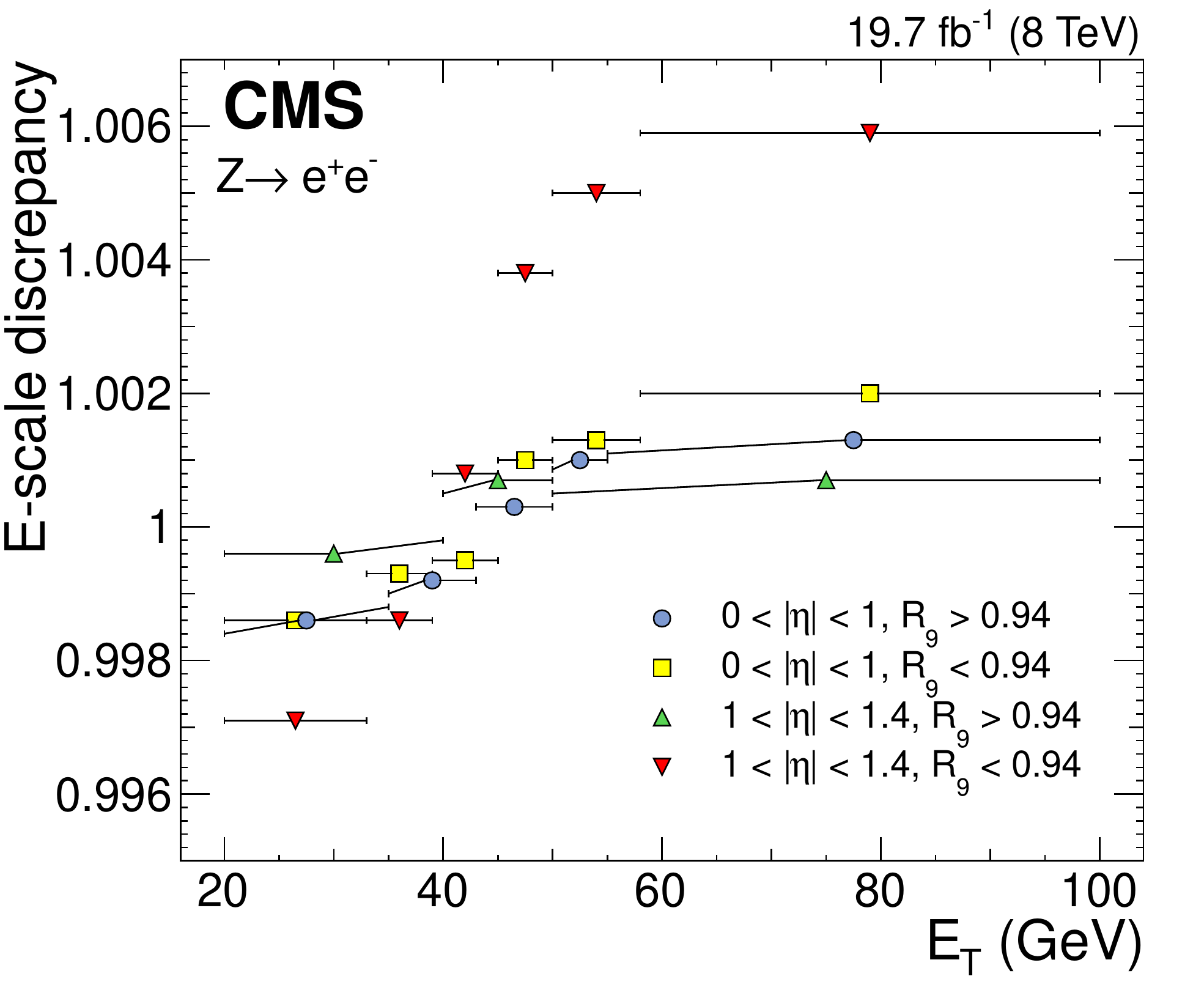}
    \caption{Residual discrepancies in the photon energy scale obtained for the barrel in the final step of the fine-tuning procedure, as a function of $\ET$,
                  for different $\eta$ and $\RNINE$ categories.
                  The statistical uncertainties in these values are negligible. The horizontal error bars indicate the ranges of the $\ET$ bins.
                  The reciprocals of these values are applied as corrections to the energy scale.
                  Some of the error bars have been deflected vertically to avoid overlap with others.
  }
    \label{fig:et-corr}
  \end{center}
\end{figure}

The large $\Zee$ data sample provides sufficient statistical precision for the third step to be performed in the barrel.
This step introduces $\ET$-dependent corrections to the energy scale using 20 bins defined by ranges in $|\eta|$, $\RNINE$, and $\ET$ using the smearing method as in the second step.
In this step the smearing procedure is iterated because the value of the corrections applied can change the $\ET$ bin into which a photon falls. Convergence is achieved after three iterations.
The residual discrepancies measured in this final step are shown, as a function of $\ET$, in Fig.~\ref{fig:et-corr}, and their reciprocals are applied as corrections,
with the value for the highest $\ET$ bin being used for photons with $\ET>100\GeV$.
It can be seen from the figure that the largest corrections obtained in the third and final step are for photons with $\RNINE<0.94$ and $|\eta|>1$.

The energy scale corrections finally applied to the data are the product of the corrections extracted in the steps described above.
The smearing to be applied to the simulated energy resolution, extracted in the second step for the endcaps and in the third step for the barrel,
is modelled by an amplitude and a mixing angle specifying the sharing of this amplitude between a constant term and a $1/\sqrt{E}$ term, providing thereby an extra degree of freedom to the energy resolution uncertainty.
The uncertainties and correlations from the fit contribute to the systematic uncertainty in the energy resolution.
In the endcaps, it is not possible to determine the sharing between a constant and energy dependent term, and therefore the smearing is taken to be
constant, not varying with energy.
The corrections to the resolution of the simulated photons range from $\approx$ 0.7 (1)\% to 1 (2)\% in the barrel for high (low) $\RNINE$, respectively, and from 1.6 to 2.0\% in the endcaps.
In the barrel, the uncertainties in these values are about 10\% of the values themselves.
In the endcaps the uncertainties are about 15\% for the two most relevant photon categories, and up to 50\% for the categories which contribute few event to the $\Hgg$ analysis.
The uncertainties are assessed by (i) examining the variation of the $\RNINE$ distribution as a function of $\eta$ and comparing it to what is observed for photons,
(ii) changing the $\RNINE$ value used for categorization, (iii) using an energy estimate for the electron showers based on an electron-trained regression rather than the photon regression, (iv) changing the $\pt$ threshold of the sample used, and (v) changing the identification criteria used to select the electrons.
The effect of these systematic uncertainties on the Higgs boson mass determination is ${<}10\MeV$, and they have little impact ($<1\%$) on the significance of the signal.

\subsection{Photon energy resolution}
\label{sec:resolution}

\begin{figure}[hbtp]
  \begin{center}
    \includegraphics[width=0.99\linewidth]{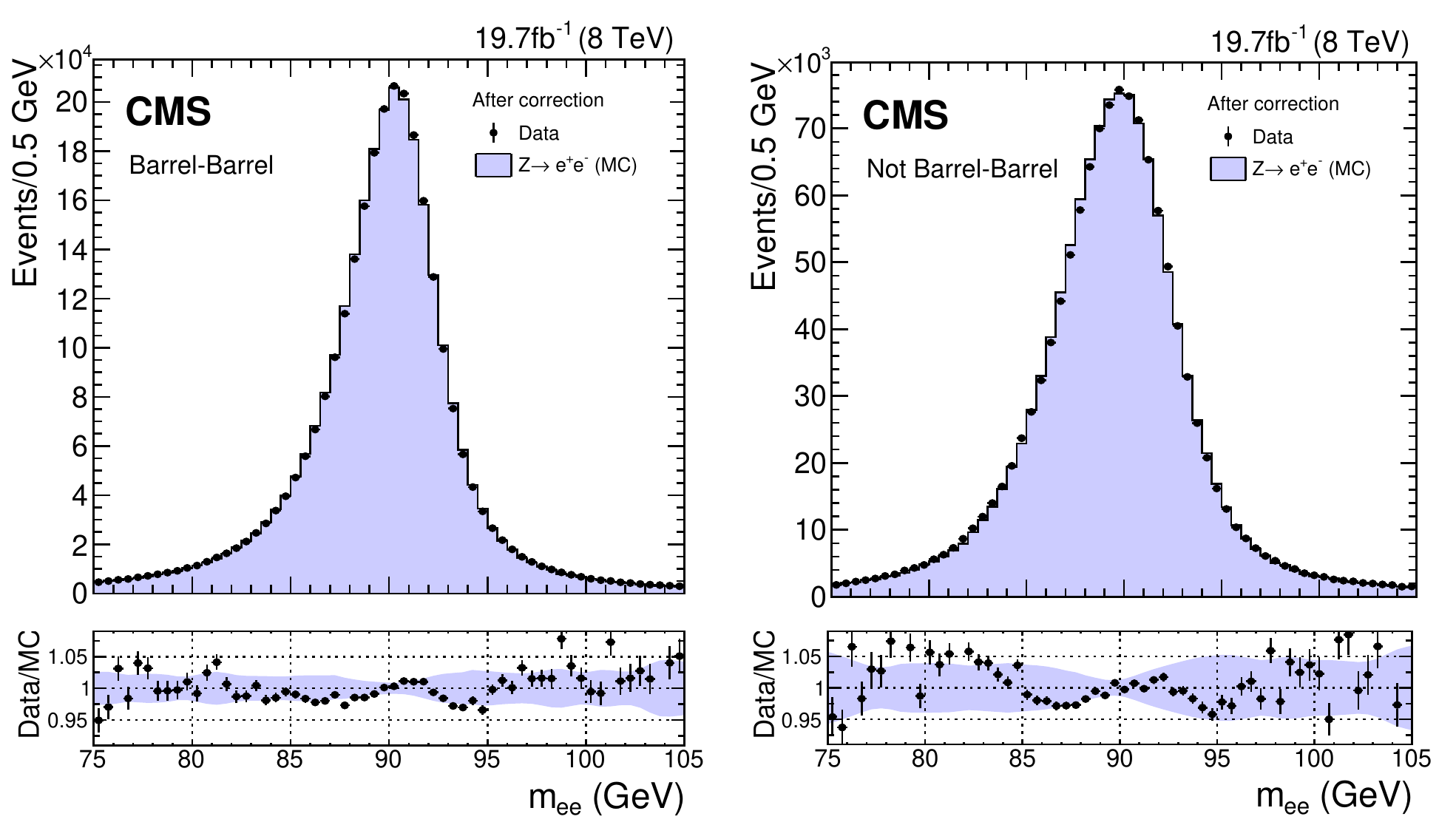}
    \caption{
Reconstructed invariant mass distribution of electron pairs in $\Zee$ events in data (points) and in simulation (histogram).
The electrons are reconstructed as photons and the full set of photon corrections and smearings are applied.
The comparison is shown for (left) events with both showers in the barrel and (right) the remaining events.
For each bin, the ratio of the number of events in data to the number of simulated events
is shown in the panels beneath the main plot.
The band shows the systematic uncertainty in the ratio originating in the systematic uncertainty in the simulated energy resolution, and in the data energy scale.
}
    \label{fig:z-mass}
  \end{center}
\end{figure}

Figure~\ref{fig:z-mass} shows the electron pair invariant mass reconstructed in $\Zee$ events in the 8\TeV data
and simulated events where the electrons are reconstructed as photons, and the full set of
photon corrections and smearings is applied.
The resulting distributions are shown separately for the case where both showers are in the barrel, and for the case
where at least one of the showers is in an endcap.
The distributions of simulated events are normalized to match the distributions in data.
In the panels beneath the main plots, the ratio of the number of events in data to the number of simulated events
in each bin is shown, together with a band obtained by propagating the uncertainties in the simulated energy resolution, and the energy scale in data, to
the dielectron masses obtained.
There is excellent agreement between the simulation and data in the cores of the distributions.
A slight discrepancy is present in the low-mass tail in the endcaps, where the Gaussian smearing
cannot account for some noticeable non-Gaussian effects.
Since the electron showers are reconstructed as photons, the mass peaks do not appear at the true Z-boson mass, both
in data and in the simulation.
This is because the fraction of the original particle energy contained in a supercluster is, on average, a little smaller for
electrons than for photons, and consequently the photon energy regression imperfectly estimates the energy of electron showers.
With respect to the uncorrected distributions, the corrections to the data shift the peak by about $-0.5\GeV$
for the case where both the showers are in the barrel, and by about $-1\GeV$ if either of the showers is in an endcap.
In addition, the distributions obtained from data are slightly narrower after the corrections.
The distributions for the simulated events after the correction procedure are wider, because of the applied smearing.

\begin{figure}[hbtp]
\begin{center}
 \includegraphics[width=0.65\textwidth]{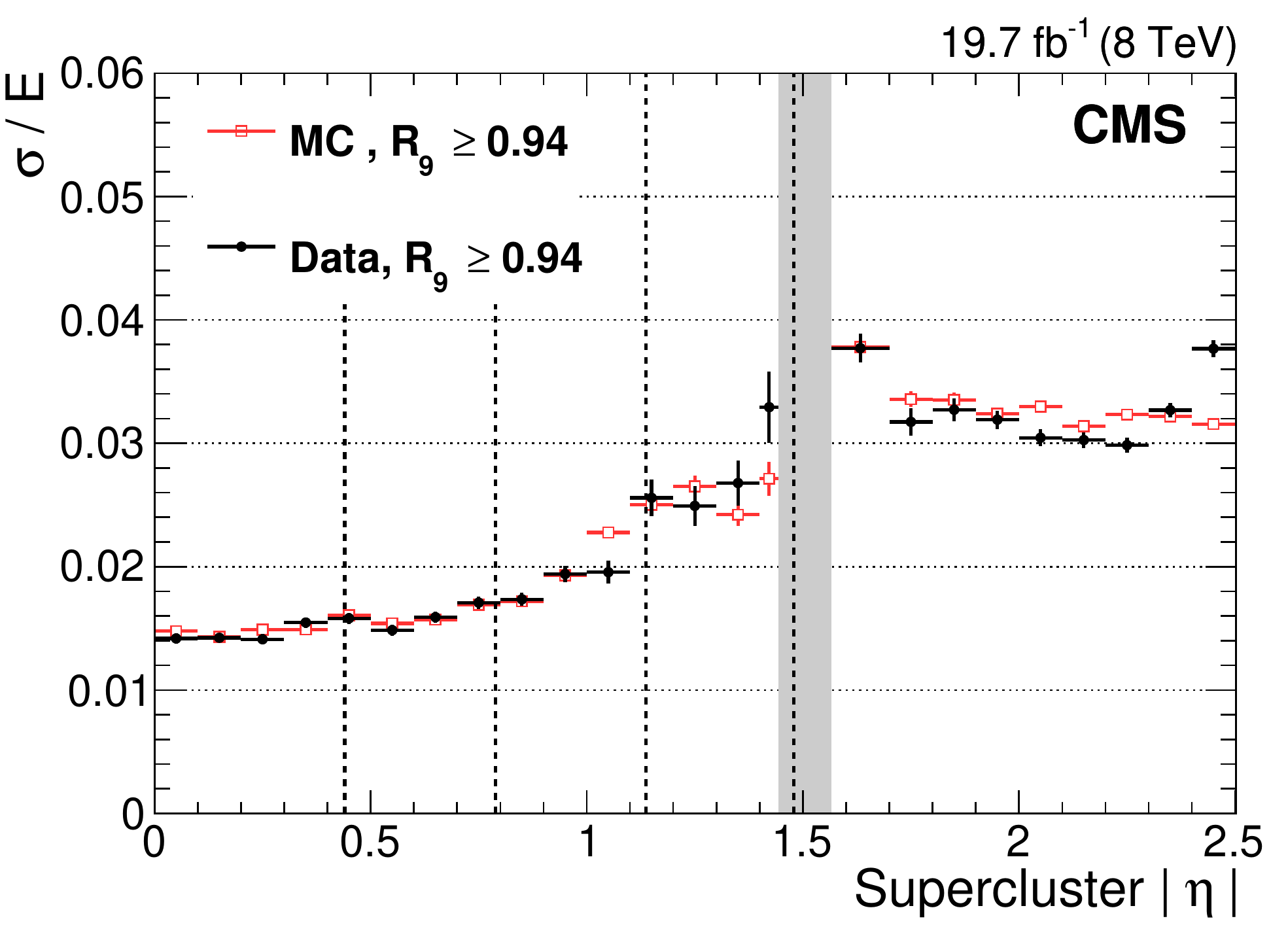}
 \includegraphics[width=0.65\textwidth]{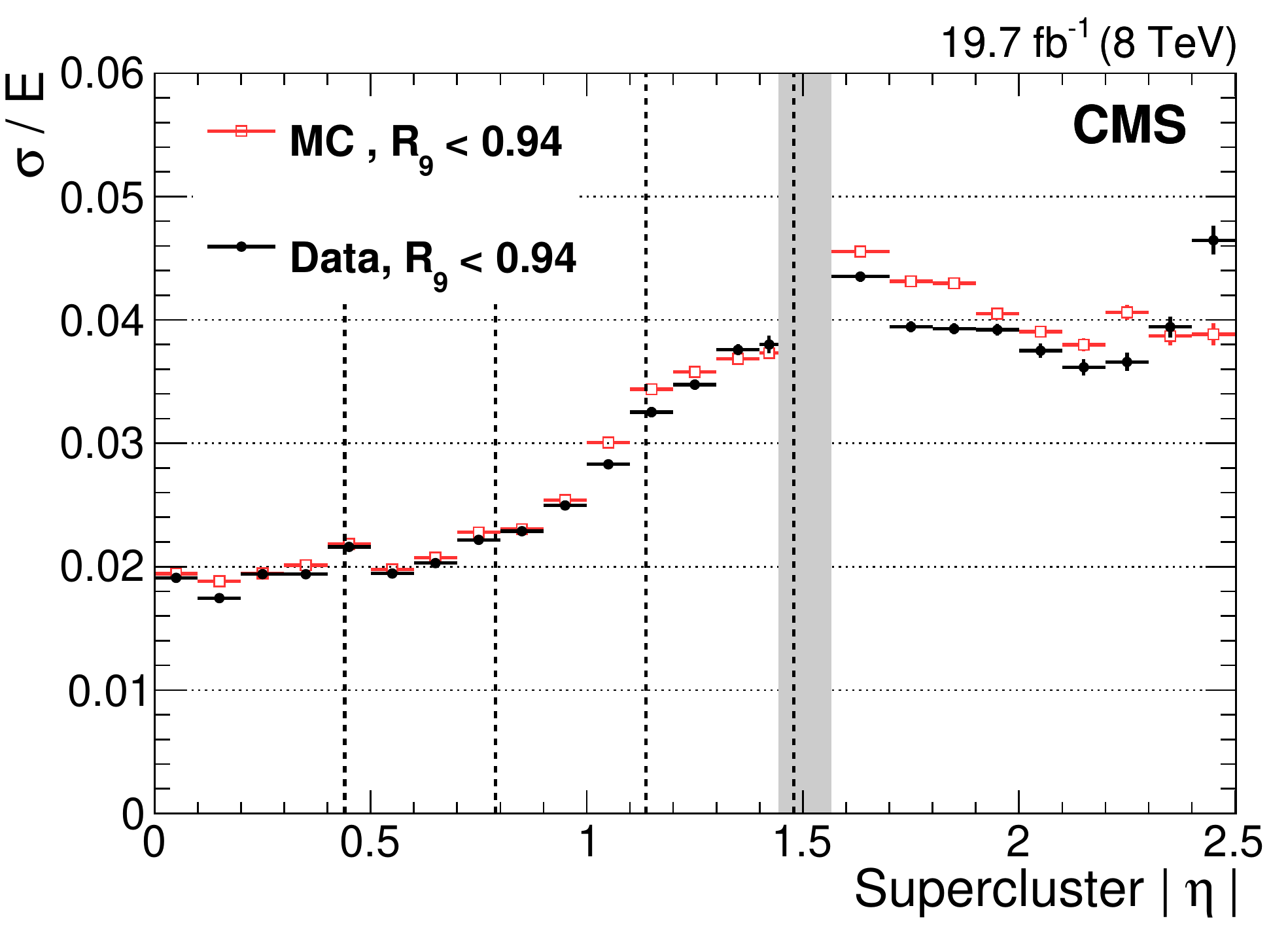}
 \caption{Relative photon energy resolution measured in small bins of absolute supercluster pseudorapidity in $\Zee$ events,
               for data (solid black circles) and simulated events (open squares), where the electrons are reconstructed as photons.
               The resolution is shown for (upper plot) showers with $\RNINE\ge0.94$ and (lower plot) $\RNINE<0.94$.
               The vertical dashed lines mark the module boundaries in the barrel, and the vertical grey band indicates the range of $\abs{\eta}$,
               around the barrel/endcap transition, removed from the fiducial region.
}
 \label{fig:Zres-vs-eta}
 \end{center}
\end{figure}

The single-photon energy resolution in $\Zee$ events where the electron showers are reconstructed as photons
has been measured in both data and simulated events using a method similar to, but independent of,
that used to obtain the corrections and smearings.
The data and simulated event samples are the same as those used to obtain the corrections and smearings.
The fitting methodology allows the resolution and energy scale for single showers to be extracted in fine bins of chosen variables,
but with the limitation that the energy resolution for each bin is parameterized as a Gaussian distribution.
Figure~\ref{fig:Zres-vs-eta} shows the resolution measured in small bins of $\eta$, taken as the position of the shower
in the ECAL, for showers with $\RNINE\ge0.94$ and $\RNINE<0.94$, for data and simulated events.
The vertical dashed lines show the barrel module boundaries, where the resolution is somewhat degraded,
and the grey band at $\abs{\eta}\approx1.5$ marks the barrel-endcap
transition region excluded from the photon fiducial region used in the $\Hgg$ analysis.
The simulated resolution matches the resolution observed in data as a function of $\eta$ very well.
There is a small systematic difference in the endcap, particularly for the photons with $\RNINE<0.94$,
with the simulated photons showing worse energy resolution than the photons in data.
This is understood as being a result of the methodology used to determine the resolution,
which focuses on the Gaussian core of the distribution.
In this region, the Gaussian smearing added to the simulation in the fine-tuning step is larger than elsewhere,
and the smearing truly required here would have a non-Gaussian tail.

Figure~\ref{fig:Zres-vs-eta} demonstrates the very good agreement between simulation and data achieved for
the resolution of electron showers reconstructed as photons.
This is an important achievement, but it does not provide a measurement of the
energy resolution of photons.
Electron showers tend to have worse energy resolution than
photon showers of the same energy since all electrons radiate to some extent in the material of the tracker,
even those with high values of $\RNINE$.
Furthermore, the fitting technique used to obtain the resolution shown in Fig.~\ref{fig:Zres-vs-eta}, parameterizes the
resolution as a Gaussian distribution and thus tends to be more sensitive to
the core of the resolution function and less sensitive to its non-Gaussian tail.
Additionally, it is of particular interest to examine the energy resolution achieved for photons
resulting from the decay of Higgs bosons, which are on average more energetic than the electrons resulting from the decay of Z bosons.

\begin{figure}[hbtp]
\begin{center}
 \includegraphics[width=0.49\textwidth]{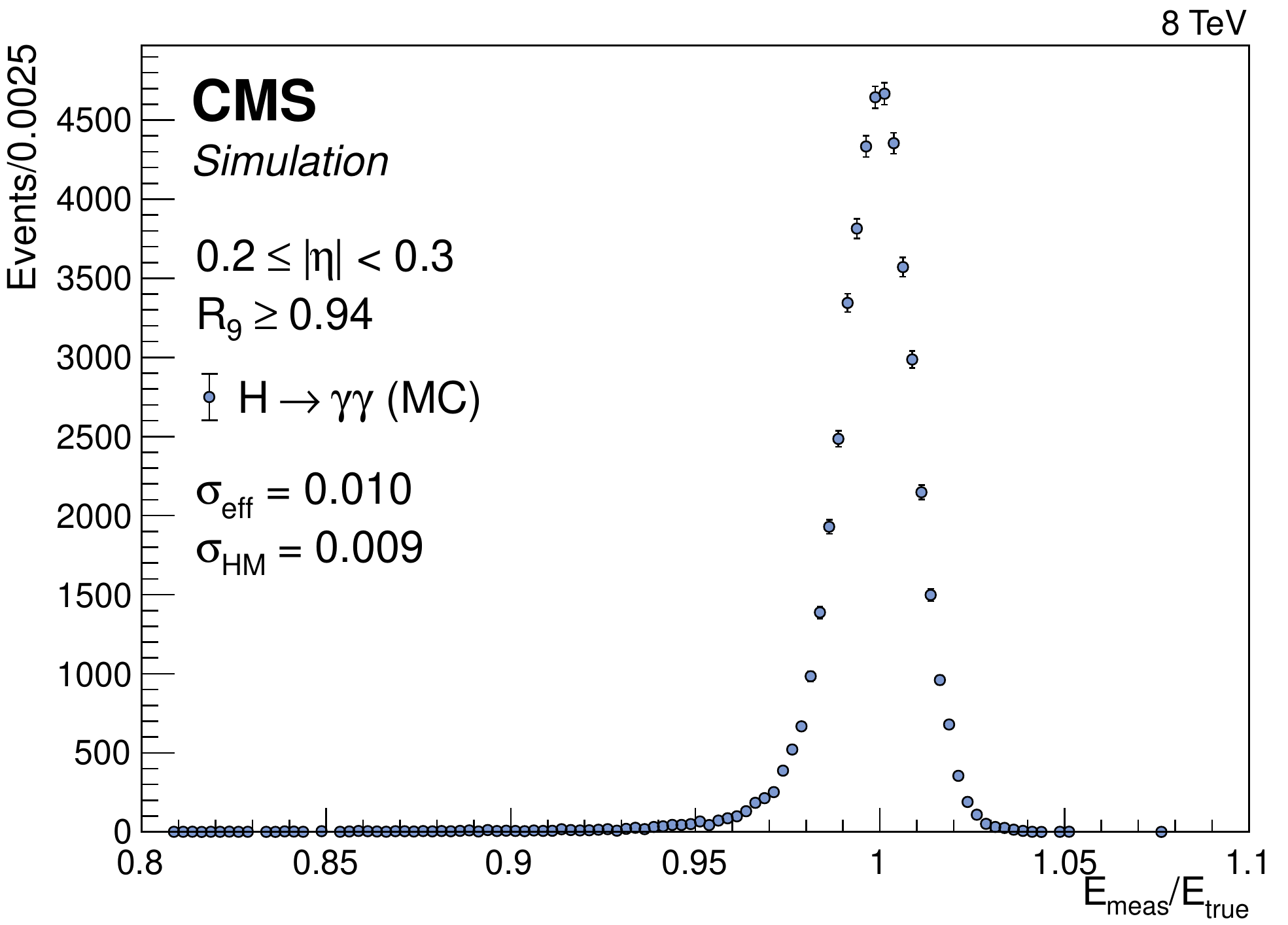}
 \includegraphics[width=0.49\textwidth]{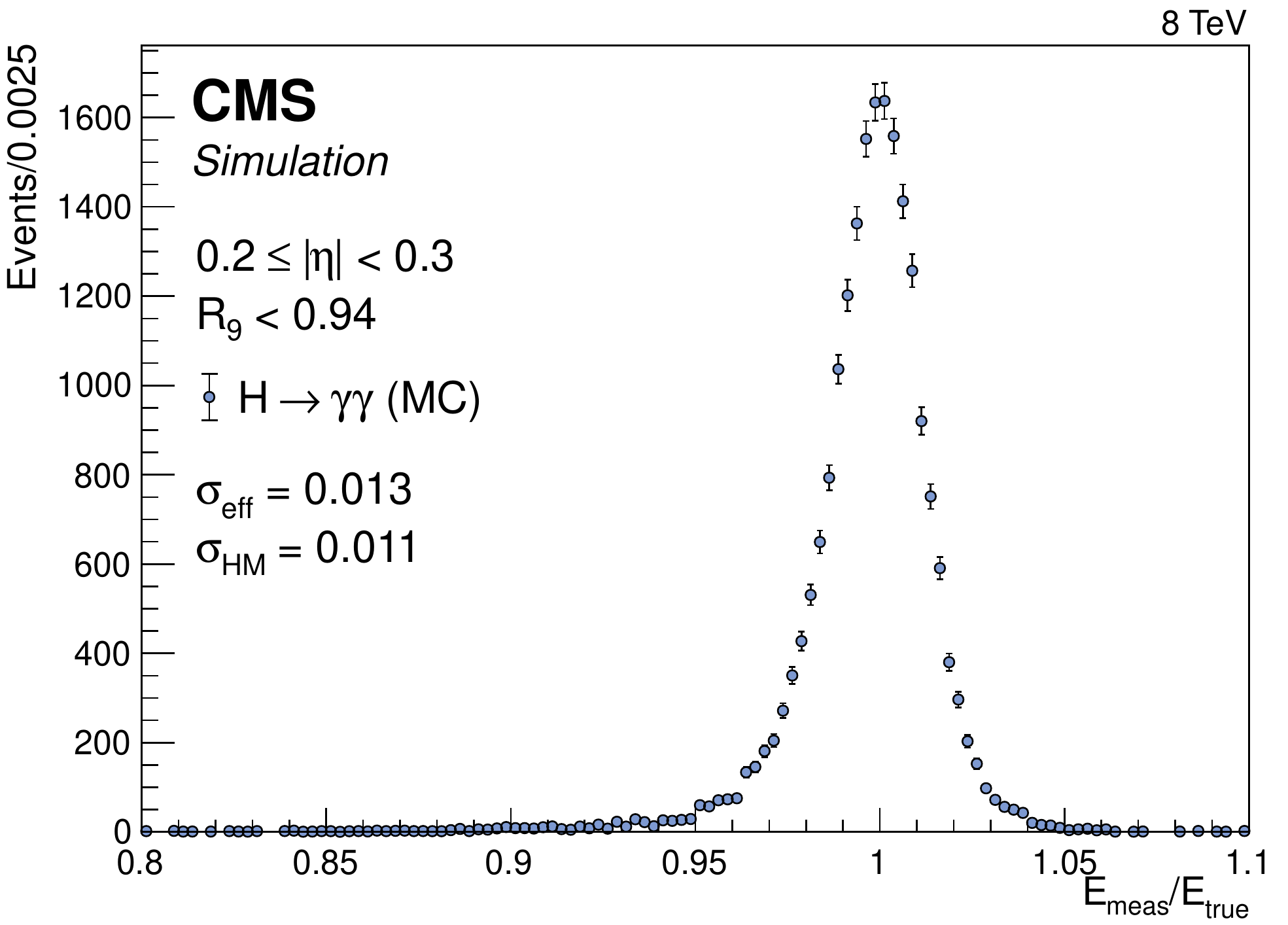}
 \caption{Distribution of measured over true energy, $E_\text{meas}/E_\text{true}$, for photons in simulated $\Hgg$ events,
                in a narrow $\eta$ range in the barrel, $0.2<\abs{\eta}<0.3$, (left) for photons with $\RNINE\geq 0.94$, and (right) $\RNINE<0.94$.
}
 \label{fig:HggResDist}
 \end{center}
\end{figure}

\begin{figure}[hbtp]
\begin{center}
 \includegraphics[width=0.75\textwidth]{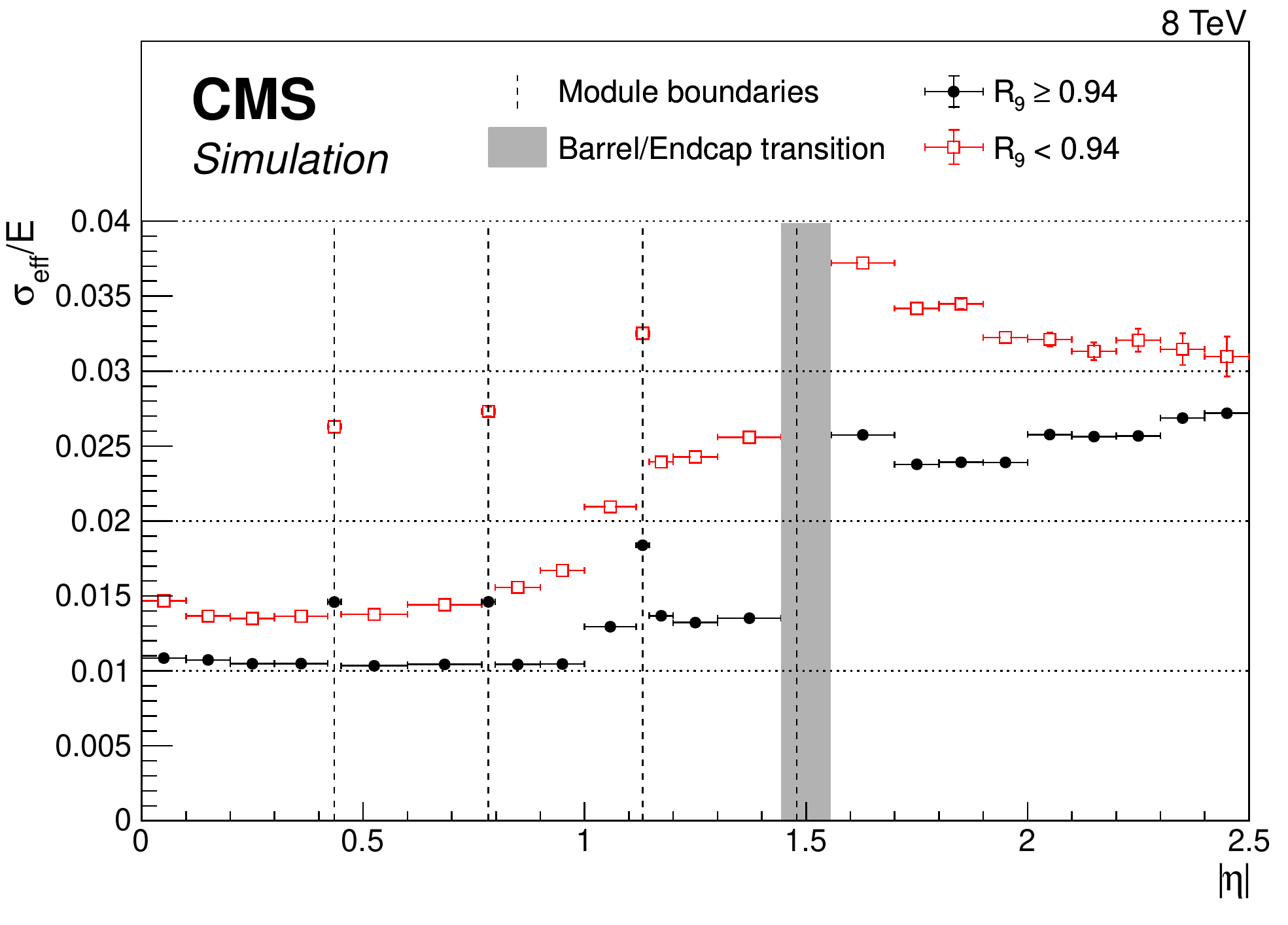}
 \caption{Relative energy resolution, $\sigma_\text{eff}/E$, as a function of $\abs{\eta}$,
in simulated $\Hgg$ events, for photons with $\RNINE\geq 0.94$ (solid circles) and photons with $\RNINE<0.94$ (open squares).
               The vertical dashed lines mark the module boundaries in the barrel, and the vertical grey band indicates the range of $\abs{\eta}$,
               around the barrel/endcap transition, removed from the fiducial region.
}
 \label{fig:HggRes}
 \end{center}
\end{figure}

Since there is excellent agreement between data and simulation for electron showers,
the energy resolution of photons in simulated events provides an accurate estimate of their resolution in data.
Figure~\ref{fig:HggResDist} shows the distribution of reconstructed energy divided by the true energy, $E_\text{meas}/E_\text{true}$, of photons
in simulated $\Hgg$ events that pass the selection requirements given in Ref.~\cite{legacy-paper},
in a narrow $\eta$ range in the barrel, $0.2<\abs{\eta}<0.3$.
The distribution for photons with $\RNINE\geq 0.94$ is shown on the left, and that for photons with $\RNINE<0.94$ is shown on the right.
The width of the distribution is parameterized in two ways: by the half-width of the narrowest interval containing 68.3\% of the
distribution, $\sigma_\text{eff}$, and by the full-width-at-half-maximum of the distribution divided by 2.35, $\sigma_\mathrm{HM}$.
These parameters are both equal to the standard deviation in the case of a purely Gaussian distribution.
Since $\sigma_\mathrm{HM}$ measures the width of the Gaussian core of the distribution, the values are smaller,
particularly where non-Gaussian tails make a larger contribution: for example, for $\RNINE<0.94$ and at the intermodule boundaries.
Figure~\ref{fig:HggRes} shows the fractional energy resolution, parameterized as $\sigma_\text{eff}/E$, as a function of $\eta$,
in simulated $\Hgg$ events that pass the analysis selection requirements.
A bin size of 0.1 in $\eta$ has been used, with adjustments to allow a small bin of width 0.03 centred on the barrel module boundaries
where it can be seen that the resolution is locally degraded.

\subsection{Energy scale uncertainty}
\label{sec:E-uncertainty}

The photon energy scale has been checked with photons in $\Zmmg$ events.
After a selection of events ensuring a pure and unbiased sample of photons, there is agreement between the measured photon energy and that
predicted from the known Z-boson mass and measured muon momenta.
The overall energy scale difference between data and simulation found with the $\Zmmg$ events (using the fine-tuning corrections,
obtained as described in Section~\ref{sec:fine-tune}) is $0.25\%\pm0.11\%\stat\pm0.17\%\syst$.
The study is made for photons with $\pt>20\GeV$, and the mean \pt of the photons selected is 28\GeV.
When binned in \pt (so as to probe possible nonlinearities), and in $\RNINE$ and $\eta$ (according to the known dependencies of the ECAL),
the agreement of the measurements with the defined
energy scale remains good, although the uncertainties in individual bins are, at best, between 0.2 and 0.3\%.
Thus this check does not provide a very strong constraint on the uncertainty in the Higgs boson mass arising from the uncertainty in the photon energy scale.
An additional limitation is that the check is for a range of photon energies that has only a limited overlap
with that used in the Higgs boson analysis.
For these reasons the uncertainty in the Higgs boson mass arising from the uncertainty in the photon energy scale
has been analysed as described below.

There are three main sources of systematic uncertainty in the energy scale that is defined by
the fine-tuning described in Section~\ref{sec:fine-tune}.
These uncertainties are the main contributions to the systematic uncertainty in the measured mass of the Higgs boson in the diphoton decay channel~\cite{legacy-paper}.
The largest uncertainties are due to the possible imperfect simulation of
(i) differences in detector response to electrons and photons, and
(ii) energy scale nonlinearity.
Finally there is an uncertainty resulting from the procedure and methodology described in Section~\ref{sec:fine-tune}.
These uncertainties are discussed in detail in Ref.~\cite{legacy-paper} and summarized below together with additional
results and information.

\begin{figure}[hbtp]
\begin{center}
 \includegraphics[width=0.60\textwidth]{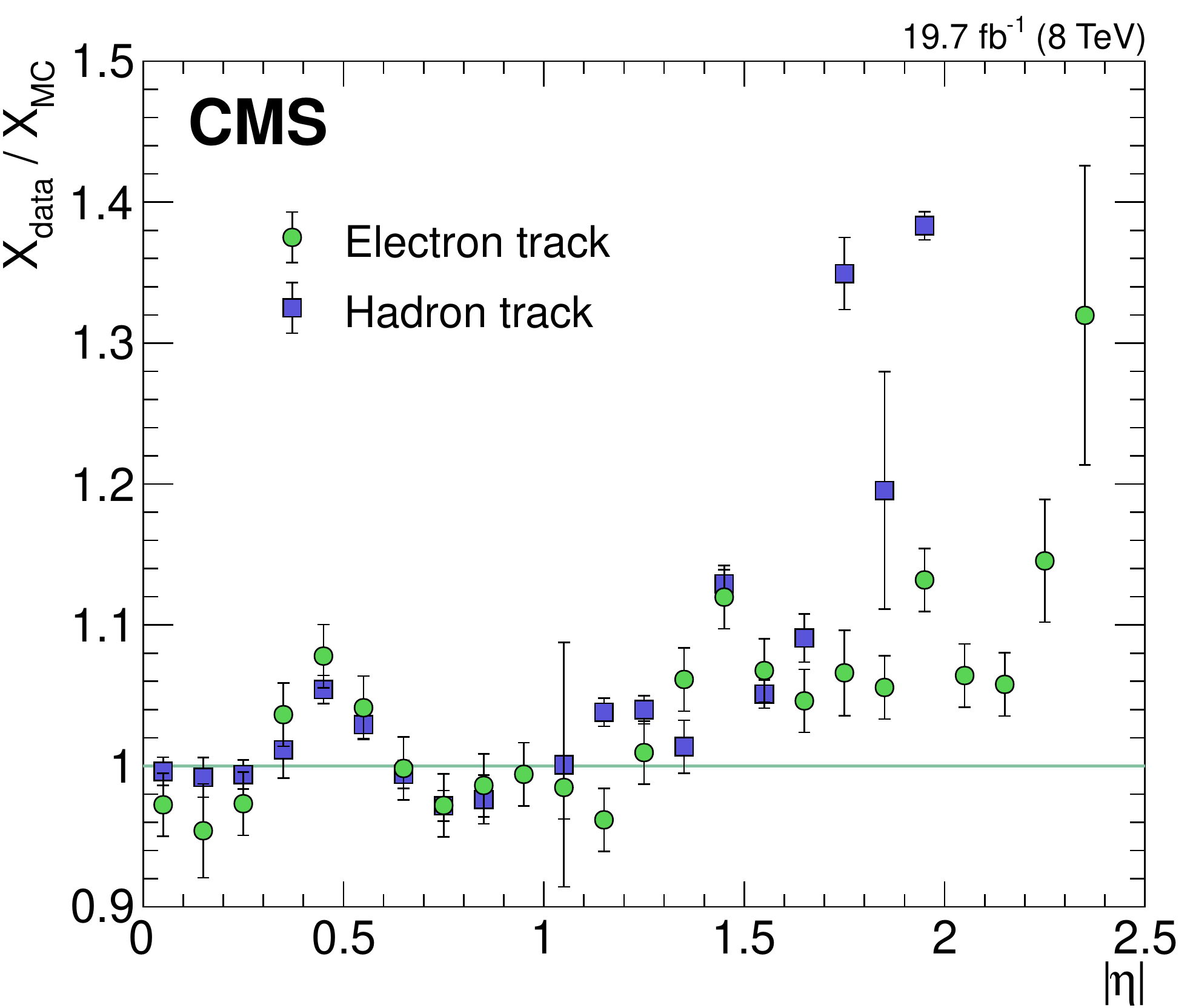}
 \caption{Tracker material thickness (in terms of radiation lengths) inferred in the data, $X_\text{data}$, relative to that
inferred in simulated events, $X_\mathrm{MC}$, as a function of $\abs{\eta}$,
using electrons in $\Zee$ events (circles), and low-momentum charged hadrons (squares).
}
 \label{fig:material}
 \end{center}
\end{figure}

Since the energy scale has been obtained using electron showers reconstructed as photons, an important
source of uncertainty in the photon energy scale is the imperfect modelling of the difference between
electrons and photons by the simulation.
The most important cause of the imperfect modelling is an inexact description of the material between the interaction point and the ECAL.
Figure~\ref{fig:material} shows the thickness of the tracker material in terms of radiation lengths, as inferred from data, relative to what is
inferred from simulated events, as a function of $\abs{\eta}$.
The two methods used to infer the material thickness employ the energy loss of electrons in $\Zee$ events
and the energy loss of low transverse momentum, $0.9<\pt<1.1\GeV$, charged-hadron tracks,
where the momentum loss is computed from the change in the track curvature between the beginning and end of the track.
The measurement using low-$\pt$ charged hadrons is difficult to implement in the regions of the tracker at large $\eta$, and no
values are available beyond $\abs{\eta}=2$, but for $\abs{\eta}<1.6$ the two methods give results that are in good agreement.
In addition, there is no charged-hadron measurement for the bin centred at $\abs{\eta}=0.95$ where the transition between the tracker
barrel and endcap results in few tracks with the number of hits required to make a good measurement.

The difference between data and simulation in the material thickness of the tracker is almost certainly due to mismodelling
of specific structures and localized regions.
This hypothesis is supported by studies
of the location of low-\pt (down to $\pt\approx1\GeV$) photon conversion vertices, as shown in Ref.~\cite{PAS-TRK-10-003}.
The results shown in Fig.~\ref{fig:material}, however, assume a simple scaling of the overall thickness.
The effect of changes in the amount of tracker material on the relative difference between the electron and photon energy scales
has been studied with events simulated using tracker models where the amount of material is increased uniformly by 10, 20, and 30\%.
Mismodelling of localized structures may affect the measurements used to infer thickness in Fig.~\ref{fig:material} somewhat differently
from the way it affects the relative difference between the electron and photon energies.
Therefore it is necessary to be rather conservative in the assignment of a systematic uncertainty.
It is assumed that the effects on the energy scale are covered by a 10\% uniform deficit of simulated material
in the region $\abs{\eta}<1.0$ and a 20\% uniform deficit for $\abs{\eta}>1.0$.
The resulting uncertainty in the photon energy scale has been assessed using the simulated samples in which the tracker material is increased
uniformly, and ranges from 0.03\% in the central ECAL barrel up to 0.3\% in the outer endcap.

Since the longitudinal profiles of energy deposition of electrons and photons differ,
a further difference in response between electrons and photons which would result from imperfect simulation, is related to
modelling of the varying fraction of scintillation light reaching the photodetector as a function
of the longitudinal depth in the crystal at which it was emitted.
Ensuring adequate uniformity of light collection was a major accomplishment in the development of the crystal calorimeter
and was achieved by depolishing one face of each barrel crystal.
However, an uncertainty in the achieved degree of uniformity remains and, in addition,
the uniformity is modified by the radiation-induced loss of transparency of the crystals.
The uncertainty results in a difference in the energy
scales between electrons and unconverted photons that is not present in the standard simulation.
The effect of the uncertainty, including the effect of radiation-induced transparency loss, has been studied.

A scaling as a function of depth, measured from the front face of the crystal, is applied to the deposited energy.
In the standard simulation this scaling is uniformly equal to unity, \ie flat, for all except
the rearmost 10\cm of the crystal.
To simulate nonuniformity of light collection, an appropriate slope is introduced based
on laboratory light-collection efficiency measurements made on the crystals, and measurements
of its dependence on crystal transparency.
The slope of the light collection efficiency as a function of depth, at the time when the ECAL was constructed, is taken to be
$-0.14\pm0.08\%/X_0$~\cite{Paramatti,Auffray}, for the front half of the crystal (``front non-uniformity'').
The change of this slope, $\Delta\mathrm{F}$, is parametrized as a function of the
absorption coefficient induced by irradiation
measured in m$^{-1}$, $\Delta\mu$, and is given by $\Delta\mathrm{F}=0.4\%\times\Delta\mu/X_0$~\cite{Tedaldi:2012zs}.
Finally, the induced absorption coefficient is related to the light-yield (LY) loss measured by the laser monitoring system, $\Delta(\mathrm{LY}/\mathrm{LY}_0)$, through $\Delta\mu=k\times\Delta(\mathrm{LY}/\mathrm{LY}_0)$, where $k=0.02\%\mathrm{/m}$
(\ie taking the average value of the measurements reported in Refs.~\cite{Huhtinen:2005wt} and~\cite{Timlin}).

The uncertainty in the slope is taken as the difference between the flat response used in the standard simulation and the
average slope measured at the time of ECAL construction plus the slope change resulting from the maximum
radiation-induced light loss in the barrel.
The resulting magnitude of the uncertainty in the photon energy scale in the barrel is 0.04\% for photons with $\RNINE>0.94$ and 0.06\% for those with $\RNINE<0.94$, but the signs of the energy shifts are opposite since unconverted photons penetrate deeper into the crystal than electrons, whereas converted photons share their energy between two electrons, whose showers thus penetrate the crystal less than a single electron shower.
In the endcaps, the magnitude of the uncertainty in the photon energy scale is taken to be the same as in the barrel, and the effect of the longitudinal uniformity has not been studied in detail, firstly because the uncertainty in the energy scale due to other effects is larger there, and secondly because these studies were done in the context of the $\Hgg$ analysis where uncertainties in the endcap energy scale had very little impact on the overall mass scale uncertainty.
For the diphoton mass in the $\Hgg$ analysis the two anticorrelated uncertainties result in an uncertainty of about 0.015\% in the mass scale.
The effect of the tracker material uncertainty on this value, where a changed tracker material budget would change the number of photons that convert in the tracker material, is negligible.

In assessing the systematic uncertainties for the $\Hgg$ mass measurement, differences between MC simulation
and data in the extrapolation from shower energies typical of electrons from $\Zee$ decays
to those typical of photons from $\Hgg$ decays, were also investigated.
The linearity of the energy response was studied in two ways: by examining the dependence
of the energy-momentum ratio, $E/p$, of isolated electrons from $\cPZ$ and $\PW$ boson decays as a function of $\ET$,
and by looking at the invariant mass of dielectrons from $\cPZ$ boson decays as a function of the
scalar sum of the transverse energies of the two electron showers, $\HT$.
In both cases, the energy or transverse energy of the electrons and the invariant mass of the dielectron,
are those obtained when the ECAL showers are reconstructed as photons.
The showers are required to satisfy $\ET>25\GeV$ and the photon identification requirements of the $\Hgg$ analysis (with the electron veto removed).
The $E/p$ distributions, obtained from simulated events for a number of bins in $\ET$, and the dielectron invariant mass distributions,
obtained for a number of bins in $\HT$, were fitted to the corresponding distributions obtained from events in data.
A scale factor was extracted from each fit, whose difference from unity measures the residual discrepancy
of the energy response in data relative to that in simulated events.
As a cross-check, an iterated truncated-mean method was used to estimate the $E/p$ or dielectron invariant mass peak positions
and gave consistent results.

\begin{figure}[hbtp]
\begin{center}
 \includegraphics[width=0.80\textwidth]{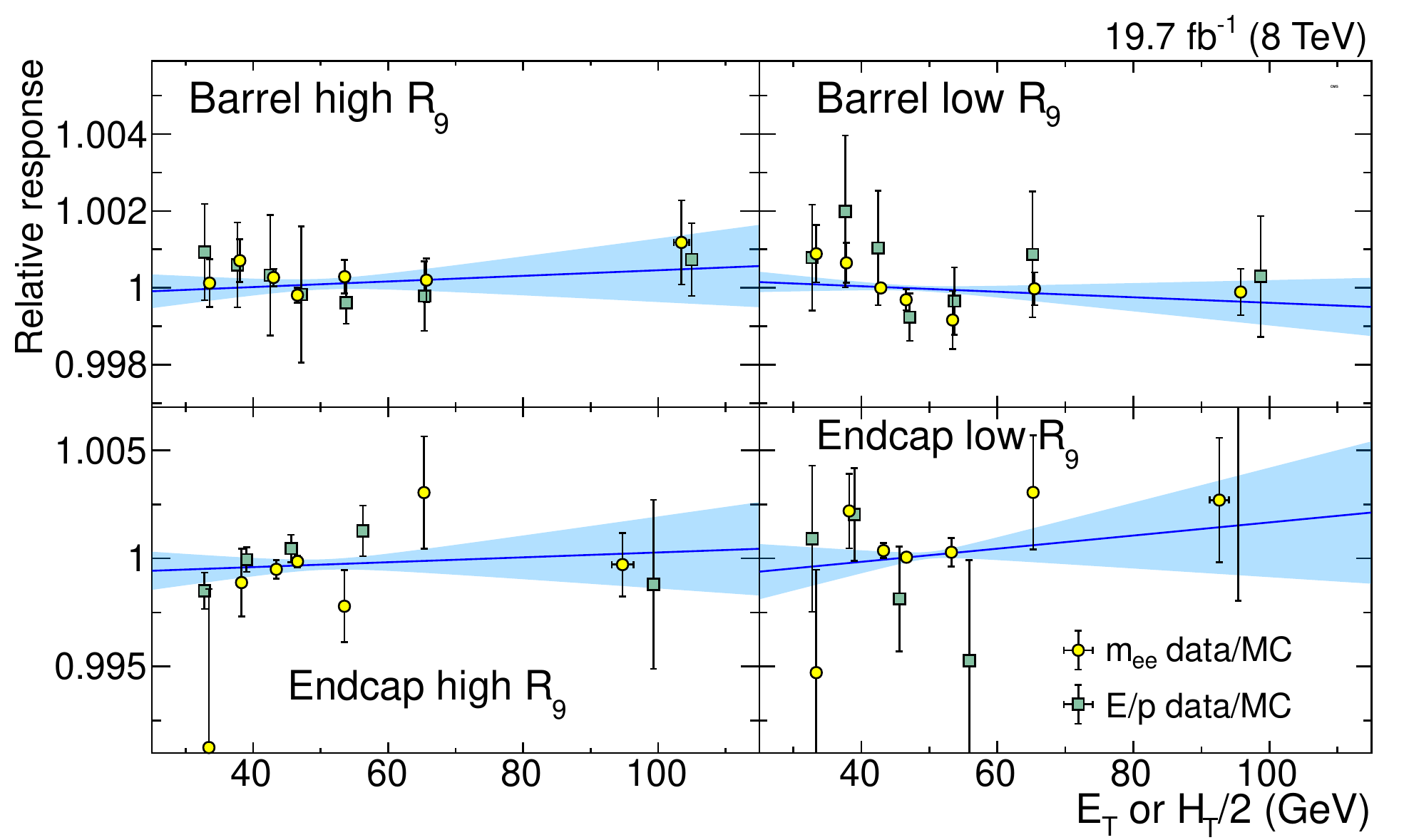}
 \caption{Residual discrepancy of the energy response in data relative to that in simulated events as a function of transverse energy
(for the $E/p$ analysis, squares) and of $\HT/2$ (for the dielectron mass analysis, circles) in four $\eta$ and $\RNINE$ categories.
The dielectron analysis is restricted to events where both the electron showers fall in the same $\eta, \RNINE$ category.
The uncertainties in the fit parameters of a linear response model are shown by bands---further details are given in the text.
}
 \label{fig:nonlinearity}
 \end{center}
\end{figure}

The results are shown in Fig.~\ref{fig:nonlinearity} for both the $E/p$ and the dielectron invariant mass analyses.
The points coming from the analysis of the dielectron mass are plotted as a function of $\HT/2$.
The four panels show results for different $\eta$ and $\RNINE$ categories,
with the dielectron analysis restricted to events where both electron showers fall in the same category.
The $\eta$ categories correspond to the barrel and endcap regions.
The horizontal error bars indicate the uncertainty in the mean $\ET$ or $\HT/2$ for the bin, but for most
bins that uncertainty is negligible and hidden behind the plotted central value marker.
In the endcaps for low $\RNINE$ the point corresponding to $\ET=95.4\GeV$ for the $E/p$ analysis has a value of 1.0146
which does not fit in the plot scale, although the lower vertical error bar, extending down below 1, can be seen.
The differential nonlinearity is estimated from a linear fit through the points (shown by the lines).
The uncertainties in the fit parameters of a linear response model, shown by the bands, are extracted after scaling the uncertainties
such that the $\chi^2$ per degree of freedom of the fits is equal to unity.
The stability of the result has been checked by removing the points of the dielectron mass analysis that have
very small statistical uncertainties (\ie where $\HT/2$ is about half the Z-boson mass).

A value of 0.1\% was assigned to the uncertainty in the effect of
differential nonlinearity for a diphoton mass around 125\GeV in all events except
those in the class in which the diphoton transverse momentum is particularly high,
so that the highest transverse momentum photon in the event typically has $\pt>100\GeV$.
For this event class the uncertainty is set at 0.2\%.

The digitization of the ECAL signals uses 12-bit analogue-to-digital-converters (ADCs) and, to increase the dynamic range,
three different preamplifiers with different gains are used for each crystal, each with its own ADC,
and the largest unsaturated digitization is recorded together with two bits coding the ADC number~\cite{Chatrchyan:2008aa}.
The possibility that imperfect matching between the different ``gain ranges'' introduces an uncertainty in the energy of the measured photons was investigated.
The effect of switching preamplifiers for digitizing large signals,
$E\gtrsim200\GeV$ in the barrel and $\ET\gtrsim80\GeV$ in the endcaps, was found to be negligible
for photons from Higgs boson decays.
The fraction of photons for which the lower-gain preamplifiers are used is small ($<$2\%) and the lower-gain preamplifiers appear to be very well
calibrated to the high-gain preamplifiers.

A further small uncertainty arises from imperfect electromagnetic shower simulation.
A simulation made with a shower description using the Seltzer--Berger model for the bremsstrahlung energy spectrum~\cite{Seltzer:1974zz},
which represents an improvement over \GEANTfour~version~9.4.p03,
changes the energy scale for both electrons and photons.
The much smaller changes in the \textit{difference} between the electron and photon energy scales, although mostly consistent with zero,
are interpreted as a limitation on our knowledge of the correct simulation of the showers, leading to a further uncertainty of 0.05\%
in the mass of the Higgs boson.

The statistical uncertainties in the measurements used to set the energy scale are small,
but the methodology, which is described in Section~\ref{sec:fine-tune},
has a number of systematic uncertainties related to the imperfect agreement between data and MC simulation.
The uncertainties range from 0.05\% for unconverted photons in the ECAL central barrel to 0.1\% for converted photons in the ECAL outer endcaps.

Accounting for all the contributions, the uncertainty in the photon energy scale at $\pt\approx m_\cPZ/2$, where $m_\cPZ$ is the Z boson mass,
is about 0.1\% in the central barrel, 0.15\% in the outer barrel, and 0.3\% in the endcaps.
These uncertainties are largely correlated.
The exact values, their correlations in two $\RNINE$ times four $\eta$
bins, together with the contribution from the residual nonlinearity and
from the uncertainties on the energy and mass resolution have been
propagated to the signal model of the $\Hgg$ analysis.
Together with similar, and not entirely correlated, uncertainties in the
7\TeV data they contribute 0.14\GeV to the systematic uncertainty of
0.15\GeV in the Higgs boson mass measurement~\cite{legacy-paper}.
\section{Conversion track reconstruction}
\label{sec:photon-conversion}

\begin{figure} [hbtp]
  \begin{center}
      \includegraphics[width=0.50\textwidth]{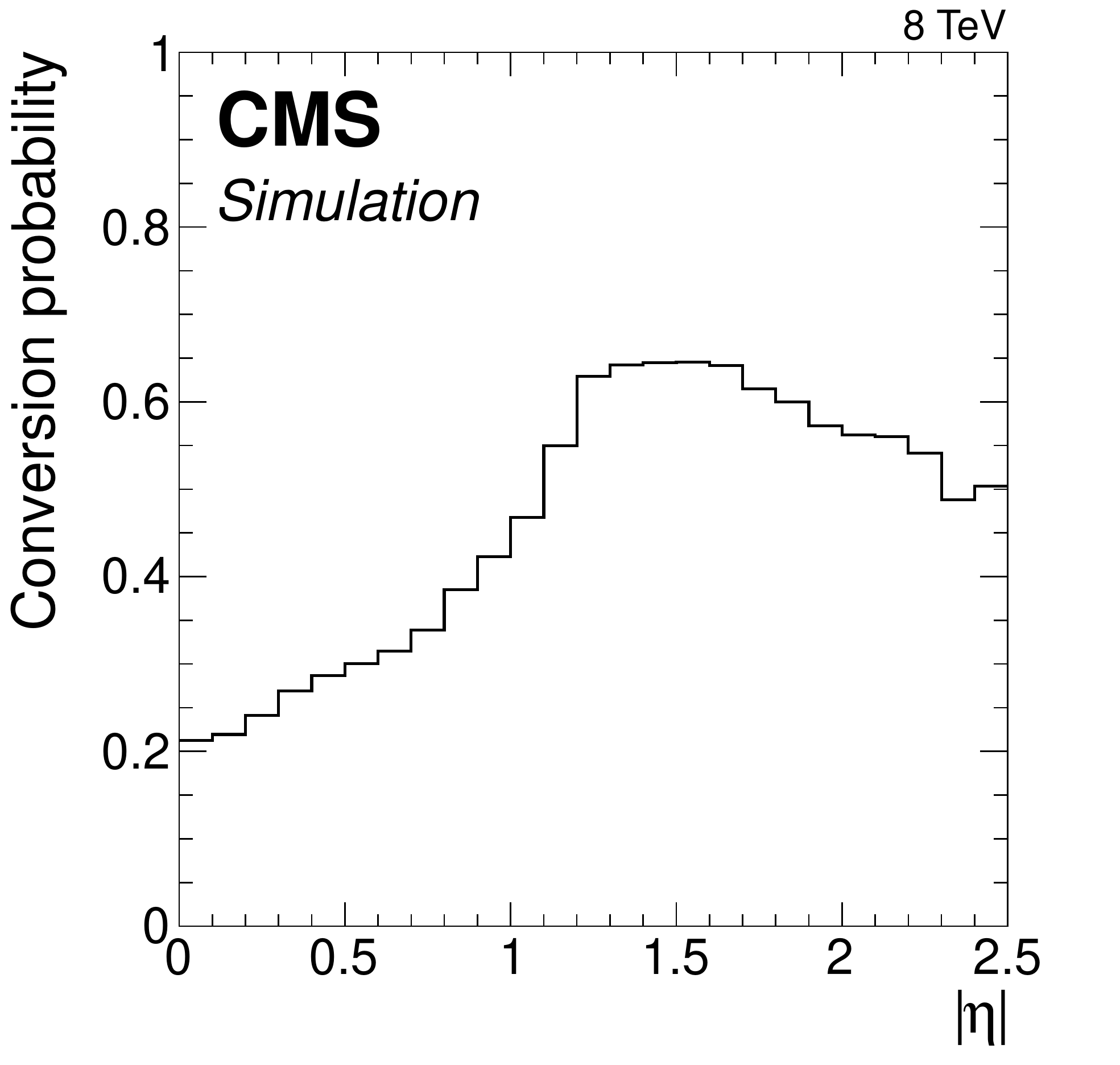} 	
      \caption{Fraction of photons converting before the last three layers of the tracker as  function of absolute pseudorapidity as measured in a simulated sample of $\Hgg$ events. The conversion location is obtained from the simulation program.}
      \label{fig:trackerMaterial}
 \end{center}
\end{figure}

Photons traversing the CMS tracker have a sizeable probability of converting into electron-positron pairs.
Although converted photons are fully clustered in the ECAL as described in Section~\ref{sec:photon-reco}, and
identified with good approximation by the $\RNINE$ shower-shape variable,
additional useful information is gained by reconstructing the associated $\Pep\Pem$ track pairs.
According to simulation, the fraction of photon conversions occurring before the last three layers of the tracker
(reconstruction of conversion tracks requires at least three layers)
is as high as about 60\% in the pseudorapidity regions with the largest amount of tracker material in front of the ECAL (Fig.~\ref{fig:trackerMaterial}).
Fully reconstructed conversions are used in the particle-flow reconstruction algorithm~\cite{CMS-PAS-PFT-09-001, CMS-PAS-PFT-10-001}:
the association of electron-track pairs with energy deposits in the  ECAL
avoids their being misidentified as charged hadrons, thus improving
the determination of the photon isolation, as discussed in Section~\ref{sec:photon-id}.
The direction of the electron-track pair is also exploited in assisting
the determination of the longitudinal coordinate of the interaction vertex in the $\Hgg$ analysis~\cite{legacy-paper}.
The aim of this section is to describe the methods used to reconstruct electron-track pairs and show the level of agreement between data and simulation in a very pure sample of photons.

Conversion reconstruction uses  the full CMS tracking power~\cite{TrackerPaper}.
Track reconstruction is based on an iterative tracking procedure.
The first iteration aims at finding tracks originating
from the interaction vertex while subsequent iterations aim at finding tracks from displaced (secondary) vertices at increasing distance from the primary vertex.
In addition,  tracks starting from clusters in the ECAL and propagated inward into
the tracker volume are sought, so as to reconstruct late-occurring conversions~\cite{CMS-NOTE-2006-005}.
All tracks associated to the main electron reconstruction~\cite{electron-paper},
as well as the subsample of  the standard tracks which can be  associated to energy deposits in the ECAL,
are possible electron candidates and are refitted with the Gaussian sum filter method~\cite{Adam2005}.
Tracks reconstructed as electrons are selected with basic quality requirements on the minimum number
of hits and goodness of the track fit.
Tracks are then required to have a positive charged-signed transverse impact parameter (the primary vertex
lies outside the trajectory helix).
Track-pairs of opposite charge are then filtered to remove tracks that might have resulted from conversions in the beam pipe,
or could possibly consist of electrons originating from the primary vertex.
Additional requirements on the track pair are meant to specifically identify the photon conversion topology.
Photon conversion candidates can be distinguished from massive meson decays, nuclear interactions or vertices from misreconstructed tracks by exploiting the fact that the momenta of the conversion electrons are approximately parallel since the photon is massless.
For this purpose, the angular separation of the track pair in the longitudinal plane,
measured in terms of $\Delta\cot\theta$, is required to be less than 0.1.
Also, the two-dimensional distance of minimum approach between the two tracks is required to be positive to remove intersecting helices.
Finally, the point in which the two tracks are tangent is required to be well contained in the tracker volume.

Track pairs surviving the selection  are fitted to a common vertex with a 3D-constrained kinematic vertex fit.
The 3D constraint imposes the tracks to be parallel in both transverse and longitudinal planes. The pair is retained if the vertex fit converges
and the $\chi^2$ probability is greater than a given threshold. The transverse momentum of the pair is finally refitted with the vertex constraint.

Reconstructed conversions are required to satisfy a minimum transverse momentum threshold,
meant to reduce accidental or poorly reconstructed pairs.
The threshold on the converted photon $\pt$ as measured by the tracks can vary depending on the application:
in this paper, mainly focussing on medium to high transverse momentum, the threshold is chosen to be 10\GeV.
More than one conversion track-pair candidate can be reconstructed for the same supercluster.
When such a case occurs, the optimal conversion is chosen by finding the best directional match between the momentum
direction of the track pair and the position of the supercluster.
The matching criterion is expressed in terms of the
$\Delta R=\sqrt{\smash[b]{\Delta \eta^2 + \Delta \phi^2}}$ distance between the supercluster direction and the conversion direction.
The conversion candidate with  minimum $\Delta R$ is retained if $\Delta R$ is less than 0.1.
Both the conversion and supercluster directions are redefined with respect to the fitted conversion
vertex position.

A sample of $\Zmmg$ events with a photon resulting from final-state radiation (FSR)
is selected from dimuon-triggered data, together with a corresponding sample of simulated events.
A very high photon purity (98\%) is achieved in the selection, which is not reachable in any other sample.
Events from $\Zmmg$ decays are selected by requiring the presence of two high-quality muon tracks
reconstructed with both the muon detector and the tracker within $\abs{\eta}<2.4$,
originating from the interaction vertex, and each having $\pt>10\GeV$.
Each muon track is also required to be associated to small energy deposits in the hadron calorimeter.
The dimuon invariant mass is required to be above 35\GeV.

Photon candidates are selected with loose identification criteria and with transverse momentum above 10\GeV, within $\abs{\eta}<2.5$
(excluding the ECAL barrel-endcap transition region) and added to the dimuon system.
The distance of the photon from the closest muon is required to satisfy $\DR<0.8$, while
the muon furthest from the photon must satisfy $\pt>20\GeV$.
It is required that the track of the muon closest to the photon is not reconstructed also as an electron.
Finally the three-body invariant mass, $m_{\mu\mu\gamma}$, is required to satisfy $60<m_{\mu\mu\gamma}<120\GeV$.

\begin{figure}[hbtp]
  \begin{center}
      \includegraphics[width=0.50\textwidth]{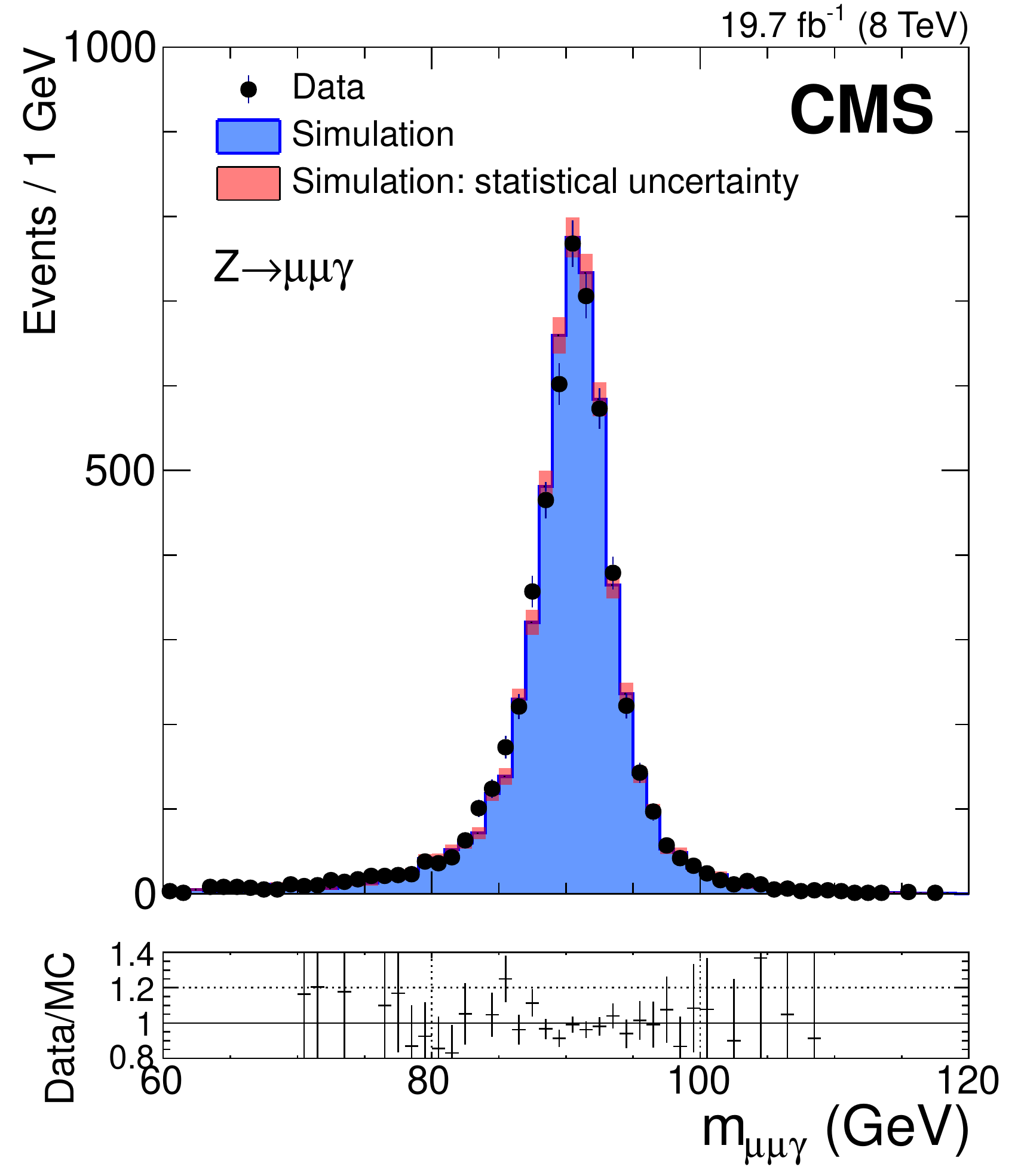}
       \caption{Invariant mass for $\Zmmg$ events in which the photon is associated to a conversion track pair in
data (points with error bars) and simulation (filled histogram).}
      \label{fig:conv_zmumugInvMass}
 \end{center}
\end{figure}

Figure~\ref{fig:conv_zmumugInvMass} shows the  $\mu \mu \gamma$ invariant mass for events in which a conversion track pair, matched to the photon, has also been reconstructed.
The invariant mass is calculated using the photon energy measured in the ECAL and taking the dimuon vertex.
The distributions are normalized to the number of candidates in data and show good agreement between data and simulation.

\begin{figure}[hbtp]
  \begin{center}
        \includegraphics[width=0.45\textwidth]{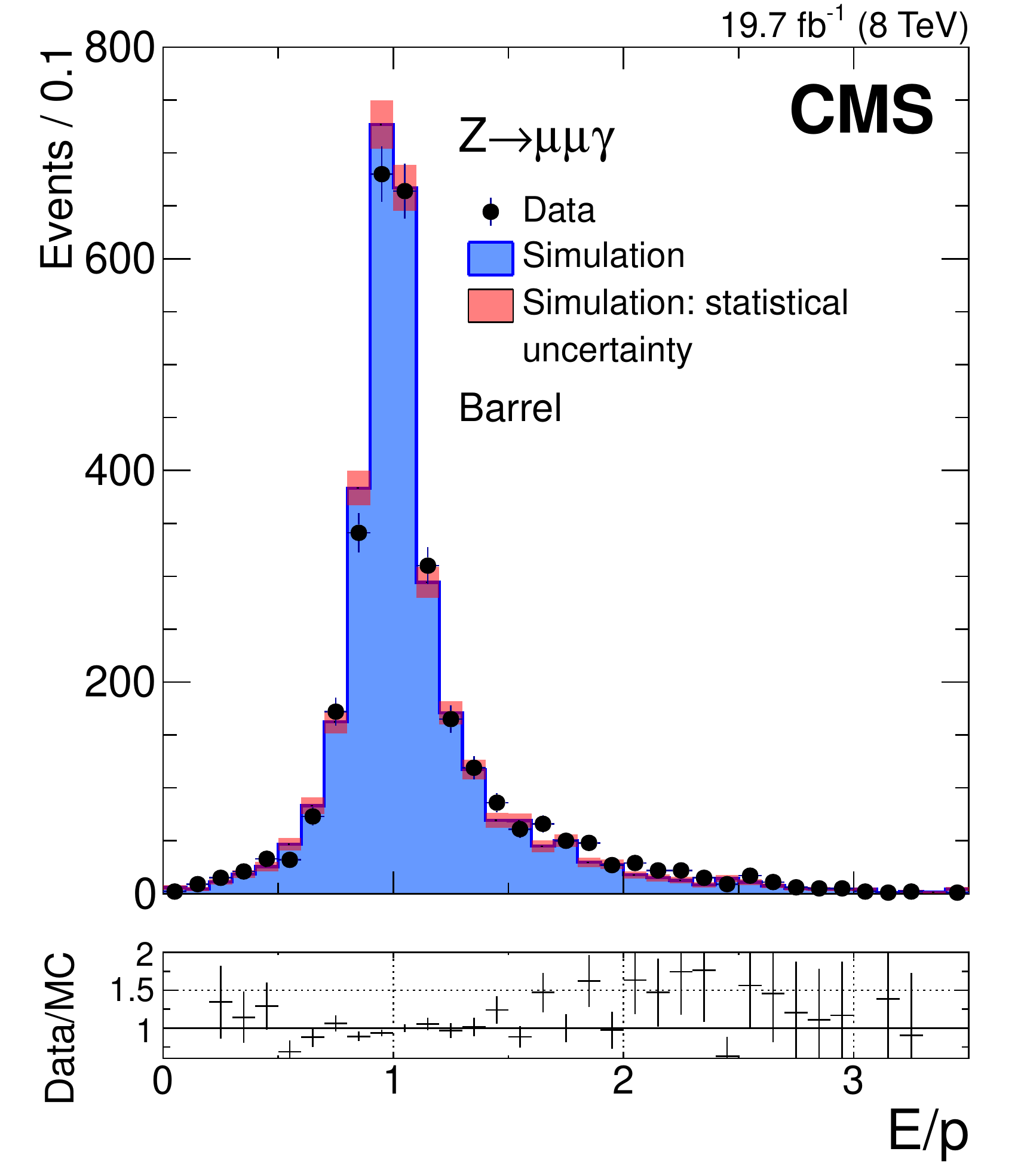}
        \includegraphics[width=0.45\textwidth]{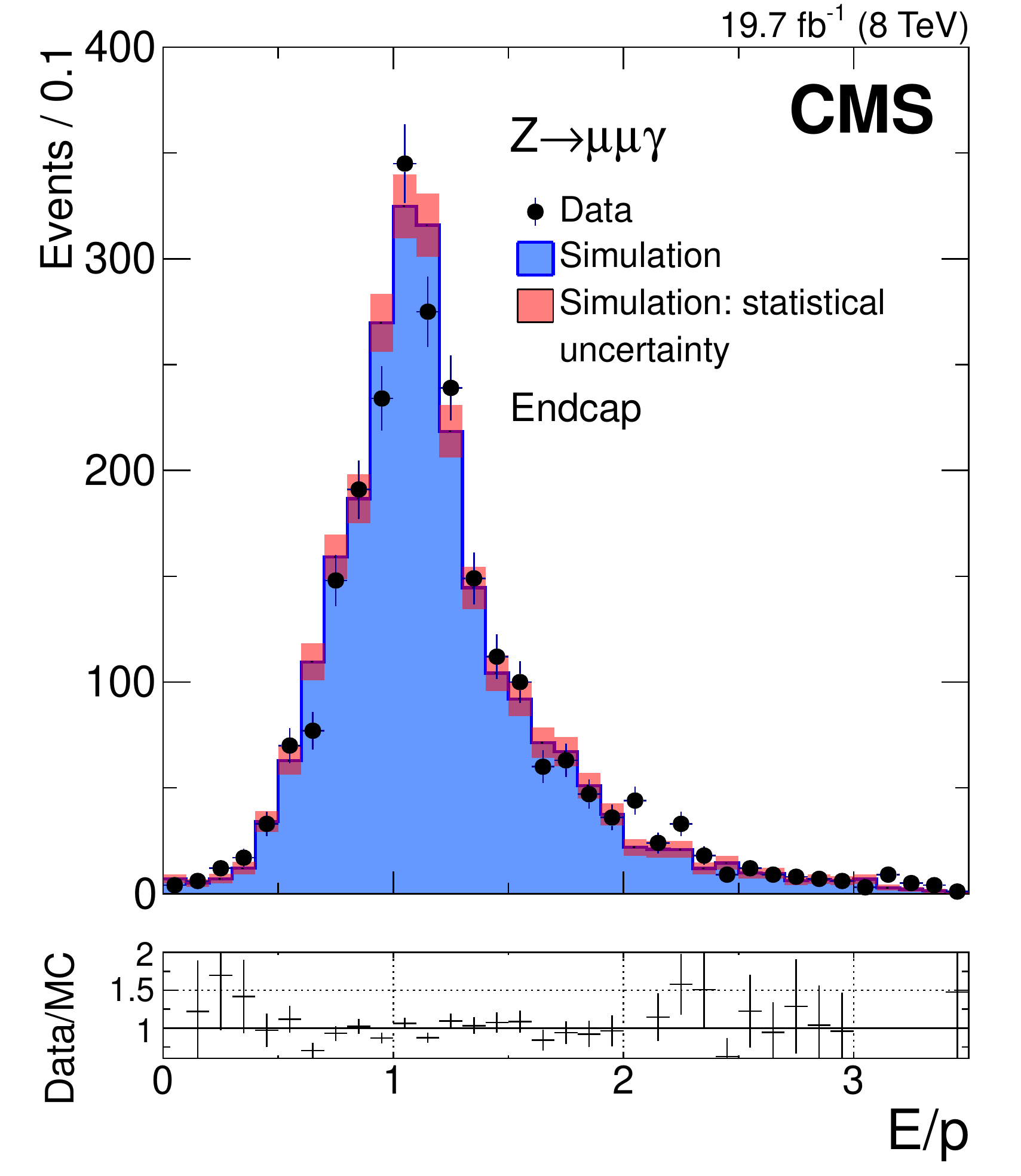}
      \caption{Distribution of the $E/p$ ratio, where $E$ is the supercluster energy measured in the ECAL and $p$ is the total momentum measured from the track
      pair refitted with the conversion vertex constraint, for photons in $\Zmmg$ events in data (points with error bars) and simulation (histograms), separately  for (left) barrel and (right) endcap.
      The simulated distributions  are normalized to the number of entries in data.}
      \label{fig:conv_EoP}
 \end{center}
\end{figure}

An estimator of the quality of the conversion reconstruction is the matching between the energy measured in the ECAL and the momentum measured from the track pair after refitting with the conversion vertex constraint.
If the track pair is correctly reconstructed and associated to the right cluster in the calorimeter the ratio $E/p$ must be close to one.
As for single electrons~\cite{electron-paper}, however, the distribution of the $E/p$ shows tails around unity,
because the electrons from conversions both emit bremsstrahlung along their trajectory through the tracker
and the total track-pair momentum does not account for the total energy collected in the calorimeter.
The distributions are shown in Fig.~\ref{fig:conv_EoP} for barrel and endcap separately, where the shape of the $E/p$ distribution in data is compared to that in simulation.
The distributions are normalized to the number of entries in data.
Converted photons from the decay of neutral mesons in jets or accidental track pairs do not exhibit a $E/p$ peak at unity.

\begin{figure}[hbtp]
  \begin{center}
    \includegraphics[width=0.45\textwidth]{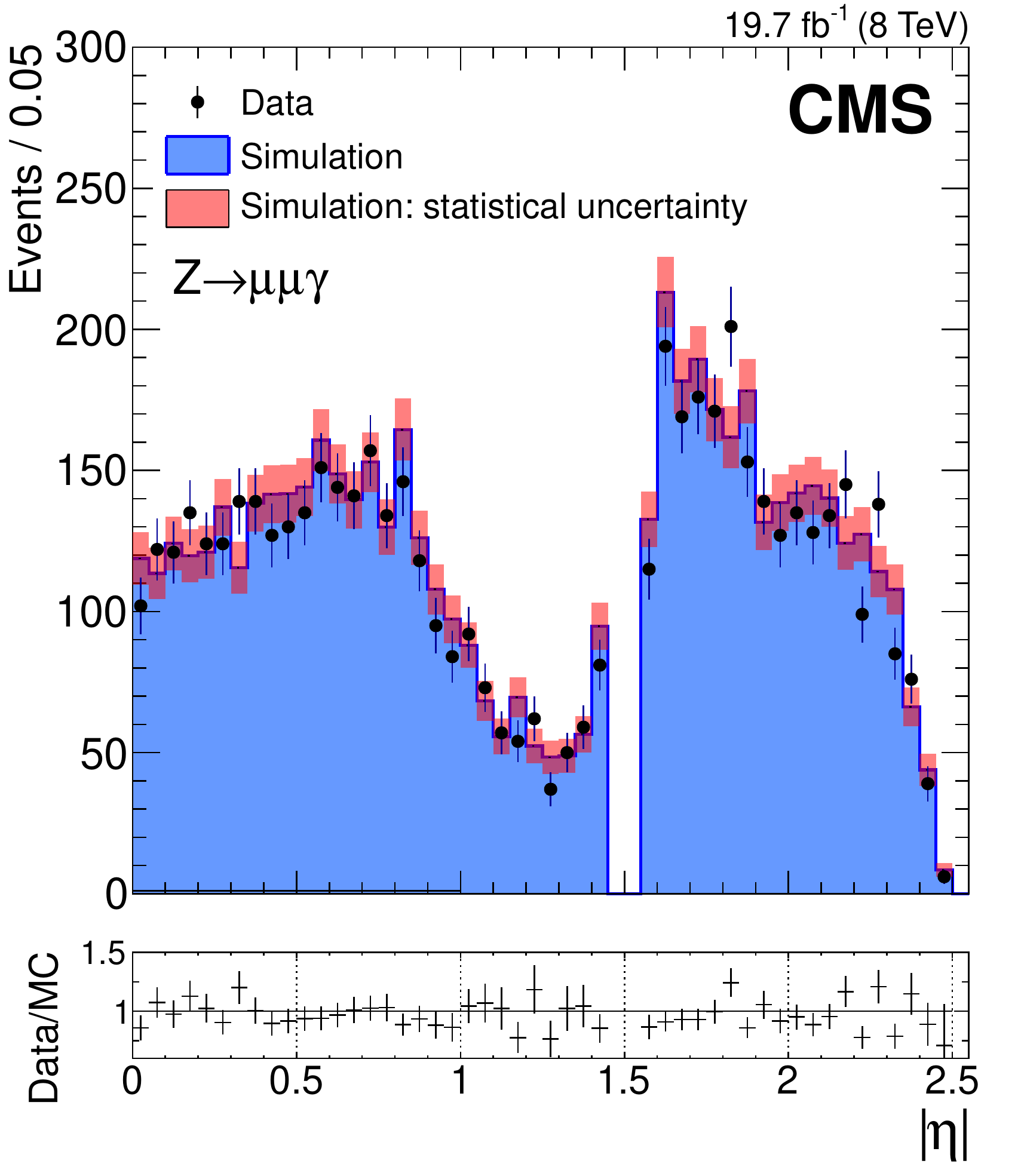}
    \includegraphics[width=0.45\textwidth]{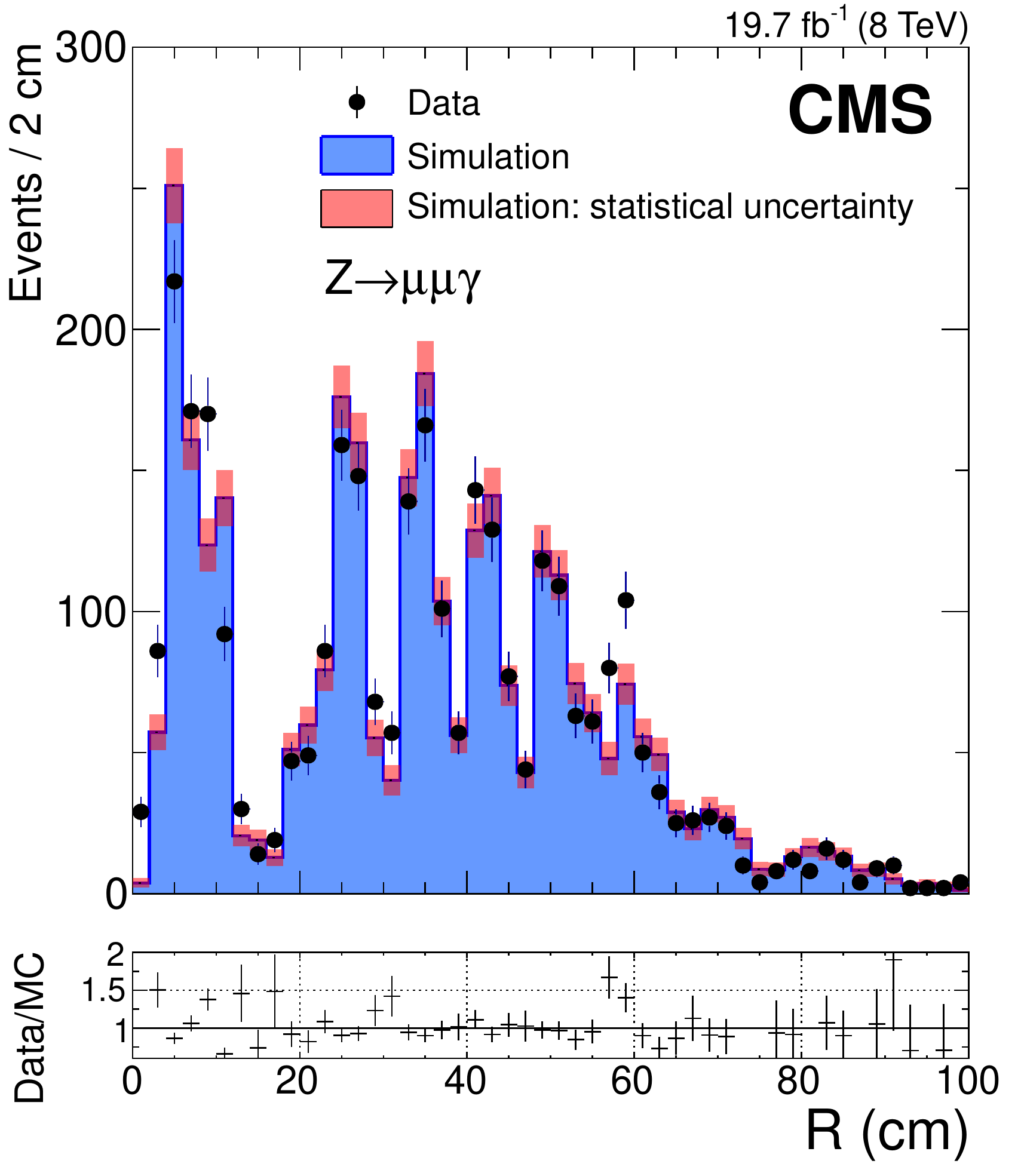}
      \caption{(Left) Distribution of the photon supercluster (absolute) pseudorapidity for events with
      reconstructed conversion vertices in data (points with error bars) and simulation (histograms).
      (Right) Distribution of the conversion vertex radius for photons in the range  $\abs{\eta}<1.4$ in data (points with error bars)
      and simulation (histograms).}
      \label{fig:conv_R}
 \end{center}
\end{figure}

The distributions of photon supercluster pseudorapidity and of photon conversion radius are shown in Fig.~\ref{fig:conv_R}.
The empty bin in the left plot, centred on $\abs{\eta}=1.5$, corresponds to the
ECAL barrel-endcap transition region in which photons are excluded from the analysis.
The radial position of the conversion vertices for  $\abs{\eta}<1.4$ in the right plot reveals the tracker structure,
as shown in Ref.~\cite{PAS-TRK-10-003} using low-\pt conversions in minimum bias events.
Data and simulation are in fair agreement.
The number of photons from $\Zmmg$ events in data is however insufficient to probe the local differences between data and simulation shown in Fig.~\ref{fig:material}.

\section{Photon identification}
\label{sec:photon-id}

In physics analyses using photon signals, a large and reducible background comes from
photon candidates that arise from neutral mesons produced in jets.
In the transverse momentum range of interest, the photons from the decay of neutral pions
are collimated and are reconstructed as a single photon---in the barrel the minimum separation
of the two photons from the decay of a $\Pgpz$ with $\pt=15\GeV$
is about the same as the crystal size.
The background tends to be dominated by $\Pgpz$'s that take a substantial fraction
of the total jet $\pt$ and are thus relatively isolated from jet activity in the detector.
Nevertheless, rejection of this background must rely heavily on isolation, particularly since
the high probability of conversion in the tracker material, followed by the separation
of the $\Pep\Pem$ pair in the 3.8\unit{T} magnetic field, means that the lateral shower-shape patterns
in $\phi$ have little power to discriminate prompt or single photons from background,
leaving only the $\eta$ coordinate for lateral shape discrimination.
A further consequence of the high probability of conversion in the tracker material is that
the $\RNINE$ distributions of signal and background differ for two independent reasons: firstly,
the showers from $\Pgpz$'s tend to have lower $\RNINE$ values because of the two separated
decay photons; and secondly, there is a higher chance that at least one of two photons from a $\Pgpz$ converts.

Two photon identification algorithms are used in CMS to select against candidate photons originating in jets:
an approach using selection requirements applied to a set of individual variables, and a multivariate technique.
Both methods include a criterion intended to reject electrons misidentified as photons.

\subsection{Electron rejection}
\label{sec:photon-id-electronveto}

The photon identification prescriptions discussed in this paper use the ``conversion-safe electron veto'' to reject electrons.
This veto requires that there be no charged-particle track with
a hit in the inner layer of the pixel detector not matched to a reconstructed conversion vertex, pointing to the photon cluster in the ECAL.
The ``hit in the inner layer'' is computed as a hit in the first layer where a hit is possible,
accounting for the small number of inoperative sensors, and for geometrical configurations where a
track can pass between the first layer of sensors without leaving a hit.
The photon inefficiency is thus reduced, almost entirely, to that resulting from photons converting in the beam pipe.

The conversion-safe electron veto is appropriate where electrons do not constitute a significant background,
as for example in the $\Hgg$ analysis, both because the invariant mass range of interest is sufficiently far from the Z boson mass,
the largest source of prompt electron pairs,
and because there are two photons to which the requirement can be applied, providing a powerful rejection of an
electron pair being identified as a photon pair.
A more severe rejection of electrons can be achieved by rejecting any photon for which a ``pixel track seed'' consisting
of at least two hits in the pixel detectors suggests a charged-particle trajectory that would arrive at the ECAL
within some window defined around the photon supercluster position.
The efficiencies for photons or electrons to pass either of these requirements, as measured in 8\TeV data, are shown in
Table~\ref{tab:eveto} separately for the barrel and the endcap.
The efficiencies are obtained from photons in $\Zmmg$ events and from electrons in $\Zee$ events,
for photons or electrons that have passed all criteria in the loose photon identification based on sequential requirements (Section~\ref{sec:photon-id-cutbased}) except the electron veto.

\begin{table}[hbtp]
    \topcaption{\label{tab:eveto}Fractions of photons and electrons, in the ECAL barrel and endcap, passing the two different electron vetos.
                                            The statistical uncertainties in the values given for electrons are negligible.
      }
\renewcommand{\arraystretch}{1.1}
  \begin{center}
    \begin{tabular}{l|c c|c c}
      & \multicolumn{2}{c|}{\textbf{Barrel}} & \multicolumn{2}{c}{\textbf{Endcap}} \\
      & $\Pgg$ & $\Pe$ &  $\Pgg$ & $\Pe$  \\
\hline
Conversion-safe veto & $99.1\pm0.1\%$ & 5.3\% & $97.8\pm0.2\%$ & 19.6\% \\
Pixel track seed veto  & $94.4\pm0.2\%$ & 1.4\% & $81.0\pm0.6\%$ & 4.3\%  \\
    \end{tabular}
    \end{center}
\end{table}
\renewcommand{\arraystretch}{1.0}

\subsection{Photon identification variables}
\label{sec:id_variables}

\begin{figure}[hbt]
  \begin{center}
    \includegraphics[width=0.49\textwidth]{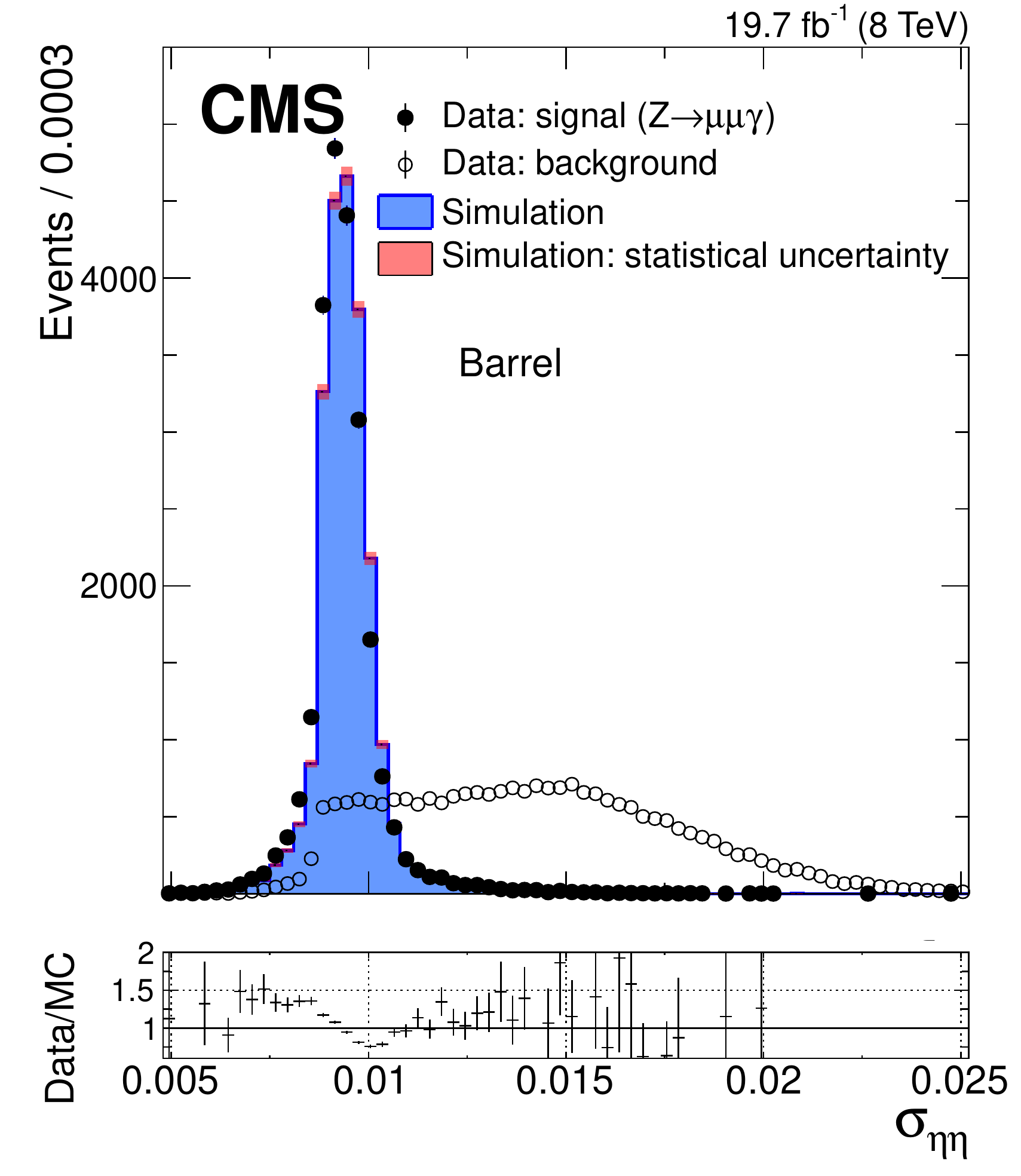}
    \includegraphics[width=0.49\textwidth]{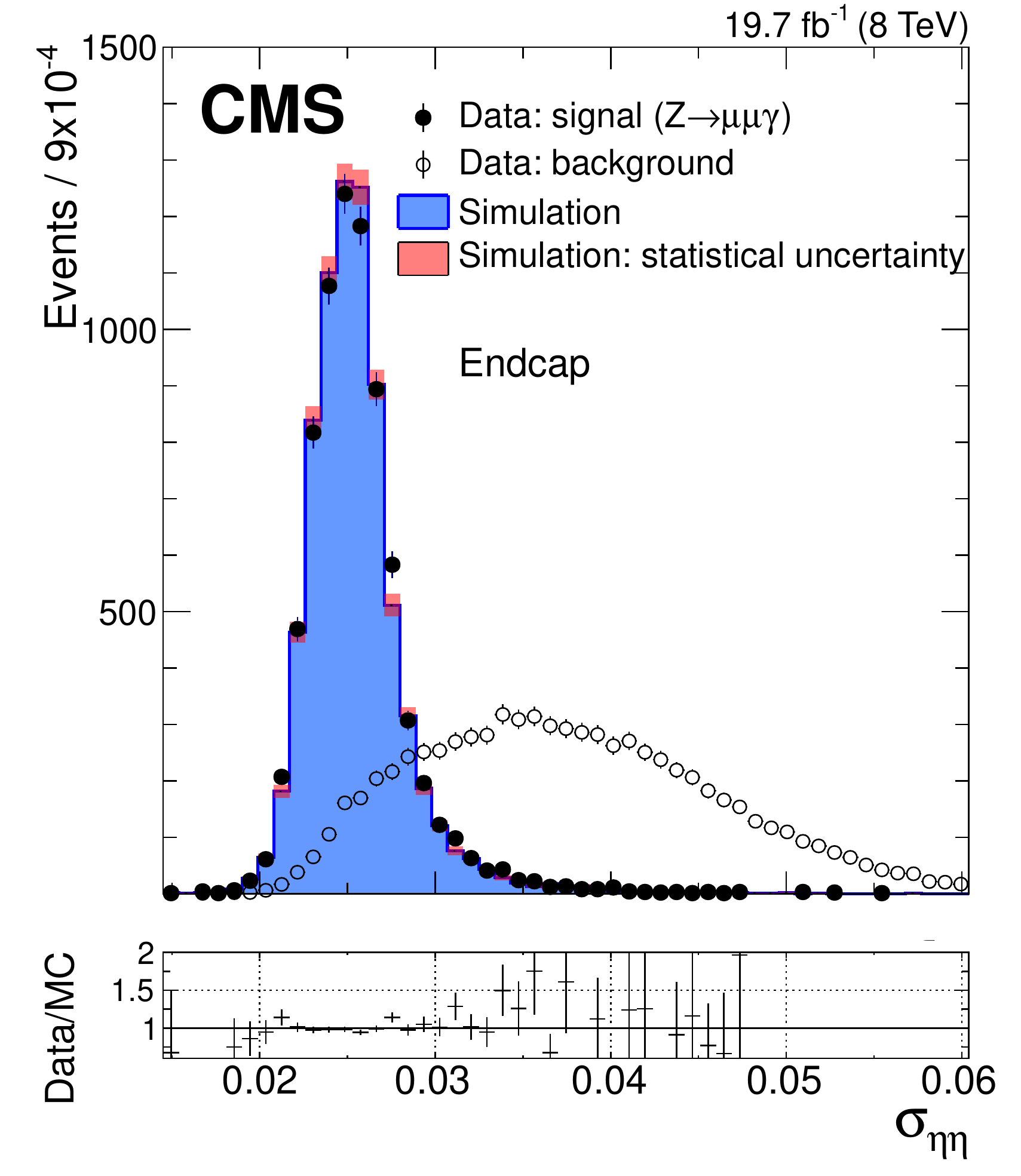}
      \caption{Distribution of the shower-shape variable, $\sigee$, for FSR photons in \Zmmg events in data (solid circles) and simulation (histogram),
and for background-dominated photon candidates in dimuon triggered events (open circles).
The barrel and endcaps are shown separately.
The simulated signal and background distributions are normalized to the number of signal photons in the data.
The ratios between the photon signal distributions in data and simulation are shown in the bottom panels.
}
      \label{fig:sieie_sigBkg}
 \end{center}
\end{figure}

Photon identification is based on two main categories of observables: shower-shape and isolation variables,
and a description is given here of those most commonly used.
The lateral extension of the shower, $\sigee$, is measured in terms of the energy weighted spread within the $5\times5$ crystal
matrix centred on the crystal with the largest energy deposit in the supercluster~\cite{electron-paper}.
This variable, like the variable $\covep$ mentioned below, is obtained by measuring position by counting crystals.
This has the advantage that the differences in the size of the voids between the crystals, particularly at the module boundaries, are ignored,
which better matches the lateral behaviour of showers.
The separation of signal from background by this variable is illustrated in Fig.~\ref{fig:sieie_sigBkg}
where the signal candidates are FSR photons in \Zmmg events.
Photon candidates are required to satisfy $\pt>20\GeV$,
$f_h<0.05$, where $f_h$ is the hadronic fraction defined in more detail below, and the conversion-safe electron veto is applied.
The  \Zmmg events are selected as in Section~\ref{sec:photon-conversion}.
Photons in data are compared with those in a simulated sample.
There is imperfect matching between data and simulation, particularly in the barrel, which has to be taken into account
when using the $\sigee$ variable.
The background-dominated photon candidates are taken from a sample of dimuon triggered events in data.
The simulated distributions are normalized to the number of signal photons in data, and the barrel and endcaps are shown separately.

The variable $\sigee$ is often used in conjunction with $\covep$,
the diagonal component of the covariance matrix constructed from the energy-weighted crystal positions within the $5\times5$ crystal array
centred on the crystal containing the largest energy.
As previously discussed in Section~\ref{sec:photon-clustering}, the $\RNINE$ variable measures the overall transverse spread of the shower.
Additional information on the shower-shape is provided by the ratio $E_{2\times2}/E_{5\times5}$, where $E_{2\times2}$ is the maximum energy sum collected in a $2\times2$ crystal array that includes
the largest energy crystal in the supercluster, and $E_{5\times5}$ is the energy collected in a $5\times5$ crystal matrix centred around the same crystal.
The energy-weighted spreads along $\eta$ ($\sigma_\eta$) and $\phi$ ($\sigma_\phi$),
calculated using all crystals in the supercluster, give further measures of the lateral spread of the shower.
In the endcap, where CMS is equipped with a preshower, the variable $\sigma_\mathrm{RR}=\sqrt{\smash[b]{\sigma^2_\mathrm{xx}+\sigma^2_\mathrm{yy}}}$ is considered, where ${\sigma_\mathrm{xx}}$ and ${\sigma_\mathrm{yy}}$
measure the lateral spread in the two orthogonal sensor planes of the detector.
The hadronic leakage of the shower, $f_h$, is defined as the ratio between the energy collected by the HCAL towers behind the supercluster and the energy of the supercluster.

Photon isolation is measured exploiting the information provided by the particle-flow event reconstruction~\cite{CMS-PAS-PFT-10-001, CMS-PAS-PFT-09-001}.
The particle-flow algorithm combines information from the tracker, the calorimeters, and the muon detectors,
and aims to reconstruct the four-momenta of all particles in the event, classifying them as
charged and neutral hadrons, photons, electrons and muons.
The photon isolation variables are obtained by summing the transverse momenta of charged hadrons, $I_\pi$,
photons, $I_{\gamma}$, and neutral hadrons, $I_\text{n}$, inside an isolation region of radius $\DR$ in the $(\eta, \phi)$ plane around the photon direction.
Since the reconstruction of the signal photons and the particle-flow objects is not (yet) optimally synchronized,
energy from the signal photon must be removed from the isolation sums by imposing geometrical requirements.
When calculating $I_{\gamma}$, particle-flow photons falling in a pseudorapidity slice of size $\Delta \eta = 0.015$
are excluded from the sum.
Similarly, when constructing $I_\pi$, summing the transverse momenta of charged hadrons, a region of $\DR = 0.02$ is excluded.

Charged hadrons are reliably associated with reconstructed primary vertices and thus $I_\pi$ is potentially independent of pileup.
However, the association of photons with a primary vertex is often less than certain, and
an incorrect choice of the vertex used will give a random isolation sum consistent with an isolated photon.
For this reason, two variables are defined, $I_\pi$, where the list of charged hadrons is measured with respect to the primary vertex chosen for the photon, and
 $I_\pi^\text{max}$, where the isolation sum is the largest among those calculated for all reconstructed primary vertices.

\begin{figure}[hbtp]
  \begin{center}
         \includegraphics[width=0.49\textwidth]{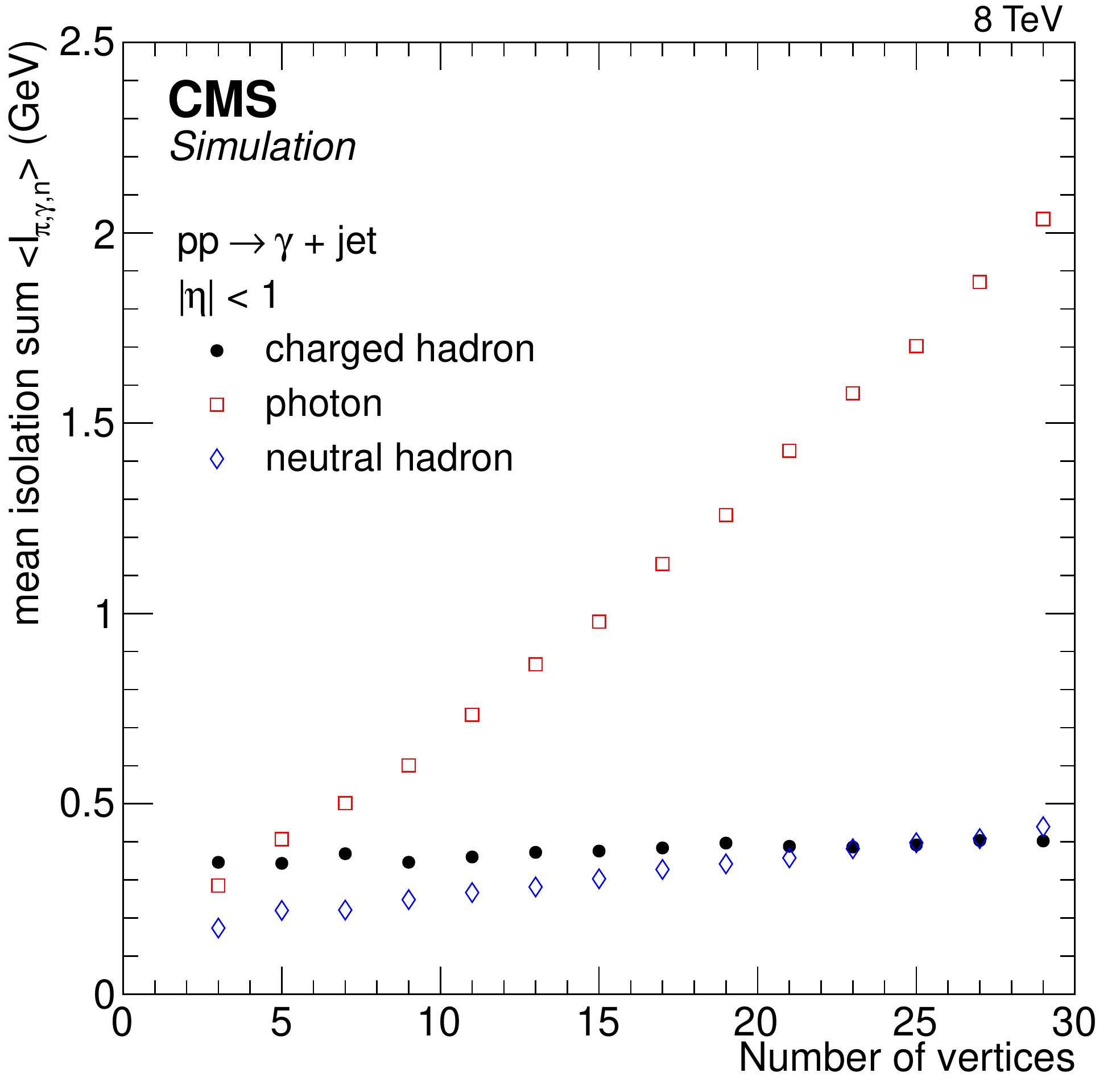}
         \includegraphics[width=0.49\textwidth]{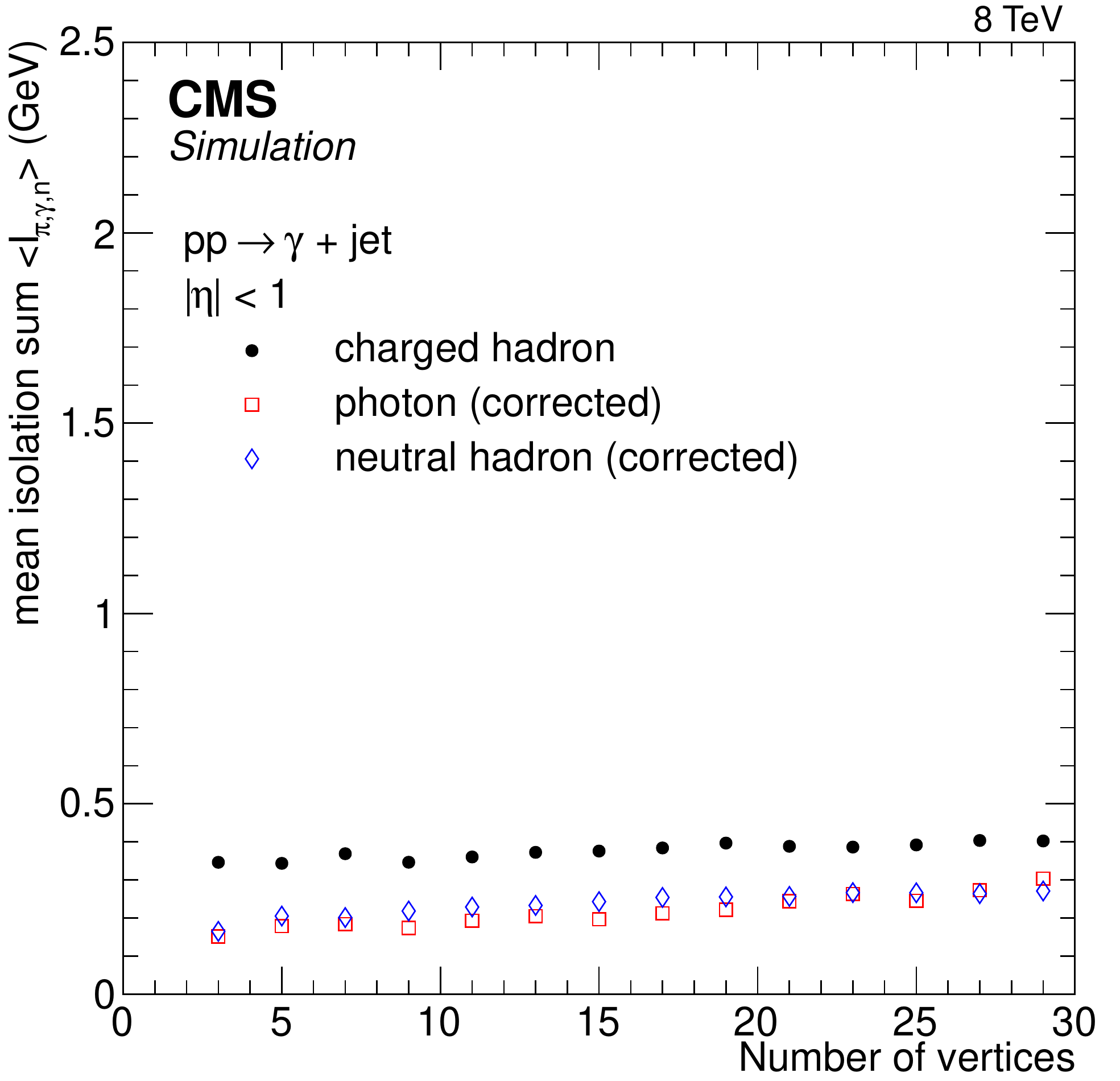}
      \caption{Mean value of the isolation variables for photons with $\pt>50\GeV$ in $\GAMJET$ events,
as a function of the number of reconstructed primary vertices, for events (left) before and (right)
after being corrected for pileup using the $\rho$ variable.}
      \label{fig:rho_corr}
 \end{center}
\end{figure}

When the charged-hadron component of the isolation is calculated from candidates compatible with the chosen primary vertex,
it is independent on the number of pileup events as shown in the left plot of Fig.~\ref{fig:rho_corr},
where the number of reconstructed primary vertices in the event is used as a measure of the number of pileup events.
This illustrative figure is made using photons in $\GAMJET$ events and requiring them to satisfy $\pt>50\GeV$, which,
by ensuring 50\GeV of recoil in the event, results in a high probability that the primary vertex of the hard interaction, and hence of the photon, is correctly identified.
The variables constructed by summing photons and neutral hadrons, inside an isolation region,
need to be corrected to remove the contribution from pileup.
The extra contribution in the isolation region is estimated as $\rho\, A_\text{eff}$,
where $\rho$ is the median of the transverse energy density per unit area in the event~\cite{Cacciari:2007fd} and $A_\text{eff}$ is the area of the isolation
region weighted by a factor that takes into account the dependence of the pileup transverse energy density on pseudorapidity.
The effective areas have been determined in $\GAMJET$ events.
When the extra contribution due to pileup, calculated using $\rho$, is subtracted from the photon and neutral hadron sums,
their dependence on the number of vertices is removed (Fig.~\ref{fig:rho_corr}, right).

\begin{figure}[hbtp]
  \begin{center}
      \includegraphics[width=0.49\textwidth]{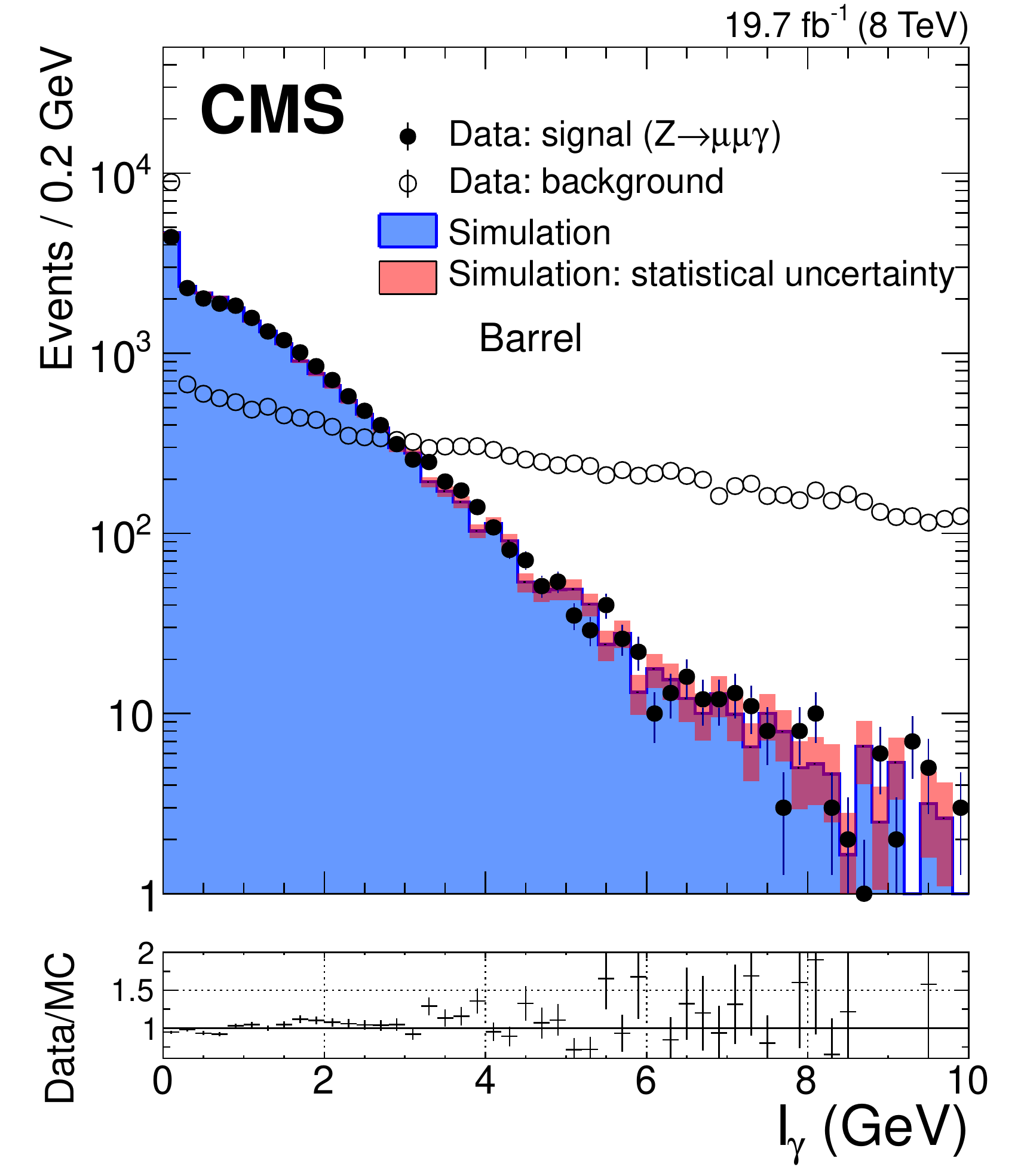}\\
      \includegraphics[width=0.49\textwidth]{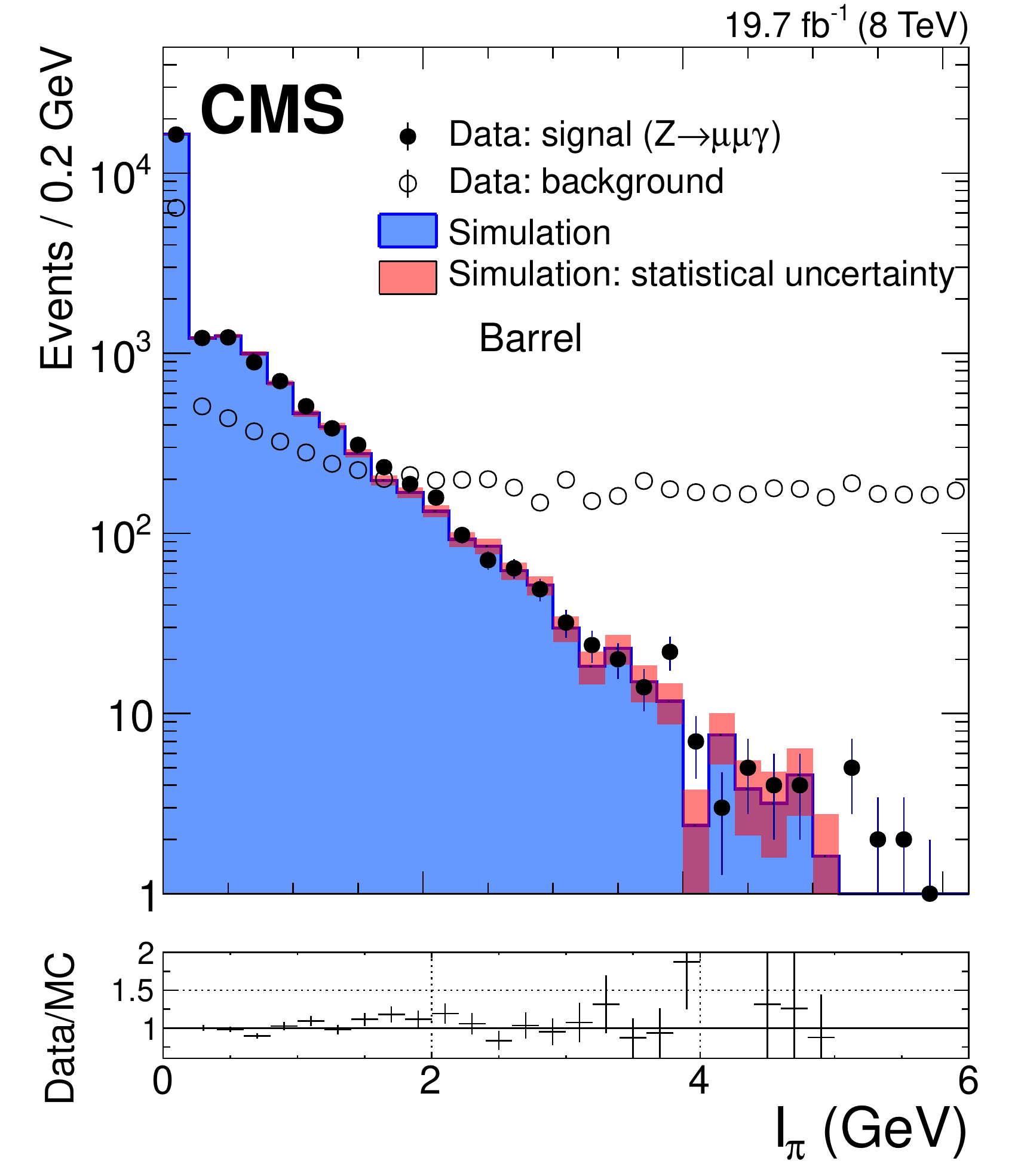}
      \includegraphics[width=0.49\textwidth]{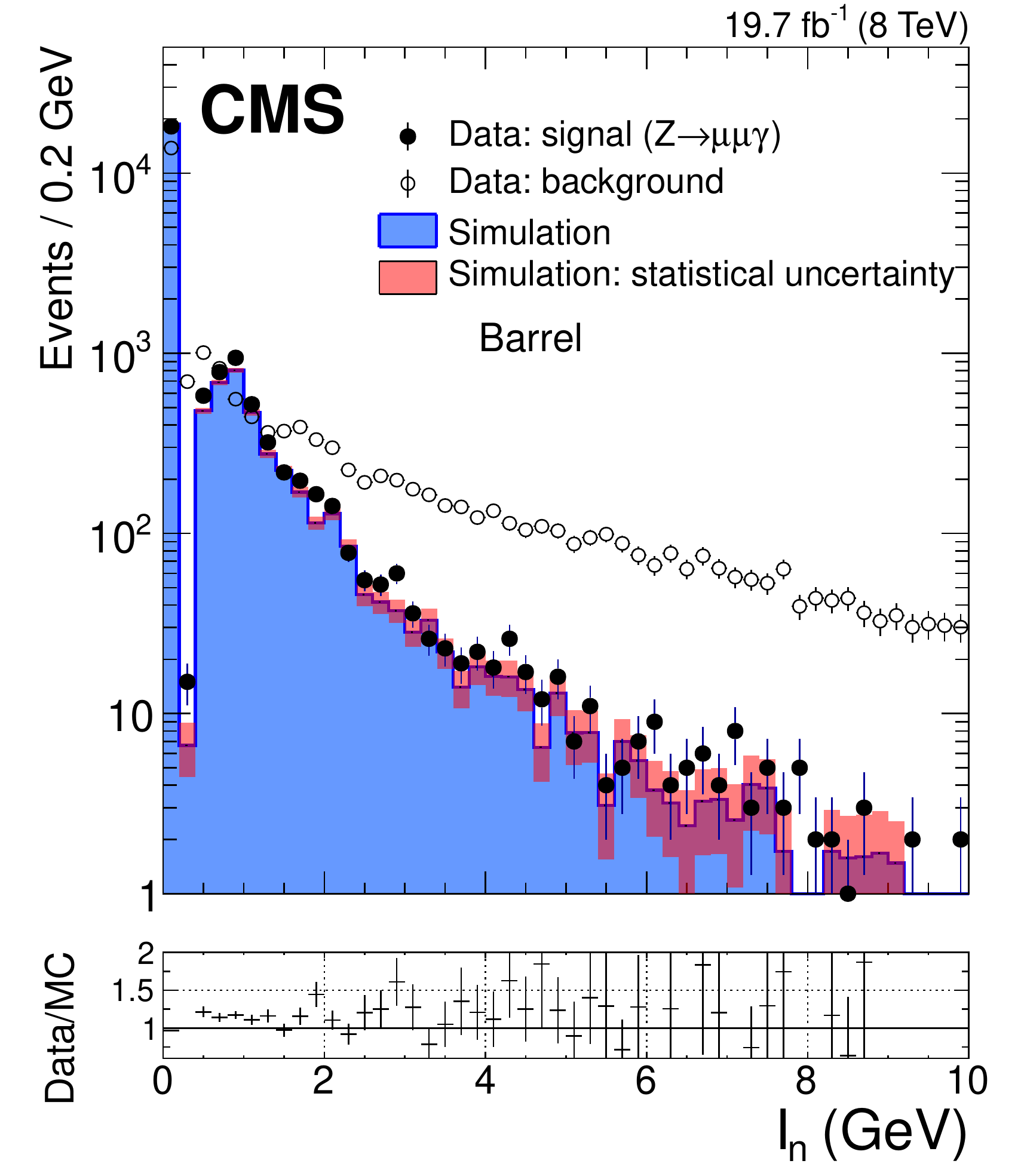}
      \caption{Distributions of the isolation variables: (top) $I_{\gamma}$, (bottom left) $I_\pi$, and (bottom right) $I_\text{n}$, constructed from particle-flow objects.
The distributions are shown for FSR photons from \Zmmg events in data (solid circles) and simulation (histogram)
and for background-dominated photon candidates in dimuon triggered events (open circles).
The simulated signal and background distributions are normalized to the number of signal photons in data.
The ratios between the photon signal distributions in data and simulation are shown in the bottom panels.
}
      \label{fig:iso_sigBkg}
 \end{center}
\end{figure}

Figure~\ref{fig:iso_sigBkg} illustrates how the three isolation variables defined above behave for
signal and background, as well as the good agreement between data and simulation for a region with radius $\DR=0.3$.
The figure shows the distribution of the variables for photons in the ECAL barrel.
Similar results are found in the endcaps.
The signal photons shown have high purity and are from \Zmmg events, and the background-dominated candidates are obtained from data, as in Fig.~\ref{fig:sieie_sigBkg}.
A value of zero is plotted for the isolation variables in those cases when the pileup subtraction results in a negative value.
For the distributions of the variables for signal photons, the ratio of values found in data and simulation is shown.

\subsection{Photon identification based on sequential requirements}
\label{sec:photon-id-cutbased}

\begin{figure}[htbp]
  \begin{center}
         \includegraphics[width=0.49\textwidth]{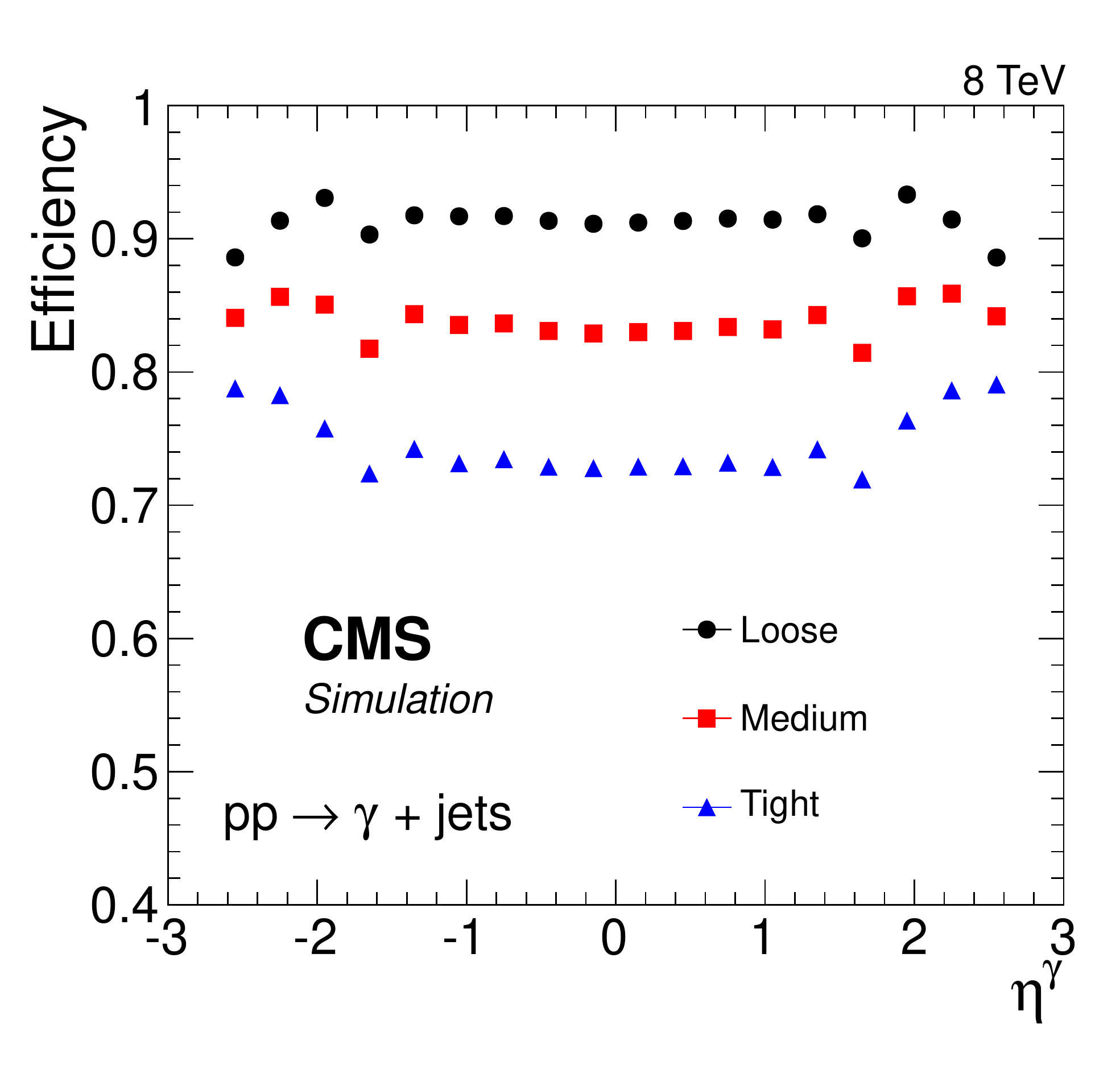} \\
         \includegraphics[width=0.49\textwidth]{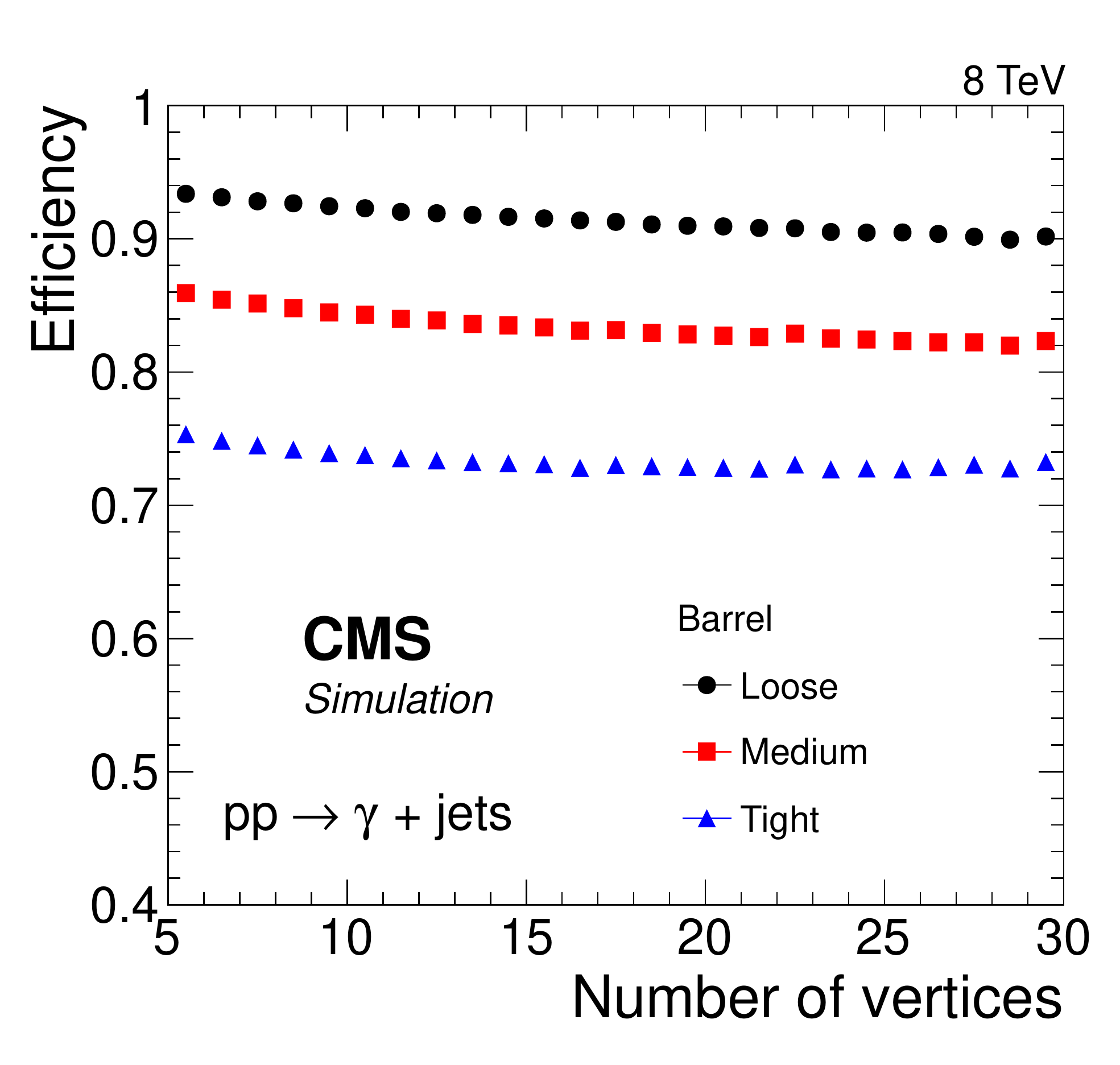}
         \includegraphics[width=0.49\textwidth]{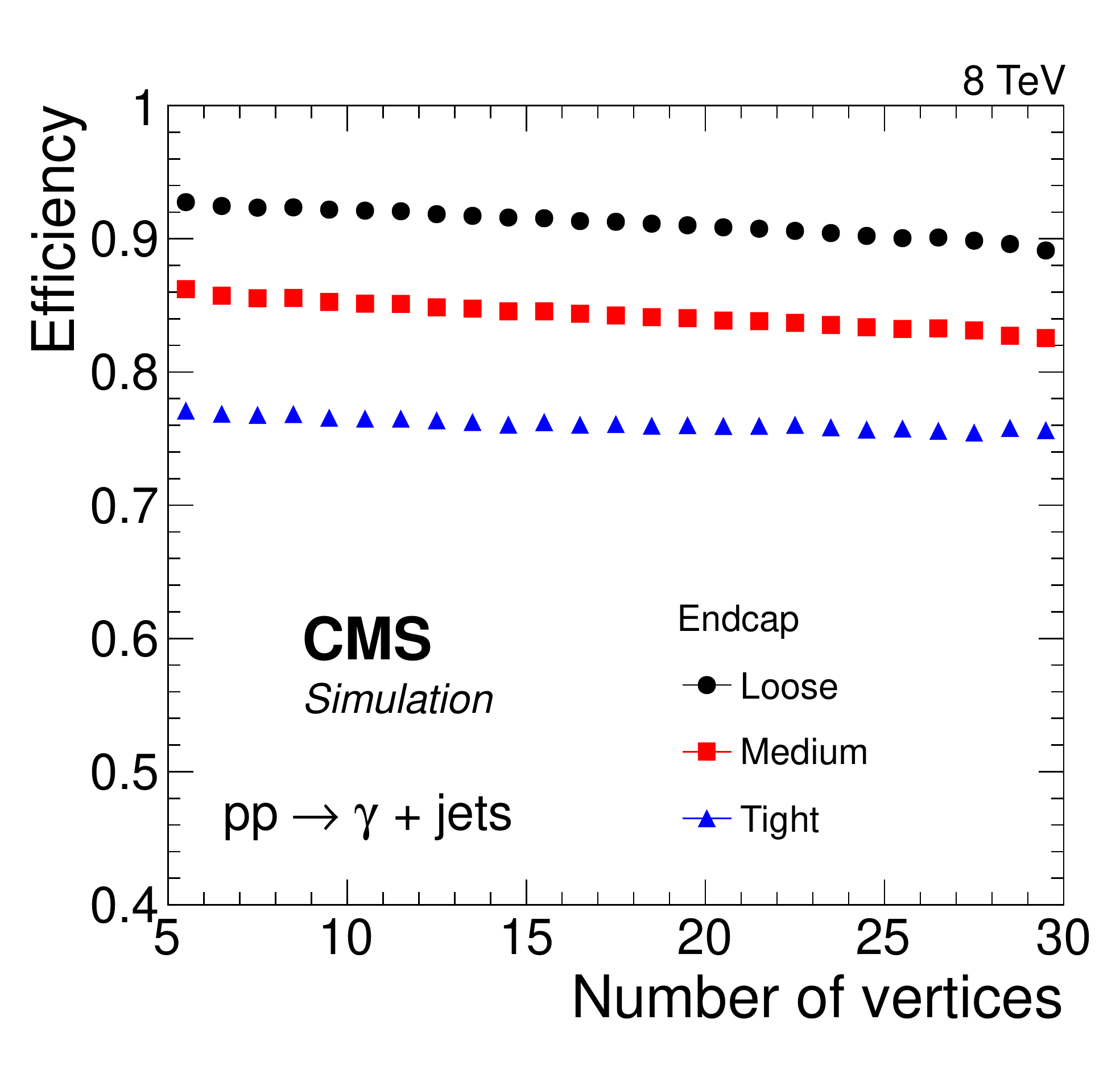} \\
         \includegraphics[width=0.49\textwidth]{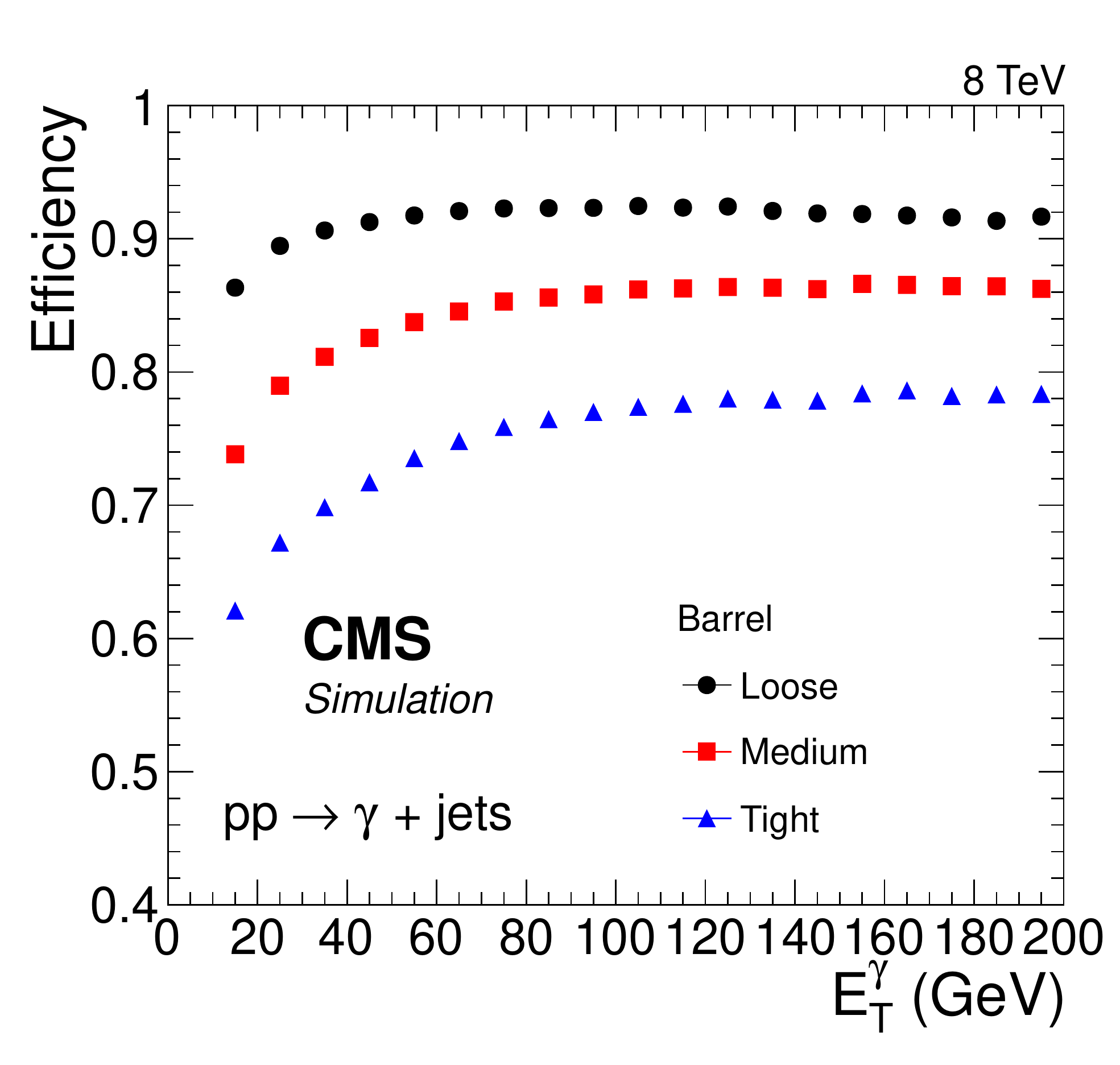}
         \includegraphics[width=0.49\textwidth]{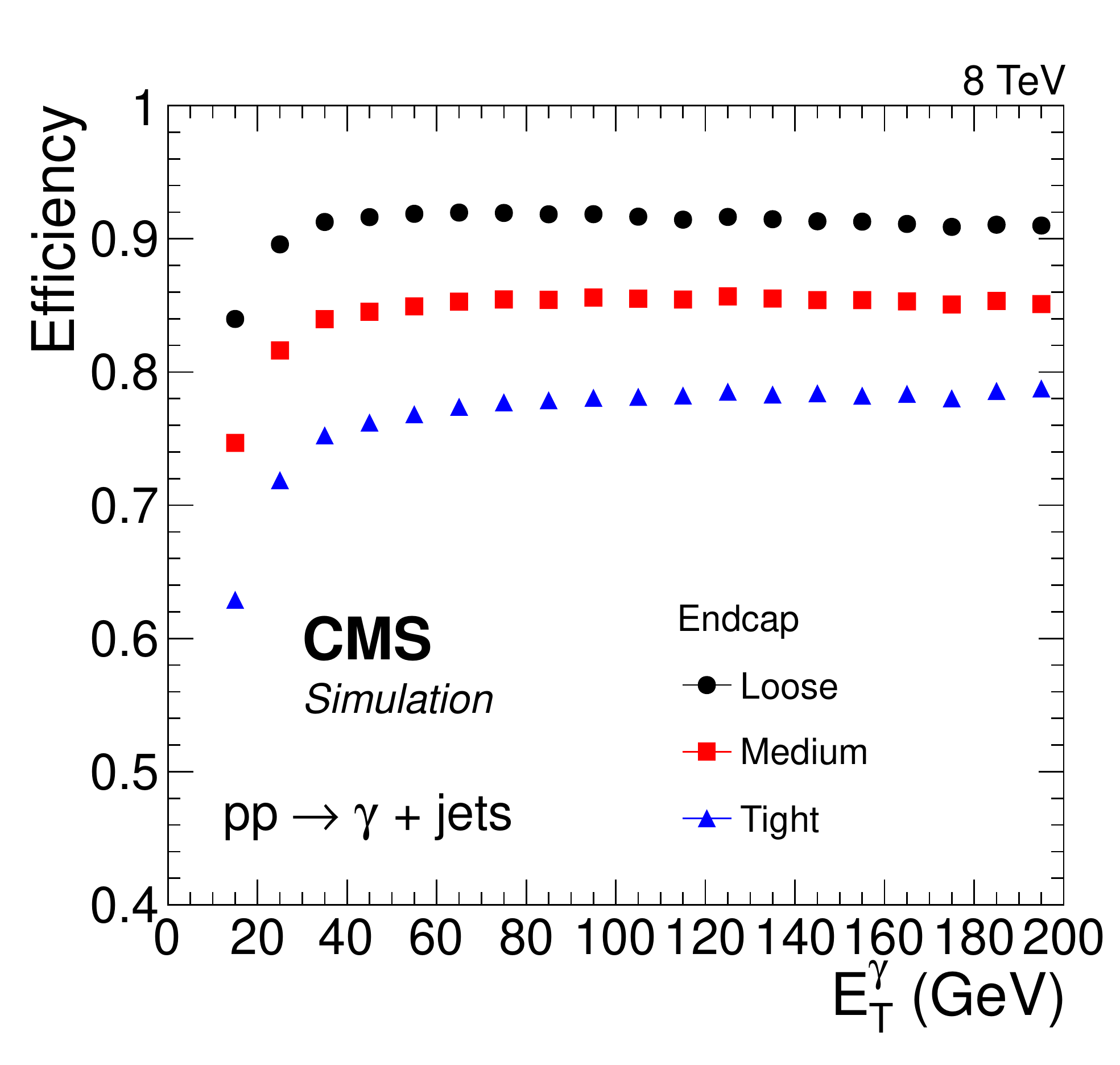}
      \caption{Efficiency of photon identification based on sequential requirements in simulated $\GAMJET$ events
       for three different working points,
       as a function of the (top) photon pseudorapidity, (middle) number of pileup vertices, and (bottom) photon transverse momentum.
      The efficiencies shown include the electron veto requirement.}
      \label{fig:cutbased_eff}
 \end{center}
\end{figure}

This section describes the identification of photons by sequential application of requirements.
Various versions have been used in different data analyses, although the basic principles remain the same.
After applying the electron veto, requirements are made on $\sigee$, $f_h$,
and the isolation sums.
In most cases, the isolation thresholds are expressed as a constant term added to a term proportional to the candidate photon transverse momentum, $\ptg$.
A summary of the standard photon identification requirements, where different combinations of requirements and thresholds
are used for the barrel and the endcap, is given in Table~\ref{table:photon-id-cutbased} for three different working points.
The working points correspond to selections of different stringency, and
the corresponding efficiency curves are shown in Fig.~\ref{fig:cutbased_eff},
for photon candidates with $\pt>15\GeV$ in a sample of simulated $\GAMJET$ events.

\begin{table}[hbtp]
\topcaption{Photon identification requirements for three working points corresponding to selections of different stringency.}
\begin{center}
\begin{tabular}{ll|r|r|r}
 & &\multicolumn{1}{c|}{Loose} & \multicolumn{1}{c|}{Medium} & \multicolumn{1}{c}{Tight}  \\
\hline
\multirow{2}{*}{ $I_{\gamma}$}&\multicolumn{1}{r|}{barrel} & $1.3\GeV+0.005\,\ptg$  & $0.7\GeV+0.005\,\ptg$ & $0.7\GeV+0.005\,\ptg$ \\
&\multicolumn{1}{r|}{endcap} & \multicolumn{1}{c|}{---} & $1\GeV+0.005\,\ptg$ & $1\GeV+0.005\,\ptg$ \\
\hline
\multirow{2}{*}{$I_\text{n}$}&\multicolumn{1}{r|}{barrel} & $3.5\GeV+0.04\,\ptg$ & $1.0\GeV+0.04\,\ptg$ & $0.4\GeV+0.04\,\ptg$ \\
&\multicolumn{1}{r|}{endcap} & $2.9\GeV+0.04\,\ptg$ & $1.5\GeV+0.04\,\ptg$ & $1.5\GeV+0.04\,\ptg$ \\
\hline
\multirow{2}{*}{$I_\pi$}&\multicolumn{1}{r|}{barrel} & 2.6\GeV & 1.5\GeV & 0.7\GeV \\
&\multicolumn{1}{r|}{endcap} & 2.3\GeV & 1.2\GeV & 0.5\GeV \\
\hline

\multirow{2}{*}{$\sigee$}&\multicolumn{1}{r|}{barrel} & 0.012 & 0.011 & 0.011 \\
&\multicolumn{1}{r|}{endcap} & 0.034 & 0.033 & 0.031 \\
\hline
$f_h$& & \multicolumn{3}{c}{0.05} \\
\hline
Electron veto & & \multicolumn{3}{c}{conversion-safe} \\
\end{tabular}
\end{center}
\label{table:photon-id-cutbased}
\end{table}

\begin{figure} [hbtp]
  \begin{center}
         \includegraphics[width=0.49\textwidth]{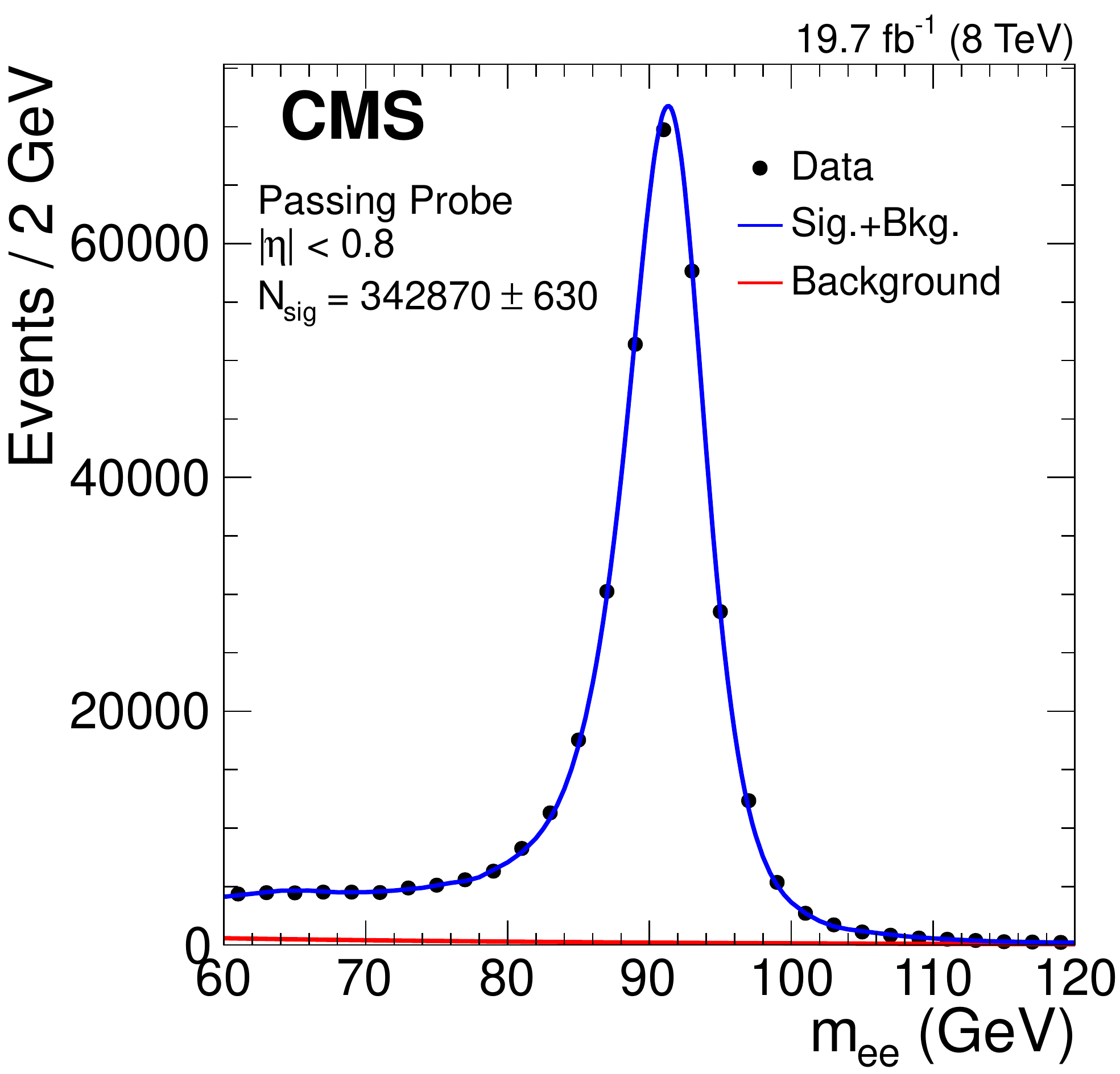}
         \includegraphics[width=0.49\textwidth]{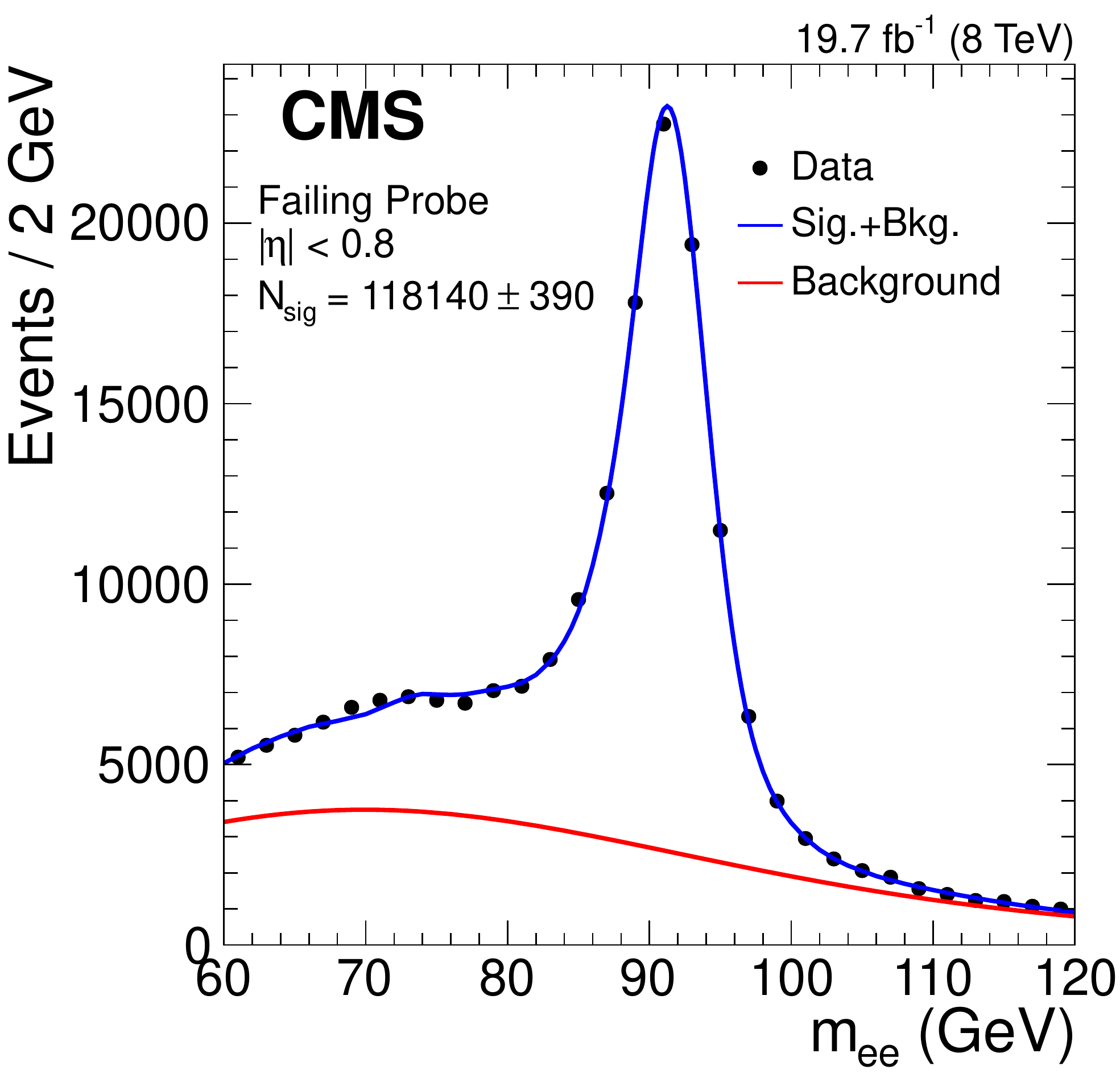}
      \caption{Example of fits to the $\Zee$ invariant mass distribution for (left) passing
  and (right) failing probes, in the transverse momentum range $20<\pt<30\GeV$ and $\abs{\eta} < 0.8$.}
      \label{fig_cutbased_TP}
 \end{center}
\end{figure}

\begin{figure} [hbtp]
  \begin{center}
        \includegraphics[width=0.49\textwidth]{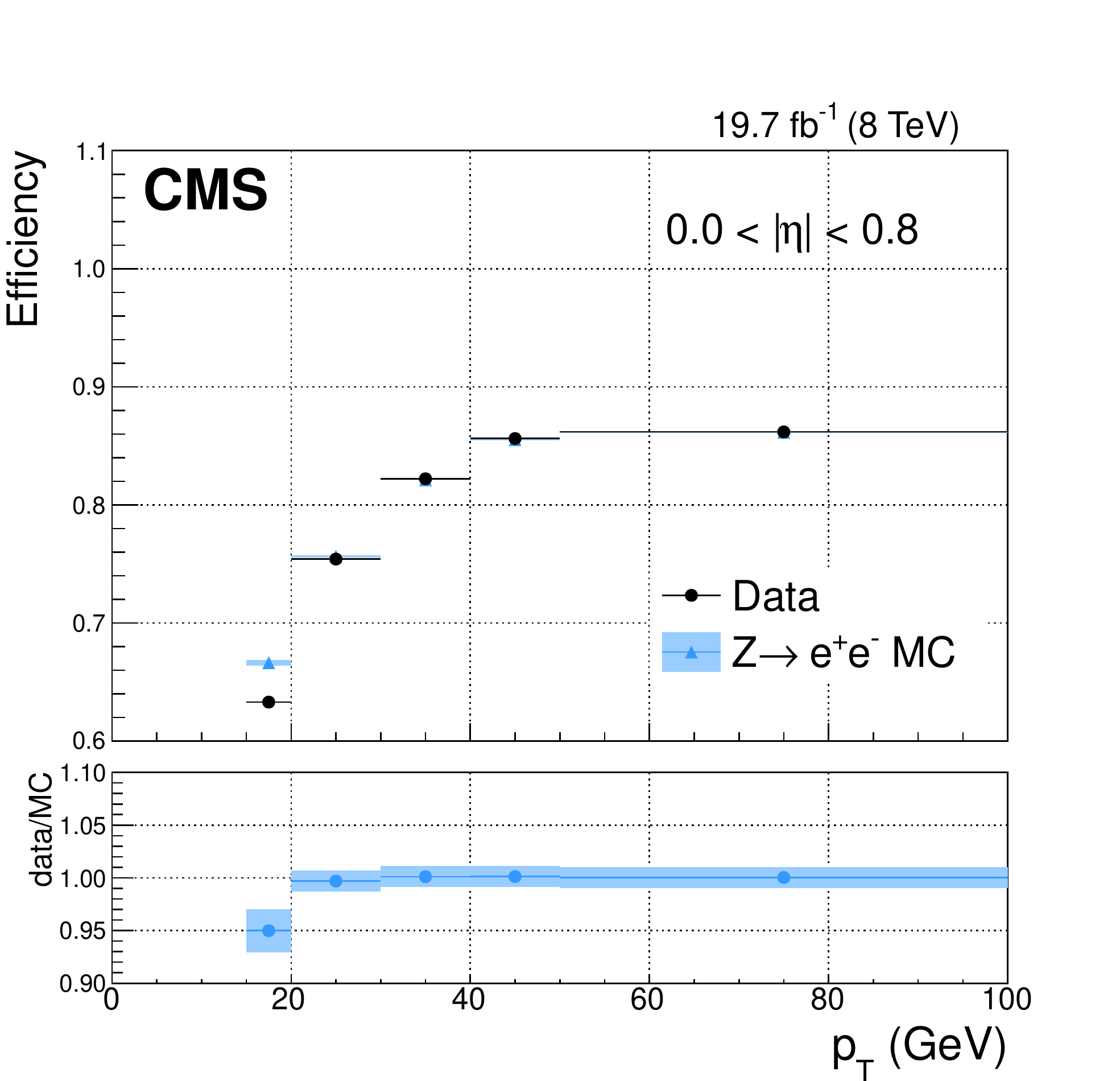}
        \includegraphics[width=0.49\textwidth]{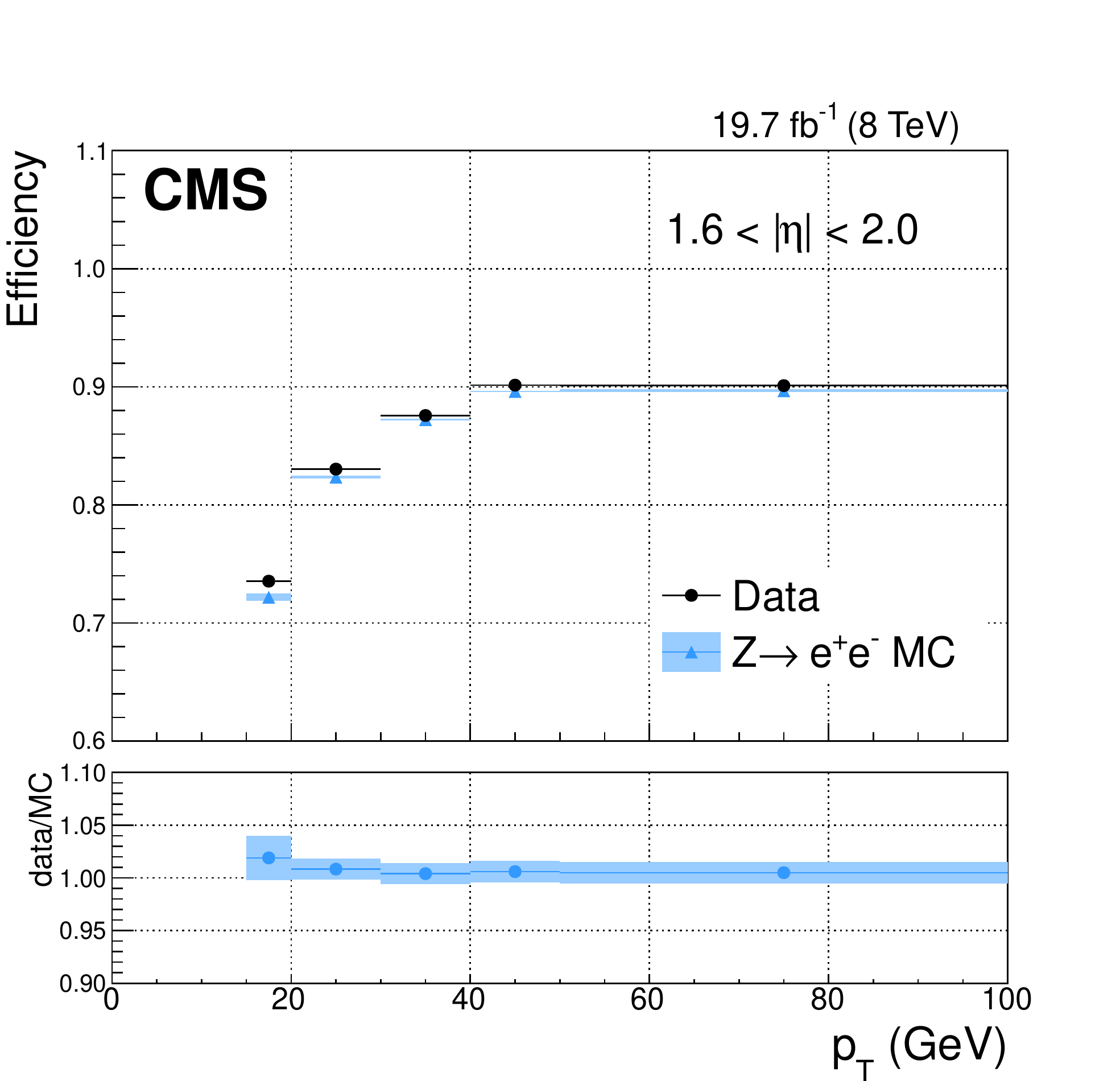}
        \caption{
Comparison of the selection efficiency as a function of photon transverse momentum in data
(circles) and simulation (triangles) for the identification based on sequential requirements for (left) $\abs{\eta} < 0.8$ and (right) $1.6 < \abs{\eta} < 2$.
Statistical and systematic uncertainties are respectively shown by the error bars and shaded bands.
The horizontal error bars mark the full width of the \pt bins in which the measurements are made, and the data points are plotted at
the centre of each bin.
         The ratios of efficiencies in data and simulation are shown in the bottom panels.}
      \label{fig_cutbased_scaleFactors}
 \end{center}
\end{figure}

Photon identification efficiencies are measured with the ``tag-and-probe'' method, as described in Ref.~\cite{T&P},
using samples of $\Zee$ events.
The results of these measurements can be used to correct the simulation for any mismodelling by evaluating the ratio of efficiencies in data and simulation.
For the results shown here, refinements to the simulation were implemented to reproduce the changes of the magnitude of
the energy-equivalent electronic noise during the data-taking period (most relevant for the barrel),
and to better simulate the effects of out-of-time pileup (more relevant for the endcaps).
These refinements have been described in Section~\ref{sec:dat-sim}.
Electrons resulting from Z-boson decays, in a data sample passing the 27\GeV single-electron trigger, are used for the measurement.
The ``tag'' candidates are required to have $\pt>30\GeV$, satisfy tight electron identification~\cite{electron-paper},
and be matched to a triggering electron.
The dielectron invariant mass is required to be in the range $60<m_{\Pe\Pe}<120\GeV$.
The ``probe'' candidates are electron showers reconstructed as photons and matched to the non-tag electron.
They are required to have $\pt>15\GeV$ and are tested for passing (or not) the photon identification criteria,
with the exception of the electron veto.
Invariant mass distributions are then made separately for the cases in which the probe photons satisfy or fail the identification requirements,
hereafter referred to as ``passing'' and ``failing'' distributions.
Simultaneous fits to the passing and failing distributions are performed to extract the identification efficiency.
The Z-boson invariant mass distribution is modelled with a template extracted from simulation and convolved with a Gaussian function.
The background is modelled with an exponential times an error function.
Figure~\ref{fig_cutbased_TP} shows an example of fits to the $\Zee$ mass peak for the central barrel region.
The transverse momentum of the probe photon is in the range $20<\pt<30\GeV$ and the identification
criteria correspond to the medium working point quoted in Table~\ref{table:photon-id-cutbased}.
The number of events in data is such that the statistical uncertainties in the data points, shown by error bars, are not
visible in the figure.
The fitted numbers of signal events in the two plots give a measured efficiency of 74\% with negligible statistical uncertainty.
The hump on the left side of failing probes is due to radiating electrons for which a fraction of energy is not collected.
Figure~\ref{fig_cutbased_scaleFactors} shows the comparison of the selection efficiency in data and simulation,
as a function of the photon transverse momentum, for barrel and endcap separately.
The values are obtained using electrons from $\Zee$ decays with a tag-and-probe technique, with the probe electron reconstructed as a photon, and the electron veto removed from the identification criteria.
The data-to-simulation ratio, showing a good level of agreement for $\pt>20\GeV$ is shown in the panels beneath the main plot.
The shaded bands represent the systematic uncertainties, which
have been evaluated by replacing, in the fits to the invariant mass distribution, the background modelling with simple exponential and polynomial functions.
The statistical uncertainties in the data measurements are too small to be visible.
Since the measurement is made for an electron sample, the electron veto is not applied,
and its efficiency (Table~\ref{tab:eveto}) and the agreement of data and simulation, are measured separately.
The different level of efficiency in the barrel and the endcap seen in Fig.~\ref{fig_cutbased_scaleFactors}, but not in Fig.~\ref{fig:cutbased_eff},
is explained by the requirement (or not) of the electron veto.

The background rejection, defined as the reciprocal of the efficiency of background photons to pass selection requirements, and the signal efficiency
have been determined for the three working points of the photon identification based on sequential requirements
in a simulated $\GAMJET$ sample.
The signal corresponds to reconstructed photons matched to simulated prompt photons and the background corresponds to reconstructed photons matched to a jet.
The signal transverse momentum distribution is reweighted to follow the background spectrum,
and photon candidates are required to satisfy $25<\pt<200\GeV$.
In Section~\ref{sec:photon-id-mva} the values found for background rejection and signal efficiency when using photon identification based on
sequential requirements are compared with the background rejection as a function of signal efficiency obtained with the multivariate photon identification.

\subsection{Multivariate photon identification}
\label{sec:photon-id-mva}

A more sophisticated photon identification technique is based on a multivariate analysis,
employing a boosted decision tree (BDT) implemented in the \textsc{tmva} framework~\cite{Hocker:2007ht} .
The technique allows the definition of a single discriminating variable characterizing each photon (the BDT score)
resulting from the combination of many variables discriminating prompt photons from background candidates.
The list of variables used as the input to the BDT includes all shower-shape and isolation variables described earlier,
plus three quantities that strengthen the discrimination of signal and background by
accounting for the dependencies in the shower-shape and isolation variables on the pileup present in the event, and the $\eta$ and $\ET$ of the candidate photon:
the median energy per unit area, $\rho$, and the $\eta$ and uncorrected energy of the supercluster corresponding to the candidate photon.

The multivariate photon identification was developed in the context of the $\Hgg$ analysis, which uses a diphoton trigger employing a loose photon selection.
To ensure the independence of the analysis from the online requirements imposed with the trigger, a preselection
is applied to photons candidates.
The preselection makes similar requirements, but somewhat more severe, to those made online by the trigger.
Simulated events are required to satisfy the same preselection requirements  listed in Table~\ref{table:preselCuts}.
Besides the variables already described in Section~\ref{sec:photon-id},
two further isolation variables are used, $I_\mathrm{HCAL}$ and $I_\text{Trk}$, which are the sums of transverse energy in the HCAL towers,
and charged-particle tracks with $\pt>1\GeV$, respectively, in regions of $\DR<0.3$ about the photon candidate.
The HCAL sum is uncorrected for pileup, and the charged-particle track sum uses the tracks associated to the vertex with the highest $\Sigma\pt^2$
of associated tracks, as is done in the trigger.
There are different requirements for photon candidates depending on whether they have a high or low value of the $\RNINE$ variable.
The value used to define this categorization, $\RNINE=0.9$, reflects the one used in the trigger.

\begin{table}[hbtp]
  \begin{center}
  \topcaption{Preselection requirements used for the $\Hgg$ analysis.}
    \begin{tabular}{l || c c | c c | c | c | c} 
 & \multicolumn{2}{c|}{$f_h$} & \multicolumn{2}{c|}{\sigee} & $I_\mathrm{HCAL}$ &  $I_\text{Trk}$  & $I_\pi$\\ 
$\RNINE$ & barrel & endcap & barrel & endcap & & & \\ \hline
      $\le0.9$ & $<0.075$ & $<0.075$ & $<0.014$ & $<0.034$ & $<4\GeV$  & $<4\GeV$  & $<4\GeV$ \\ \hline
      $>0.9$  & $<0.082$ & $<0.075$ & $<0.014$ & $<0.034$ & $<50\GeV$ & $<50\GeV$ & $<4\GeV$ \\ 
    \end{tabular}
  \label{table:preselCuts}
  \end{center}
\end{table}

The BDT is trained on a sample of simulated $\GAMJET$ events, where the photon candidates matching the prompt photon are used as signal,
and photon candidates not matching the prompt photon are used as background.
The photon candidates are required to have $\pt>20\GeV$ and to satisfy the preselection.
The photon transverse momentum and pseudorapidity in signal are reweighted to match
the corresponding distribution of background non-prompt photons, so that the input signal-to-background ratio does not depend on $\pt$.
Since the training of the BDT is performed with simulated samples, it is important to verify the quality of the modelling of all input variables.
The input variables are studied in $\Zee$ events where the electrons are reconstructed as photons and in $\Zmmg$ events.
Examples of the comparison of the distributions of input variables in data and simulated events
are shown for signal photons in Figs.~\ref{fig:sieie_sigBkg} and~\ref{fig:iso_sigBkg}.

\begin{figure}[htbp]
  \begin{center}
         \includegraphics[width=0.99\textwidth]{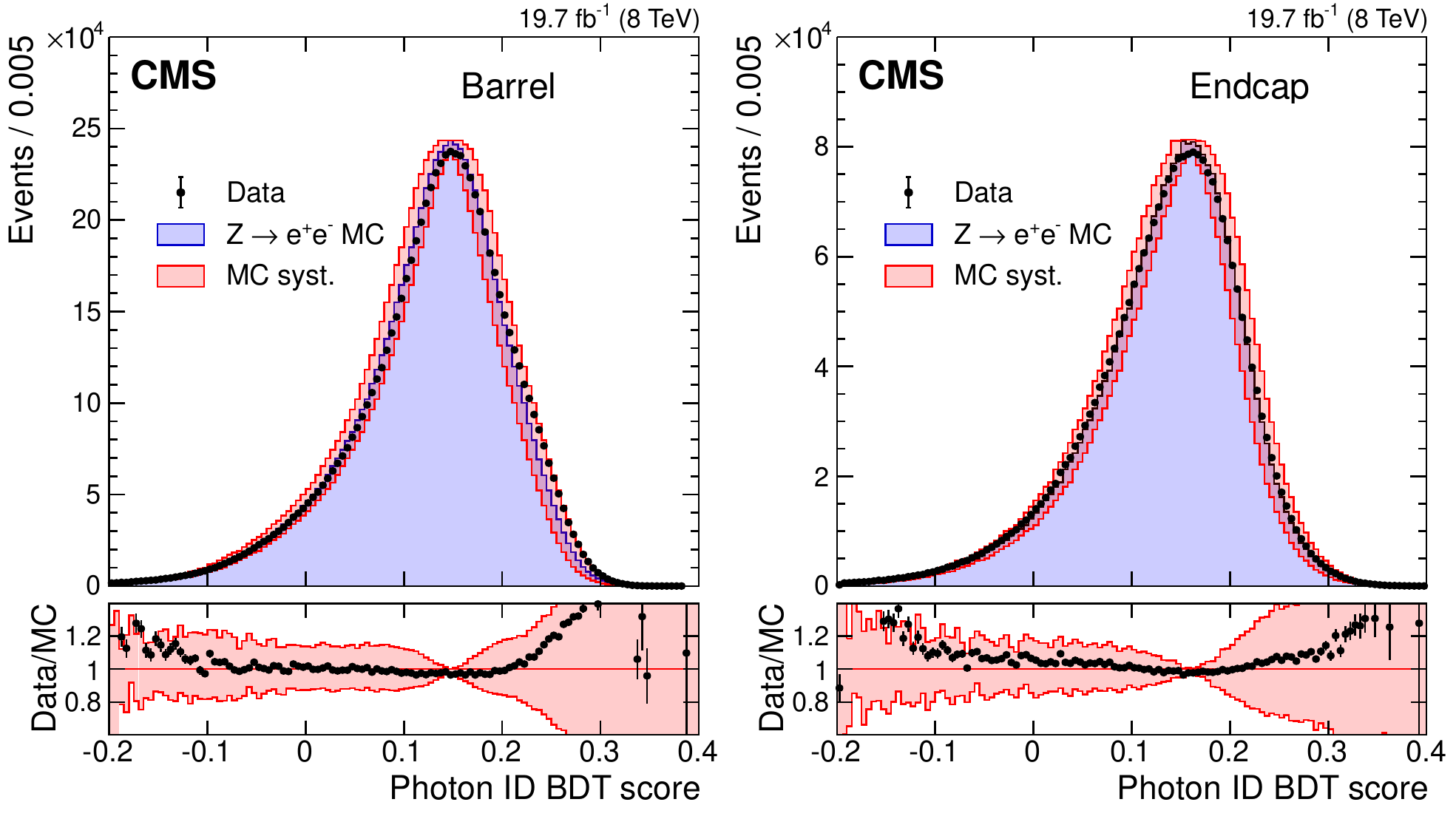}
      \caption{Photon identification BDT score for electrons from $\Zee$ reconstructed as photons in the (left) ECAL barrel and (right) endcap.
       The distributions in data are compared to those in simulated Drell--Yan events.
       The shaded bands correspond to a shift of $\pm$0.01 applied to the score in simulated events.
       The corresponding ratios of data to simulation are shown in the bottom panels.}
      \label{fig:idmva_zee}
 \end{center}
\end{figure}

\begin{figure}[hbtp]
  \begin{center}
         \includegraphics[width=0.455\textwidth]{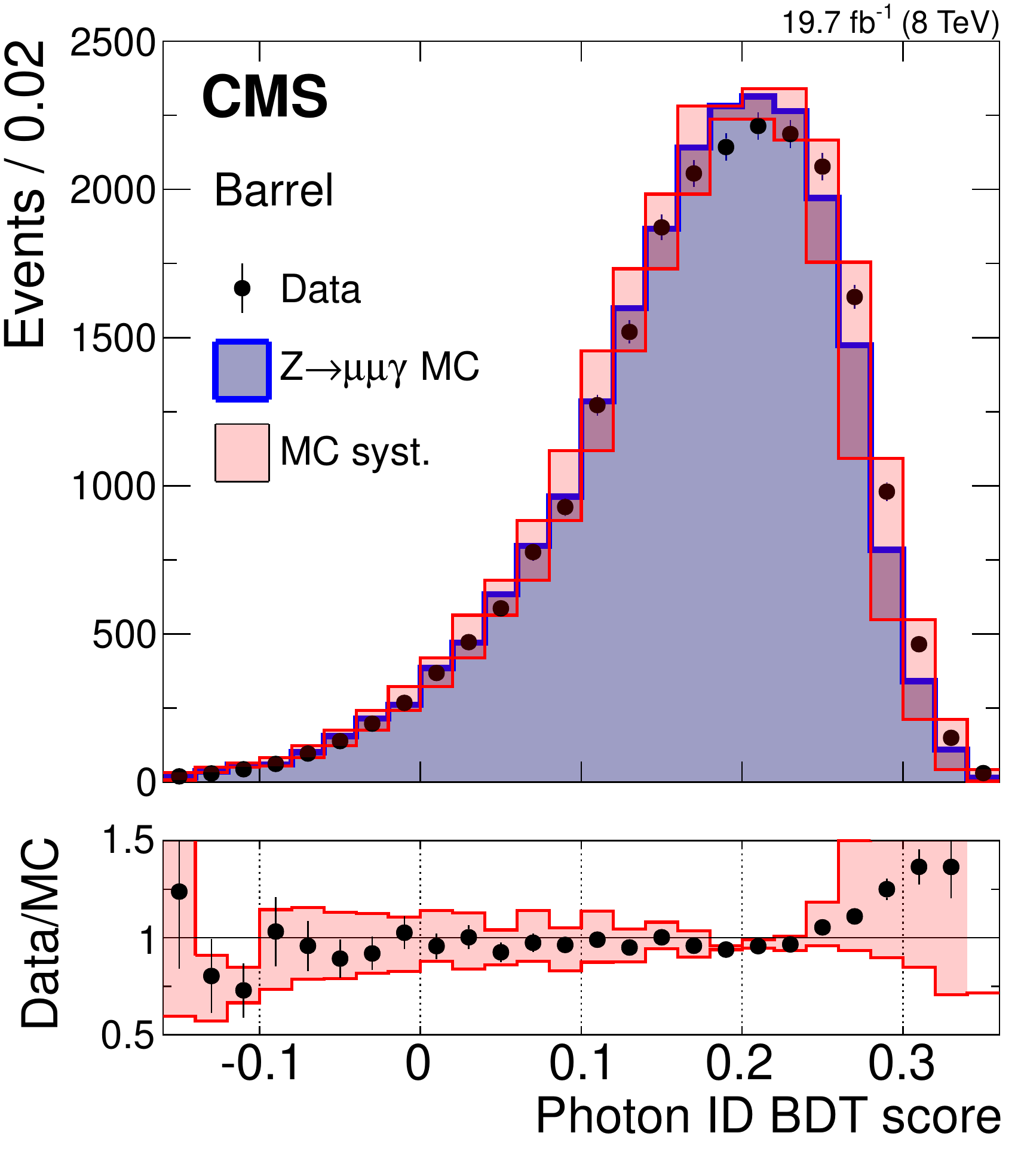}
          \includegraphics[width=0.455\textwidth]{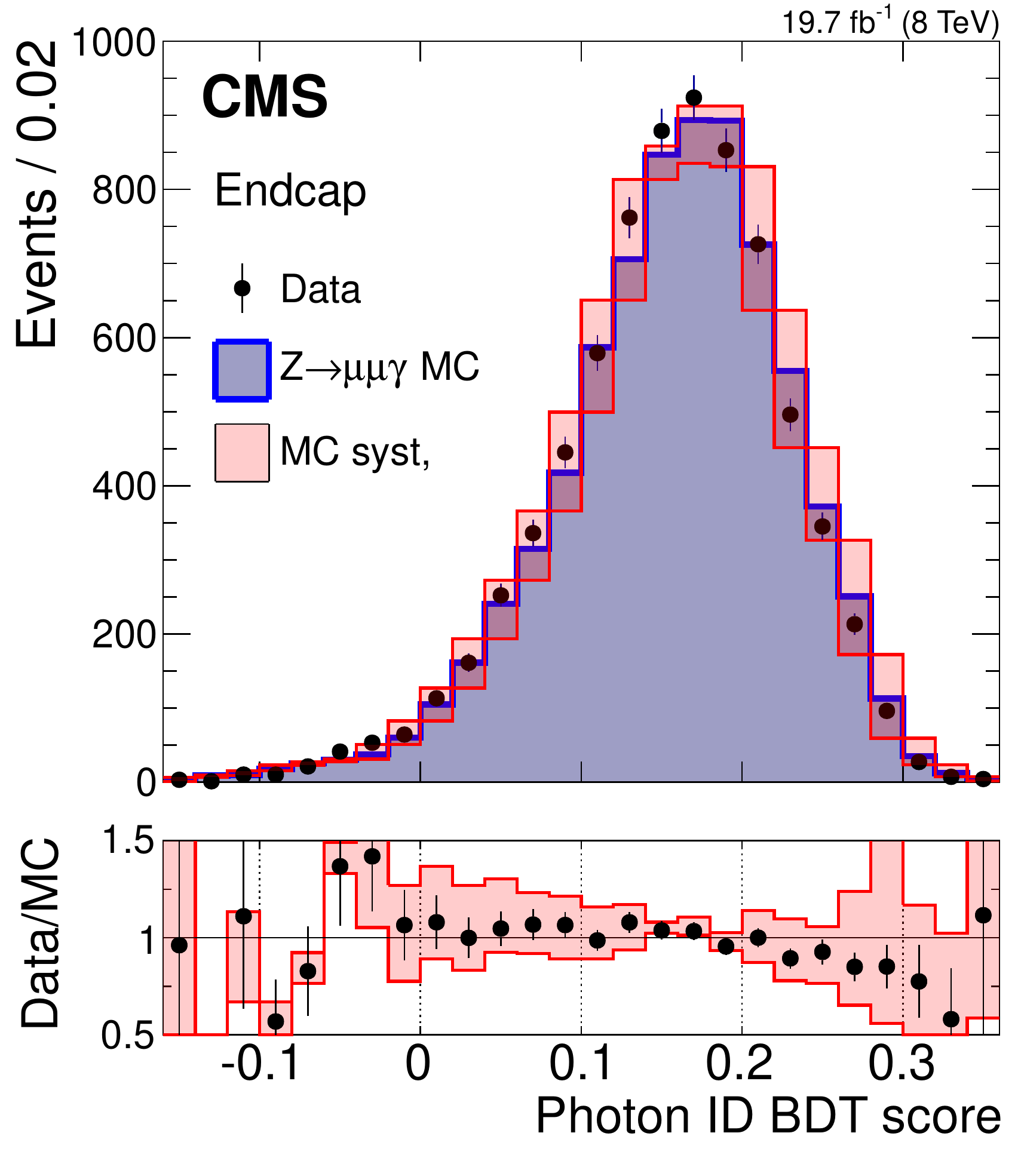}
      \caption{Identification BDT estimator for photons from $\Zmmg$ with transverse momentum above 20\GeV.
      Data (points with error bars) are compared to $\Zmmg$ events selected in Drell--Yan simulation (histograms).
      The shaded bands correspond to a shift of $\pm$0.01 applied to the estimator value in simulation.
      The corresponding ratios of data to simulation are shown in the bottom panels.}
      \label{fig:idmva_zmumug}
 \end{center}
\end{figure}

Figure~\ref{fig:idmva_zee} shows the distribution of the BDT score for $\Zee$ events, where the electrons are reconstructed as photons.
The distributions of the BDT score in data and simulation agree well.
A shift of $\pm$0.01 of the score is shown as a band in the plot.
This shift comfortably covers the small differences between the distributions in data and simulation, and
is taken as the uncertainty in the value of the photon identification BDT score predicted by simulation.
The same comparison can be made for photon candidates in $\Zmmg$ events, and Fig.~\ref{fig:idmva_zmumug} shows the
distributions of the BDT score for photons in data and in simulated events.
The agreement is again good for photons in both the barrel and the endcaps.

\begin{figure}[htbp]
  \begin{center}
         \includegraphics[width=0.65\textwidth]{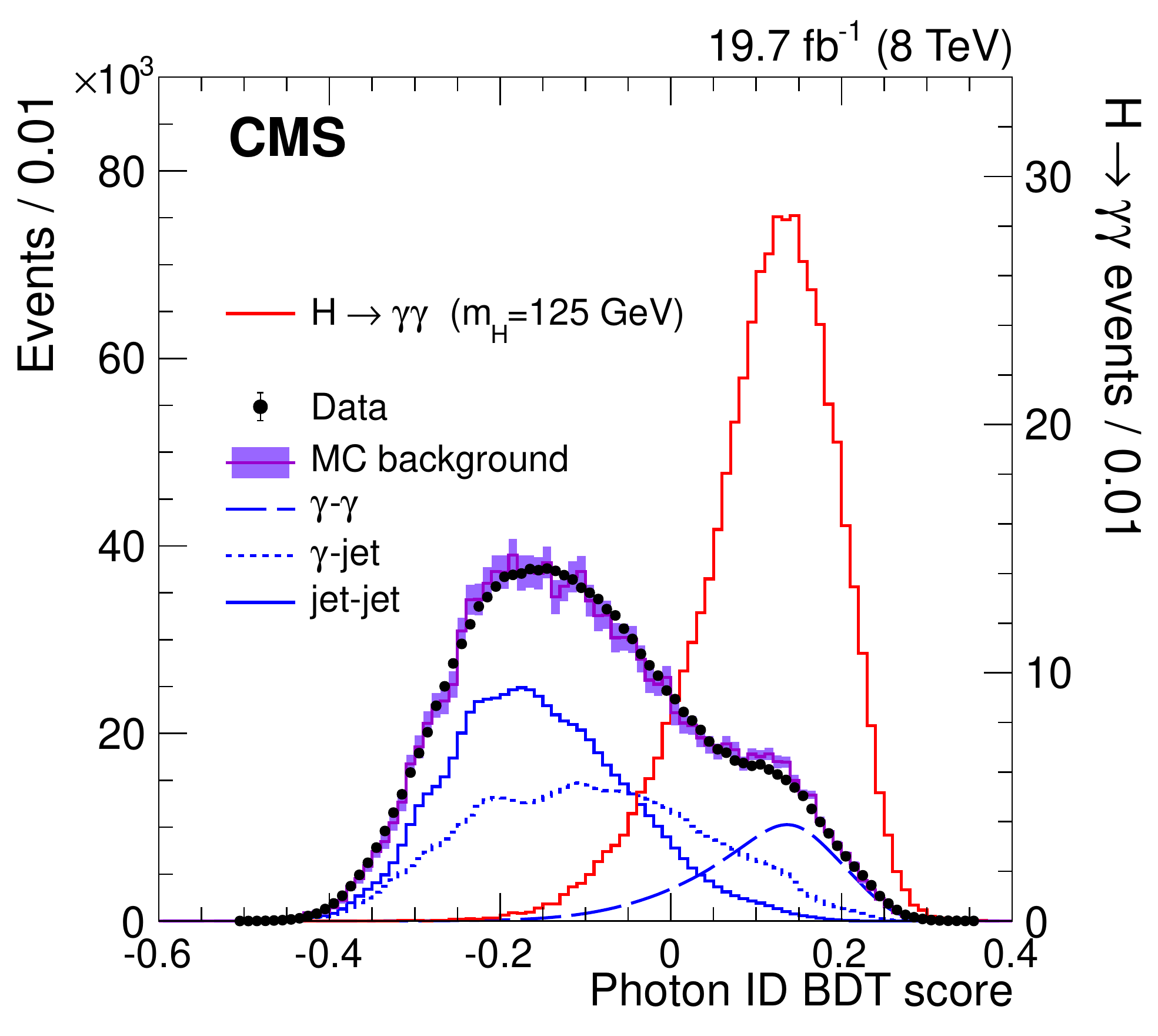}
      \caption{Photon identification BDT score of the lower-scoring photon of the diphoton pairs with invariant masses
  in the range $100<\mgg<180\GeV$, for events passing the $\Hgg$ preselection in the 8\TeV dataset (points with error bars),
  and for simulated background events (histogram with shaded error bands showing the statistical uncertainty).
  The solid line histogram on the right (righthand vertical axis) is for simulated Higgs boson signal events.
  Histograms are also shown for different components of the simulated background, in which there are either two, one, or
  zero prompt signal-like photons.
}
      \label{fig:mva_train_test}
 \end{center}
\end{figure}

The separation of signal and background can be seen in Fig.~\ref{fig:mva_train_test}.
The figure shows the photon identification BDT score of the lower-scoring photon in diphoton pairs with an invariant mass in
the range $100<\mgg<180\GeV$ for diphoton events passing the preselection in the 8\TeV dataset and for simulated background events
(histogram with shaded error bands showing the statistical uncertainty).
The relative fractions of diphoton pairs arising from $\gamma\textendash\gamma$, $\gamma\textendash$jet, and jet$\textendash$jet processes in the MC sample is the result of using the cross sections and $K$-factors described in Section~\ref{sec:dat-sim}.
The tall histogram corresponds to simulated Higgs boson events ($\mH=125\GeV$).
The distribution of the photon identification BDT score of the lower-scoring photon for simulated diphoton
background events also agrees well with the distribution seen in the data.
The bump that can be seen in both distributions at a BDT score of about 0.13 corresponds
to events where both photons are prompt and, therefore, signal-like.

\begin{figure} [hbtp]
  \begin{center}
        \includegraphics[width=0.49\textwidth]{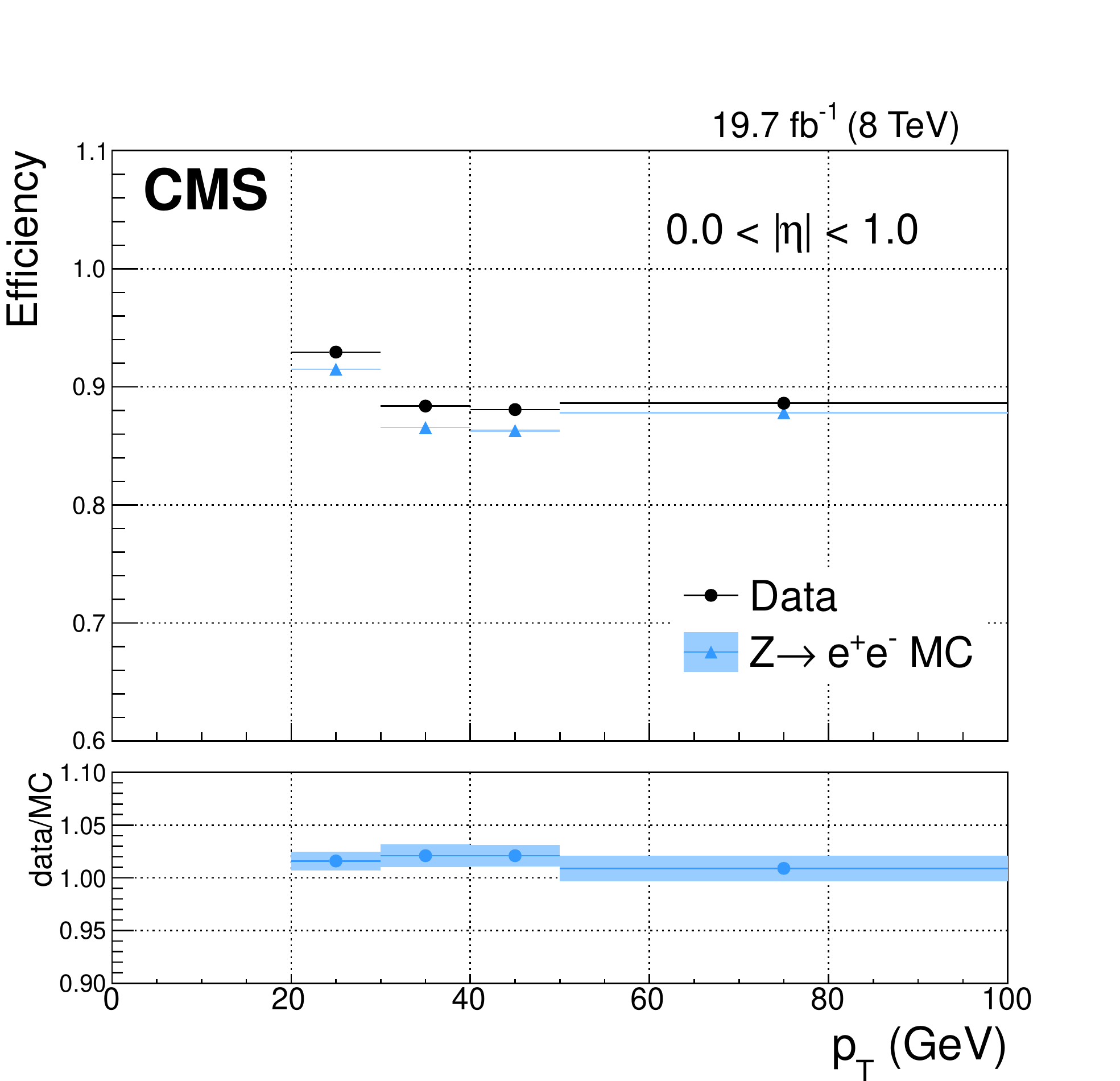}
        \includegraphics[width=0.49\textwidth]{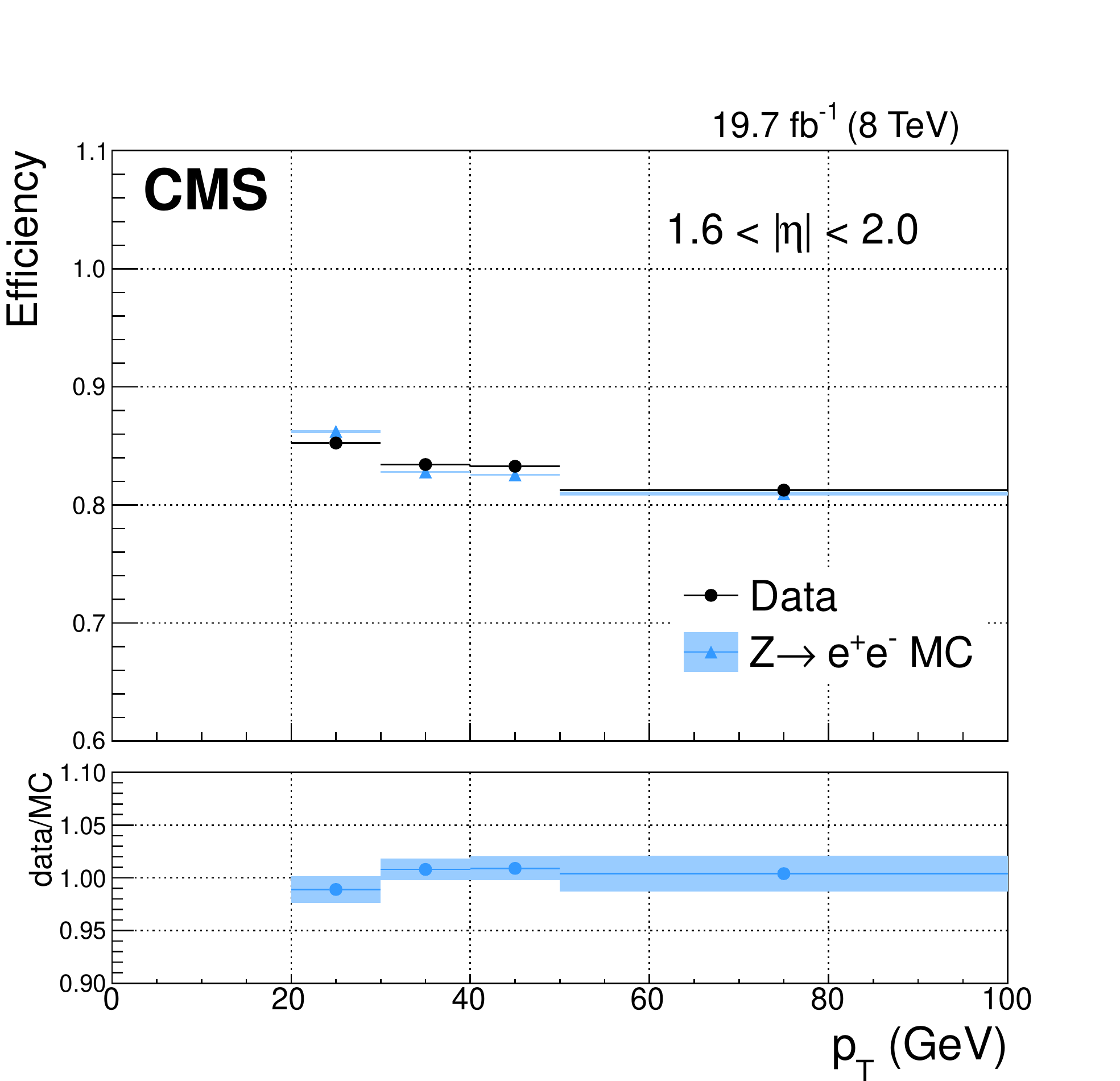}
        \caption{Selection efficiency, as a function of $\pt$, for a particular (example) requirement on the photon identification BDT score.
        The efficiency is measured with the tag-and-probe technique in \Zee events where the electrons are reconstructed as photons,
        for photons in the (left) central barrel, $\abs{\eta} < 1$, and (right) outer endcap, $1.6 < \abs{\eta} < 2$.
        The systematic uncertainties in the tag-and-probe efficiency measurements are shown by shaded bands. The error bars representing the statistical uncertainties are too small to be visible.}
      \label{fig:mva_scaleFactors}
 \end{center}
\end{figure}

If a simple requirement is made on the BDT score of photon candidates,
defining a working point with a signal efficiency of about 80\%,
the signal and background efficiencies are found to be flat as a function
of the photon transverse momentum and the number of vertices in the event,
for both ECAL barrel and endcaps.
The identification efficiency obtained by making such a requirement on the photon identification BDT score
has also been measured in data with the tag-and-probe technique in \Zee events
(reconstructing the electrons from the Z boson as photons).
The tag photon is required to have $\pt>35\GeV$ and the BDT score is required to be $>0.15$.
Figure~\ref{fig:mva_scaleFactors} shows the data-to-simulation comparison of the efficiencies as a function
of the probe photon transverse momentum for $\abs{\eta} < 1$ and $1.5 < \abs{\eta} < 2$ separately.
The $\pm$0.01 systematic uncertainty assigned to the BDT score in simulation covers, together
with the systematic uncertainty in the tag-and-probe efficiency measurements, the
residual difference observed between the efficiencies measured in data and simulation.

\begin{figure}[htbp]
  \begin{center}
         \includegraphics[width=0.49\textwidth]{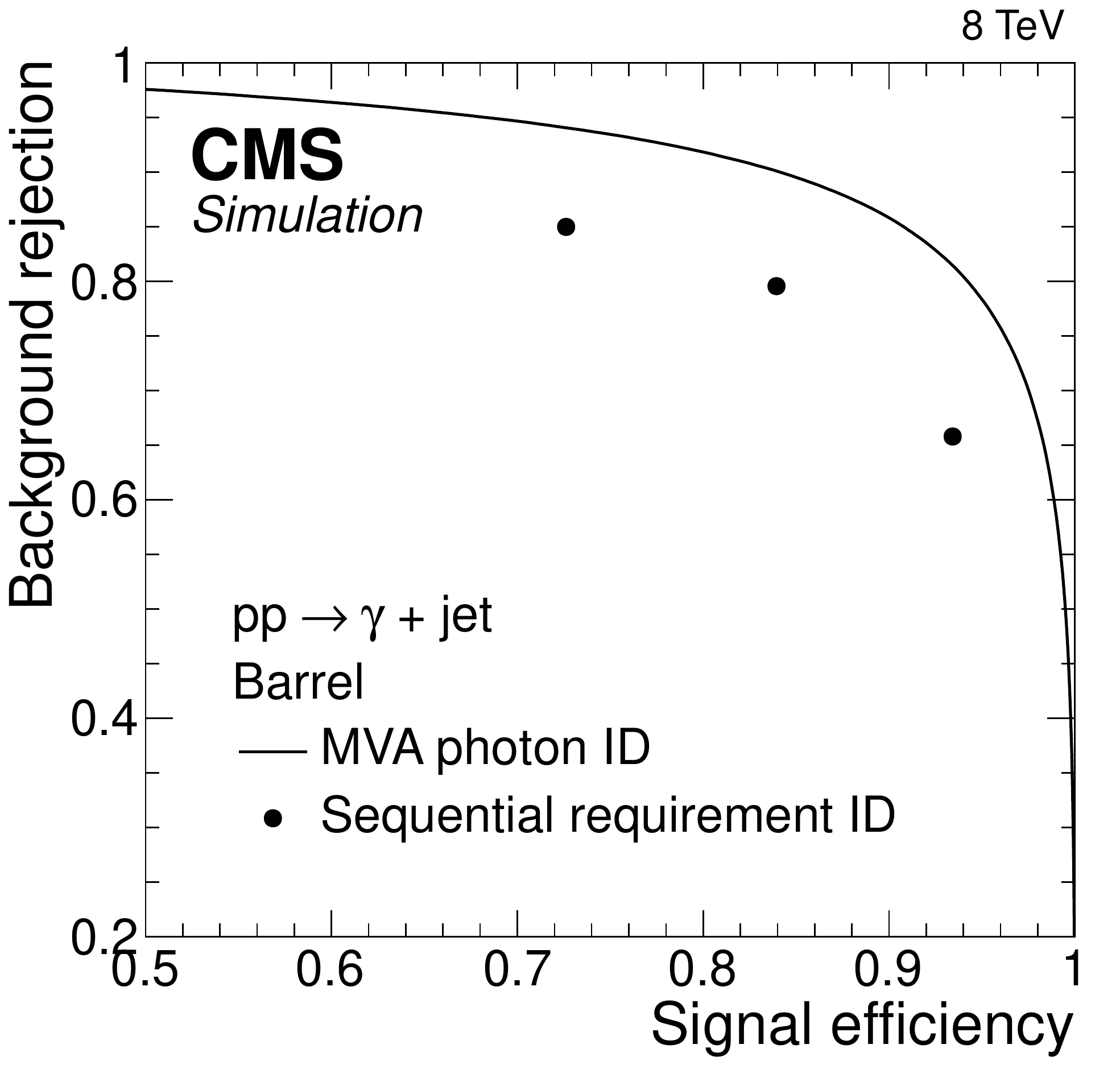}
         \includegraphics[width=0.49\textwidth]{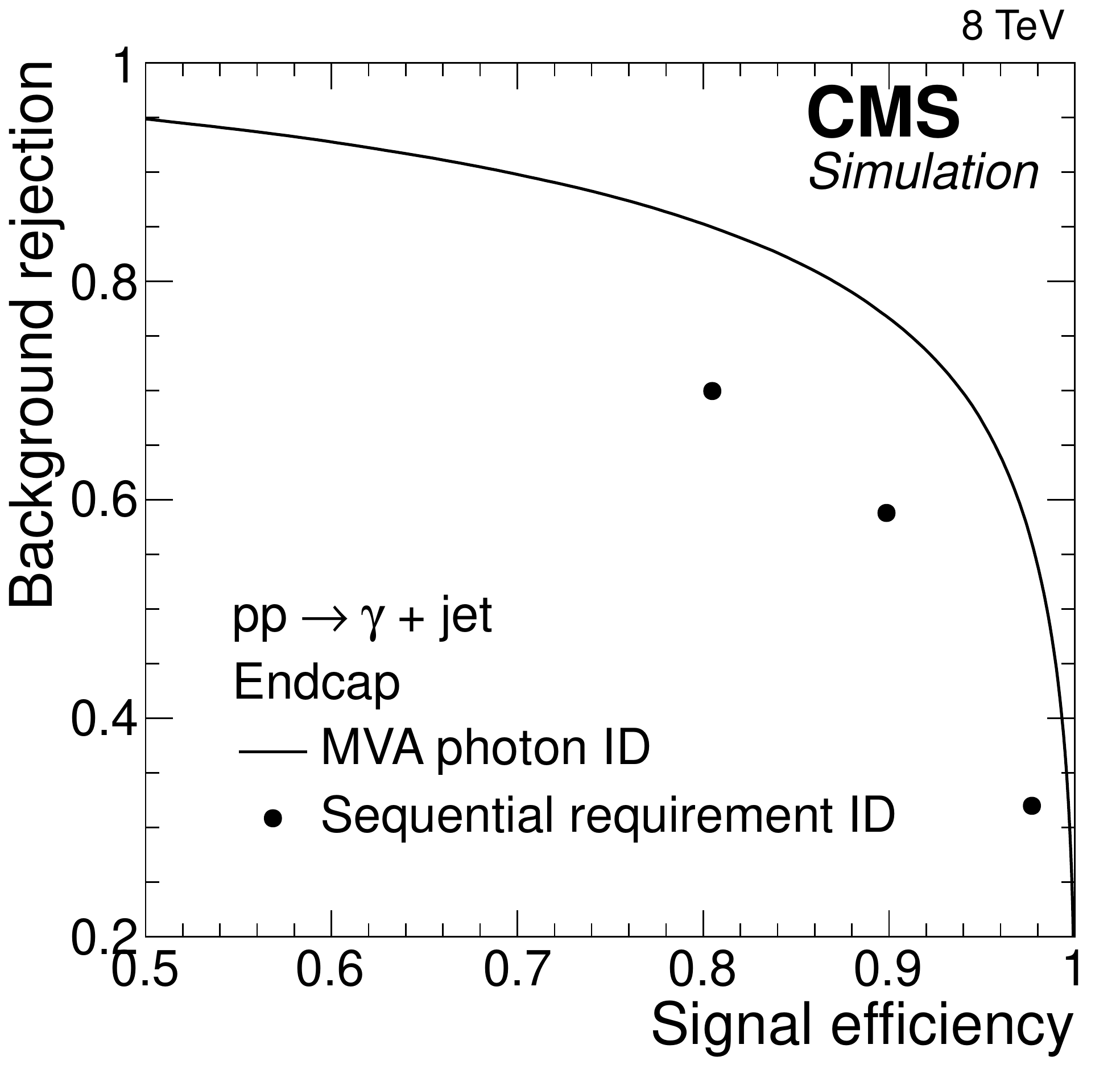}
      \caption{Background rejection versus signal efficiency for both sequential requirement (points) and multivariate (MVA) (curve)
                     identification techniques in the (left) ECAL barrel and (right) endcap for simulated $\GAMJET$ events.
                     The signal is the prompt photon and background are jets with a large electromagnetic component.
                     A loose selection is first applied to both the simulated event samples and the values of efficiency and background rejection are
                     relative to that (see text).}
      \label{fig:compare_cutbased_and_mva}
 \end{center}
\end{figure}

The dependence of background rejection on signal efficiency as the requirement
on the photon identification BDT score is varied, is shown in Fig.~\ref{fig:compare_cutbased_and_mva}.
The signals are reconstructed photons matched to prompt photons in simulated $\GAMJET$ events.
The background are photons reconstructed in simulated dijet events.
The loose preselection defined in Table~\ref{table:preselCuts} is applied to all photon candidates, so the background rejection and signal efficiency shown are relative to this preselection.
The signal transverse momentum distribution is reweighted to follow the background spectrum,
and photon candidates are required to satisfy $25<\pt<200\GeV$.
The figure also shows the background rejection and signal efficiency of the three working points of the photon identification using sequential requirements.
The multivariate selection can be seen to have better performance, arising from the use of additional information, including
the correlation among variables.

\clearpage
\section{Summary}
\label{sec:summary}
A description has been provided of the performance of the CMS detector for photon reconstruction and identification
in proton-proton collisions at a centre-of-mass energy of $8\TeV$ at the CERN LHC.
Details are given of the reconstruction of photons from energy deposits in the ECAL
and of the extraction of photon energy estimates.
The reconstruction of electron tracks from photons that convert to $\Pep\Pem$-pairs in the CMS tracker is also described,
as is the optimization of the photon energy reconstruction and its accurate modelling in simulation
for the analysis of the Higgs boson decay into two photons.
The excellent agreement between data and simulation, demonstrated for electron showers, enables
the extraction of an accurate estimate of the energy resolution of photons from $\Hgg$
decays in data.
In the barrel section of the ECAL, an energy resolution of about 1\% is achieved for unconverted or late-converting
photons arising from the $\Hgg$ decay.
The remaining barrel photons have a resolution of about 1.3\% up to $\abs{\eta}=1$, rising to about 2.5\% at $\abs{\eta}=1.4$.
In the endcaps, the resolution of unconverted or late-converting photons from the same sample
is about 2.5\%, while the remaining endcap photons have a resolution of somewhat worse than 3\%.

The photon energy scale uncertainty and its impact on the Higgs boson mass measurement are discussed in depth.
Since the scale is set using the showers of electrons from $\Zee$ decays reconstructed as photons, the largest uncertainties are due to the possible imperfect simulation of
(i) differences in detector response to electrons and photons, and (ii) energy scale nonlinearity between the energies
typical of electrons from the Z boson decay and photons from the Higgs boson decay.
Results of measurements of the material thickness of the tracker are shown, together with a comparison between data and simulated events of
the energy response as a function of $\ET$.
The uncertainty in the photon energy scale at $\pt\approx m_\cPZ/2$,
is about 0.1\% in the central barrel, 0.15\% in the outer barrel, and 0.3\% in the endcaps.

Different photon identification methods are discussed, and their corresponding
selection efficiencies in data are compared with those found in simulated events.
For the two photon identification methods considered, the agreement between data and simulation for the efficiency
as a function of photon $\pt$ is found to be good.
Comparing the background rejection as a function of signal efficiency, the multivariate selection has somewhat better performance,
resulting from the use of additional information including the correlation among variables.

\begin{acknowledgments}
\hyphenation{Bundes-ministerium Forschungs-gemeinschaft Forschungs-zentren} We congratulate our colleagues in the CERN accelerator departments for the excellent performance of the LHC and thank the technical and administrative staffs at CERN and at other CMS institutes for their contributions to the success of the CMS effort. In addition, we gratefully acknowledge the computing centres and personnel of the Worldwide LHC Computing Grid for delivering so effectively the computing infrastructure essential to our analyses. Finally, we acknowledge the enduring support for the construction and operation of the LHC and the CMS detector provided by the following funding agencies: the Austrian Federal Ministry of Science, Research and Economy and the Austrian Science Fund; the Belgian Fonds de la Recherche Scientifique, and Fonds voor Wetenschappelijk Onderzoek; the Brazilian Funding Agencies (CNPq, CAPES, FAPERJ, and FAPESP); the Bulgarian Ministry of Education and Science; CERN; the Chinese Academy of Sciences, Ministry of Science and Technology, and National Natural Science Foundation of China; the Colombian Funding Agency (COLCIENCIAS); the Croatian Ministry of Science, Education and Sport, and the Croatian Science Foundation; the Research Promotion Foundation, Cyprus; the Ministry of Education and Research, Estonian Research Council via IUT23-4 and IUT23-6 and European Regional Development Fund, Estonia; the Academy of Finland, Finnish Ministry of Education and Culture, and Helsinki Institute of Physics; the Institut National de Physique Nucl\'eaire et de Physique des Particules~/~CNRS, and Commissariat \`a l'\'Energie Atomique et aux \'Energies Alternatives~/~CEA, France; the Bundesministerium f\"ur Bildung und Forschung, Deutsche Forschungsgemeinschaft, and Helmholtz-Gemeinschaft Deutscher Forschungszentren, Germany; the General Secretariat for Research and Technology, Greece; the National Scientific Research Foundation, and National Innovation Office, Hungary; the Department of Atomic Energy and the Department of Science and Technology, India; the Institute for Studies in Theoretical Physics and Mathematics, Iran; the Science Foundation, Ireland; the Istituto Nazionale di Fisica Nucleare, Italy; the Ministry of Science, ICT and Future Planning, and National Research Foundation (NRF), Republic of Korea; the Lithuanian Academy of Sciences; the Ministry of Education, and University of Malaya (Malaysia); the Mexican Funding Agencies (CINVESTAV, CONACYT, SEP, and UASLP-FAI); the Ministry of Business, Innovation and Employment, New Zealand; the Pakistan Atomic Energy Commission; the Ministry of Science and Higher Education and the National Science Centre, Poland; the Funda\c{c}\~ao para a Ci\^encia e a Tecnologia, Portugal; JINR, Dubna; the Ministry of Education and Science of the Russian Federation, the Federal Agency of Atomic Energy of the Russian Federation, Russian Academy of Sciences, and the Russian Foundation for Basic Research; the Ministry of Education, Science and Technological Development of Serbia; the Secretar\'{\i}a de Estado de Investigaci\'on, Desarrollo e Innovaci\'on and Programa Consolider-Ingenio 2010, Spain; the Swiss Funding Agencies (ETH Board, ETH Zurich, PSI, SNF, UniZH, Canton Zurich, and SER); the Ministry of Science and Technology, Taipei; the Thailand Center of Excellence in Physics, the Institute for the Promotion of Teaching Science and Technology of Thailand, Special Task Force for Activating Research and the National Science and Technology Development Agency of Thailand; the Scientific and Technical Research Council of Turkey, and Turkish Atomic Energy Authority; the National Academy of Sciences of Ukraine, and State Fund for Fundamental Researches, Ukraine; the Science and Technology Facilities Council, UK; the US Department of Energy, and the US National Science Foundation.

Individuals have received support from the Marie-Curie programme and the European Research Council and EPLANET (European Union); the Leventis Foundation; the A. P. Sloan Foundation; the Alexander von Humboldt Foundation; the Belgian Federal Science Policy Office; the Fonds pour la Formation \`a la Recherche dans l'Industrie et dans l'Agriculture (FRIA-Belgium); the Agentschap voor Innovatie door Wetenschap en Technologie (IWT-Belgium); the Ministry of Education, Youth and Sports (MEYS) of the Czech Republic; the Council of Science and Industrial Research, India; the HOMING PLUS programme of Foundation for Polish Science, cofinanced from European Union, Regional Development Fund; the Compagnia di San Paolo (Torino); the Consorzio per la Fisica (Trieste); MIUR project 20108T4XTM (Italy); the Thalis and Aristeia programmes cofinanced by EU-ESF and the Greek NSRF; and the National Priorities Research Program by Qatar National Research Fund.
\end{acknowledgments}
\bibliography{auto_generated}   

\cleardoublepage \appendix\section{The CMS Collaboration \label{app:collab}}\begin{sloppypar}\hyphenpenalty=5000\widowpenalty=500\clubpenalty=5000\textbf{Yerevan Physics Institute,  Yerevan,  Armenia}\\*[0pt]
V.~Khachatryan, A.M.~Sirunyan, A.~Tumasyan
\vskip\cmsinstskip
\textbf{Institut f\"{u}r Hochenergiephysik der OeAW,  Wien,  Austria}\\*[0pt]
W.~Adam, T.~Bergauer, M.~Dragicevic, J.~Er\"{o}, M.~Friedl, R.~Fr\"{u}hwirth\cmsAuthorMark{1}, V.M.~Ghete, C.~Hartl, N.~H\"{o}rmann, J.~Hrubec, M.~Jeitler\cmsAuthorMark{1}, W.~Kiesenhofer, V.~Kn\"{u}nz, M.~Krammer\cmsAuthorMark{1}, I.~Kr\"{a}tschmer, D.~Liko, I.~Mikulec, D.~Rabady\cmsAuthorMark{2}, B.~Rahbaran, H.~Rohringer, R.~Sch\"{o}fbeck, J.~Strauss, W.~Treberer-Treberspurg, W.~Waltenberger, C.-E.~Wulz\cmsAuthorMark{1}
% ...
\vskip\cmsinstskip
% ...
1:~~Also at Vienna University of Technology, Vienna, Austria\\
2:~~Also at CERN, European Organization for Nuclear Research, Geneva, Switzerland\\
% ...

\end{sloppypar}
\end{document}